\documentclass[ reprint,superscriptaddress,  amsmath,amssymb,  aps,floatfix]{revtex4-1}%
\usepackage{graphicx}
\usepackage{dcolumn}
\usepackage{bm}
\usepackage{amsmath}
\usepackage{amsfonts}
\usepackage{amssymb}%
\usepackage{xcolor}

\usepackage{MnSymbol}
\usepackage{natbib}

\renewcommand{\Re}{\operatorname{Re}}
\renewcommand{\Im}{\operatorname{Im}}

\newcommand{\bra}[1]{\langle{#1}|}			
\newcommand{\ket}[1]{|{#1}\rangle}

\newtheorem{Thm}{Theorem}

\begin{document}

\title{Verifiable homodyne measurement for detecting non-local properies of light}

\author{Go Kato}%
\email{go.kato@nict.nict.go.jp}%
\affiliation{
National Institute of Information and Communications Technology,\\
4-2-1, Nukui-Kita, Koganei, Tokyo, 184-8795 Japan}%

\author{Kiyoshi Tamaki}%
\affiliation{
Faculty of Engineering, University of Toyama,\\
Gofuku 3190, Toyama, Toyama 930-8555, Japan
}%

\author{Masaki Owari}%
\affiliation{
Faculty of Informatics, Shizuoka University,\\
3-5-1, Johoku, Naka-ku, Hamamatsu, Shizuoka 432-8011, Japan}%

\author{Koji Azuma}%
\affiliation{
NTT Basic Research Laboratories, NTT Corporation,\\
3-1, Moronisato Wakamiya, Atsugi, Kanagawa 243-0198, Japan}%
\affiliation{
NTT Research Center for Theoretical Quantum Physics, NTT Corporation,\\
3-1 Morinosato-Wakamiya, Atsugi, Kanagawa 243-0198, Japan}
\date{\today}%

\begin{abstract}%

The homodyne detection is one of the most basic tools for identifying the quantum state of light.
It has been used to detect useful non-local properties, such as entanglement for the quantum teleportation and 
distillability of a secret key in quantum key distribution.
In so doing, the detection scheme employs a bright optical pulse, called the local oscillator (LO) pulse, and the LO pulse is usually transmitted along with the signal pulses.
The LO pulse is presumed to be a coherent state with an infinite intensity.
 However, it is difficult in practice to hold this presumption owing to noise in the optical transmission channels or an intervention by a malicious third party. 
As a result, the implementation may no longer be the homodyne detection, and those outcomes may merely disguise successful detection of entanglement or a secret key.
Here, we present an alternative scheme that 
works as the homodyne detection to detect the non-local properties of light in a verifiable manner,
without any presumption for the LO pulses.
This scheme is essentially based on the same setup as the conventional implementation for the homodyne detection.
This result contributes to close any possible loophole in the homodyne detection caused by the deviation from the ideal LO pulses.
\end{abstract}

\maketitle

\section{Introduction}

The homodyne detection is implemented by making the signal light interfere with the local oscillator (LO)  pulse in a coherent state with an (ideally infinitely) large amplitude.
This way, many fundamental experiments~\cite{LR09,OPK92,BMP08,JWM09,PTS13,RAJ14,MMP15,MPB09,BCF10, JWK11,FSB98,YUS13,SHD13,HSD08,GSN11,ZFB11,CWA14,UFP15}, ranging from 
the field of quantum optics to quantum information, have successfully been performed. 
In the continuous-variable quantum key distribution (CVQKD), 
the homodyne detection is used to generate a secret key by exploiting the infinite dimensionality of light~\cite{GAW03,PML08,MUL12,JJL13,POS15}, which has already been ready for practical use~\cite{LCT14}.
Therefore, the homodyne detection is one of the most basic tools for detecting or utilizing quantum properties of light, and its implementation is the foundation of the optical quantum information processing as well as exploring quantum optical phenomena.

\begin{figure*}[tb]
\begin{center}
\includegraphics[width=16cm]{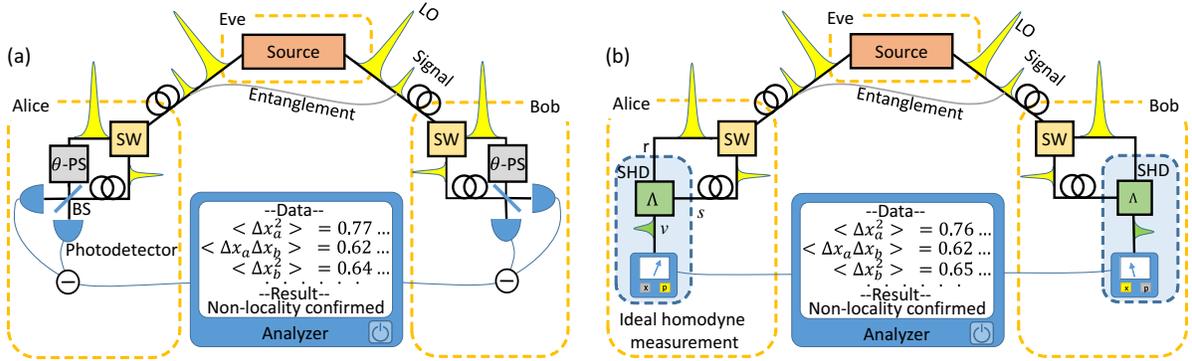}
\end{center}
\caption{
Test of a non-local property of light.
Panel (a) shows Alice and Bob who employ the conventional homodyne detection scheme to check the existence of non-locality for the input signals.
The conventional homodyne detection is composed of an optical switch (SW), a $\theta$-phase shifter ($\theta$-PS), a 50:50 beamsplitter (BS) and photodetectors.
Unfortunately,  ``the successful detection'' of the non-local property using this system may not be reliable because the states of LO pulses are deviated from the ideal ones owing to disturbances to the LO pulses.
In contrast, if Alice and Bob employ the SHD composed of a local ``squashing'' quantum operation $\Lambda_{sr\to v}$ and the ideal homodyne detection as in panel (b),
the situation becomes different. 
If the SHD informs Alice and Bob of the successful detection of the non-local property, this
is so irrespectively of disturbances for LO pulses, because the non-locality is confirmed by the ideal homodyne detection.
Here, in the squashing quantum operation $\Lambda_{sr\to v}$, the signal pulse $s$ and the LO pulse $r$ are the input systems.
}
\label{fig:ideal_LHD}
\end{figure*}

Implemented exactly as the theory requires, the homodyne detection can faithfully accomplish the tasks as we expect. 
Unfortunately, however, 
 the implementation may be deviated from the ideal homodyne detection, especially when we use it for detecting non-local properties of light. To see this, let us consider an example of a conventional experiment to detect a non-local property of light in Fig.~\ref{fig:ideal_LHD}(a)~\cite{OPK92}.
 Here, separated parties, Alice and Bob, argue that a non-local property of light sent by a third party (Eve) is confirmed from the data obtained by the conventional implementation of the homodyne detection.
This argument is true if the LO pulses from Eve are not disturbed at all and the states are in the ideal states, that is, a coherent state with an infinitely large amplitude.
However, this presumption is difficult to satisfy or even impossible to verify in practice, owing to noise in the transmission of the LO pulses or an intervention to the LO pulses by an eavesdropper (perhaps by Eve herself).
Hence, the implementation is not robust against such a disturbance and an imperfection, which could lead to a loophole of the experimental demonstrations of entanglement detection and CVQKD~\cite{MSJ13}.

To make the implementation more robust, in~\cite{PhysRevX.5.041009} they proposed to employ another LO pulse locally generated in a coherent state at the receiver side.
This local LO pulse is made phase-locked to the received LO pulse through an interferometric measurement, with an active feedforward technique.
Thanks to this phase-locked local LO pulse, this implementation guarantees that the employed LO pulse is in a coherent state. 
However, besides the complication in implementing the active feedfoward, 
the intensity of the locally generated LO pulse is still finite in contrast to what the theory requires.  
To make matters worse, this gap between the implementation and the ideal homodyne measurement is not quantified, opening up the room of disguising the successful detection of entanglement or a secret key.

In this paper, we present an implementation of the homodyne detection for detecting non-local properties of light without making any unverifiable presumption for the LO pulses.
Our main idea 
is explained as follows.
We start with introducing
 an idealized detection scheme referred to as ``squashing 
 homodyne detection (SHD)'', which is purely a theoretical 
measurement model
 and we do not need to implement in reality. 
The SHD is composed of a ``squashing'' quantum operation $\Lambda_{sr\to v}$ which squashes two input modes $sr$, the signal light $s$ and the LO pulse $r$, into a single mode $v$, followed by the ideal homodyne detection on the single mode $v$. Thanks to the ideal homodyne detection in the SHD and the monotonicity of entanglement under the squashing operation, it can faithfully detect a non-local property of light (see Fig.~\ref{fig:ideal_LHD}(b)). The implementation of the SHD itself is challenging, but fortunately, it turns out that an experimental setup close to the SHD can be implemented by slightly modifying the conventional setup for the homodyne detection. This approximated implementation is enough for our purpose 
because our theory developed here allows us to rigorously estimate the statistics which we could have obtained if we had performed the SHD.
This way, we can estimate and confirm the non-local property using the data obtained from our modified experimental setup.
We also derive analogous results applicable to implement the heterodyne measurement by using LO pulses, rather than homodyne measurement.
Therefore, our work closes any loophole in LO pulses by using practical devices, which paves a way to accomplishing quantum information processing in an unconditional manner.

This manuscript is structured as follows. In the next section, we define the 
SHD, and in Sec. \ref{sec:imp}, we show how to implement it with practical devices. In Sec. \ref{sec:clo}, we present analytical formula that estimates how close the 
SHD and the implemented SHD are.  In Sec. \ref{sec:app}, we 
present applications of our technique. The last section is devoted to discussion and conclusions.

\section{Definition of SHD}
\label{sec:def}
As we have mentioned, the SHD performs a squashing operation $\Lambda_{sr\to v}$ which 
squashes the input of two modes $sr$, the signal light $s$ and the LO pulse $r$, in a state $\hat {\rho}_2$ into a single mode $v$,
and then applies to the single mode $v$ the ideal homodyne measurement to measure a quadrature $\hat{x}_v(\theta):=\hat{x}_v\cos \theta + \hat{p}_v\sin \theta$, where $\hat{x}_v$ and $\hat{p}_v$ are quadratures for mode $v$ with $[\hat{x}_v,\hat{p}_v]=i/2$. Hence, the probability with which the single mode $v$ is found in an eigenstate $\ket{ x(\theta)}_v$ of the quadrature $\hat{x}_v(\theta)$ by the homodyne measurement is given by ${}_v\bra{  x(\theta)}\Lambda_{sr\to v}(\hat \rho_2)\ket{ x(\theta)}_v$.
The squashing operation $\Lambda_{sr\to v}$ is a 
completely positive trace-preserving (CPTP) map defined by 
\begin{align}
&\Lambda_{sr\to v}(\hat{\rho_2})
\nonumber\\
&\quad=\sum_{m=0}^\infty \sum_{n,n'=0}^m 
{}_{sr}\bra{n,m-n} \hat{\rho}_2\ket{n',m-n'}_{sr}\ket{n}_{v} {}_{v}\bra{n'},
\label{eq:lambda}
\end{align}
where
$\{\ket {l,m}_{sr}\}_{l,m=0,1,\cdots}$ and $\{\ket{n}_{v}\}_{n=0,1,\cdots}$  represent the number states of the input two modes $sr$ and the output single mode $v$, respectively.

\begin{figure}[tb]
\begin{center}
\includegraphics[width=8cm]{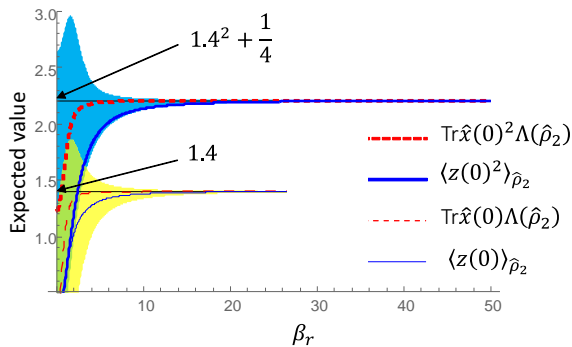}
\end{center}
\caption{
Closeness of the ideal homodyne detection, the SHD, and our implementation of the SHD. We evaluate 
the expected value of an output and its square for the three detectors
when the input two-mode state is
 $\hat{\rho}_2=\ket{\alpha_{ s}}\bra{\alpha_{ s}}_{ s}\otimes \ket{\beta_{ r}}\bra{\beta_{ r}}_{ r}$, 
where
a signal pulse is in a coherent state $\ket{\alpha_{ s}}_{ s}$ with $\alpha_{ s}=1.4$ and
 a LO pulse is in a coherent state $\ket{\beta_{ r}}_{ r}$.
Two black horizontal lines correspond to the case of the ideal homodyne detection, i.e., 
${\rm Tr}[(\hat x_s(0)\otimes {\hat I})\hat \rho_2]=1.4$ and ${\rm Tr}[(\hat x_s(0)^2\otimes {\hat I})\hat \rho_2]=1.4^2+1/4$. 
The thin (thick) red dotted line indicates the values for the SHD, i.e., ${\rm Tr}[\hat x_v(0)\Lambda_{sr\to v}(\hat \rho_2)]$ (${\rm Tr}[\hat x_v(0)^2\Lambda_{sr\to v}(\hat \rho_2)]$).
The thin (thick) blue line represents
the values for our implemented SHD, i.e.,  
 $\langle z(0)\rangle_{\hat{\rho}_2}$ ($\langle z(0)^2\rangle_{\hat{\rho}_2}$).
The yellow (skyblue) region describes the region $[\langle z(0)\rangle_{\hat{\rho}_2}-0.525\langle d_{\rm hom} \rangle_{\hat{\rho}_2}, \langle z(0)\rangle_{\hat{\rho}_2}+0.525\langle d_{\rm hom} \rangle_{\hat{\rho}_2}]$ ($[\langle z(0)^2\rangle_{\hat{\rho}_2}-0.162\langle d_{\rm hom} \rangle_{\hat{\rho}_2}, \langle z(0)^2\rangle_{\hat{\rho}_2}+1.085\langle d_{\rm hom} \rangle_{\hat{\rho}_2}]$), which is depicted by using our implemented SHD, and the region must sandwich the target ${\rm Tr}[\hat x_v(0)\Lambda_{sr\to v}(\hat \rho_2)]$ (${\rm Tr}[\hat x(0)^2\Lambda_{sr\to v}(\hat \rho_2)]$) according to Eq.~(\ref{res:LH_1}) (Eq.~(\ref{res:LH_2})). This figure implies that, if the photon number in the LO pulse is more than a few hundred,
the first and second moments of the ideal homodyne detection are approximated by those of the SHD, which are tightly bounded by expected values given by the implementation of the SHD.
}
\label{fig:relation}
\end{figure}

When the input state $\hat{\rho}_2$ is equal to $\ket{\psi}\bra{\psi}_{s}\otimes \ket{\beta_{ r}}\bra{\beta_{ r}}_{r}$ where $\ket{\beta_{ r}}_{r}$ is a coherent state with a positive amplitude $\beta_{ r}$ in mode $ r$ and an arbitrary state $\ket{\psi}_{ s}=\sum_{n=0}^{\bar n} \nu_n \ket{n}_{s}$ with $|\beta_{ r}|^2\gg \bar n$ for a certain integer $\bar n$,  the output $\Lambda_{sr\to v}(\hat\rho_2)$ is close to $\ket{\psi}\bra{\psi}_{ s}$.
Therefore,
the SHD for such an input state $\hat\rho_2=\ket{\psi}\bra{\psi}_{s}\otimes \ket{\beta_{ r}}\bra{\beta_{ r}}_{r}$ is approximately equivalent to the ideal homodyne detection for the state $\ket{\psi}_{s}$, as shown in Fig.~\ref{fig:relation}. 

Similarly, the SHDs for any state  $\hat{\rho}_{2N}$  consisting of $N$ pairs of signal pulses and LO pulses can be defined. 
That is, the probability with which the $j$-th pair is found in an eigenstate $\ket{x(\theta_j)}_{v_j}$ of a quadrature $\hat{x}_{v_j}(\theta_j):=\hat{x}_{v_j} \cos \theta_j + \hat{p}_{v_j} \sin \theta_j$---where $\hat{x}_{v_j}$ and $\hat{p}_{v_j}$ are quadratures for the single output mode of the squashing operation $\Lambda_{s_jr_j\to v_j}$ for the $j$-th pair, with $[\hat{x}_{v_j},\hat{p}_{v_{j'}} ]=i\delta_{j,j'}/2$ for the Kronecker's delta $\delta_{j,j'}$---is given by 
$(\bigotimes_j{}_{v_j}\bra{{x}(\theta_j)}) \Lambda^{\otimes N} (\hat{\rho}_{2N})(\bigotimes_j\ket{{x}(\theta_j)}_{v_j})$ where $\Lambda^{\otimes N}:=\bigotimes_j\Lambda_{s_jr_j\to v_j}$.

\section{An implementation of SHD}
\label{sec:imp}
\begin{figure*}[tb]
\begin{center}
\includegraphics[scale=0.4]{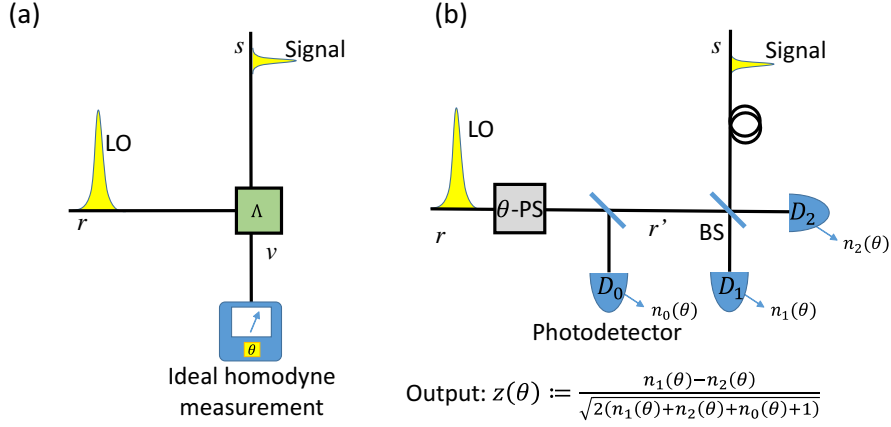}
\end{center}
\caption{
The squashing 
homodyne detection. The panel (a) indicates the ideal case, that is the SHD. The input state, a pair of the signal pulse and the reference pulse, is squashed into a single-mode state with the squashing quantum operation $\Lambda_{sr\to v}$ followed by ideal homodybe detection. 
The panel (b) indicates our implementation for the SHD. This is composed of 50:50 beamsplitters (BSs), a $\theta$-phase shifter ($\theta$-PS), and photodetectors.
Without any assumption on the input state,
we can confirm the 
closeness in term of the statistics of the  outcomes
 between the SHD and its implementation.
}
\label{fig:implementation}
\end{figure*}

Our implementation close to the SHD (see Fig.~\ref{fig:implementation}(b)), which is characterized by the parameter $\theta\in [0,2\pi)$, is based on the experimental setup composed of 50:50 beamsplitters (BSs), a phase shifter, and photodetectors.
First, the LO pulse is subjected to a $\theta$-phase shift,
 and it splits into two pulses by a BS, one of which is directly measured with a photodetector $D_0$.
 Then, the signal pulse and the other half of the LO pulse interfere with each other by the other BS, followed by the detection with photodetectors $D_1$ and $D_2$.
Let $n_k(\theta)$ be the number of photons detected by $D_k$ ($k\in\{0,1,2\}$) 
for a selection of $\theta$. The 
 outcome~\footnote{Even if the $-\theta$-phase shift is performed on the signal pulse instead of $\theta$-phase shift on the LO pulse, the output of the implementation is unchanged.  
This is so because the measurement commutes with the operator of the total photon number in the signal pulse and  the LO pulse.} of our implementation is represented by 
\begin{equation}
z(\theta):= \frac {n_1(\theta)-n_2(\theta)}{\sqrt{2(n_1(\theta)+n_2(\theta)+n_0(\theta)+1)}}. 
\label{eq:x_phi}
\end{equation}
In the followings, the expected value of the outcome for the two-mode input $\hat\rho_2$ is expressed by $\left<z(\theta)\right>_{\hat \rho_2}$.

Having finished defining the SHD as well as explaining its implementation, we have a remark below.
The outcome of  this implementation for $\ket{\psi}_{s}\otimes \ket{\beta_{ r}}_{ r}$ is 
similar to that of the ideal homodyne detector for the signal state $\ket{\psi}_{ s}$
if the LO pulse is in a coherent state $\ket{\beta_{ r}}_{ r}$ with $\beta_{ r}\gg0$. In particular, for large $\beta_{ r}$, the difference $\hat{n}_1(\theta) - \hat{n}_2(\theta)$ of the number operators $\hat{n}_1(\theta)$ and $\hat{n}_2(\theta)$ for the pulses into the detectors $D_1$ and $D_2$ is close to $\beta_{ r} (\hat{a}_{ s}e^{- i\theta}+\hat{a}_{ s }^{ \dag}e^{i\theta})/\sqrt{2}=\sqrt{2} \beta_{ r} \hat{x}_{s}(\theta)$ with the annihilation operator $\hat{a}_{ s}$ and a quadrature $\hat{x}_s(\theta):=(\hat{a}_{ s}e^{- i\theta}+\hat{a}_{ s }^{ \dag}e^{i\theta})/2$ for the signal pulse. On the other hand, $\hat{n}_1(\theta)+\hat{n}_2(\theta)+\hat{n}_0(\theta)+1$ is approximated to $\beta_{ r}^2$ because $\hat{n}_1(\theta)+\hat{n}_2(\theta)+\hat{n}_0(\theta)$ is the total number of photons of the signal pulse and the LO pulse, which can be approximated to $\beta_r^2$ when the intensity of the LO pulse is large. 
 Therefore, the expected value of $\langle z(\theta) \rangle_{\hat{\rho}_2}$ is close to
 ${\rm Tr}\left[(\hat{x}_s(\theta)\otimes \hat I)\hat \rho_2\right]$ in the case of $\hat{\rho}_2=\ket{\psi}\bra{\psi}_s \otimes \ket{\beta_r}\bra{\beta_r}_r$
with a strong LO pulse (see Fig.~\ref{fig:relation}).

The SHDs on $N$ pairs can be implemented in a similar manner.
That is,
the setup for the implementation is just to apply the SHD for a single pair repeatedly to every pair, and the outcome for the $j$-th pair is defined as
\begin{equation}
z^{(j)}(\theta_j):= \frac {n_1^{(j)}(\theta_j)-n_2^{(j)}(\theta_j)}{\sqrt{2(n_1^{(j)}(\theta_j)+n_2^{(j)}(\theta_j)+n_0^{(j)}(\theta_j)+1)}} ,
\label{eq:x_phi_j}
\end{equation}
where $n_k^{(j)}(\theta_j)$ is the number of photons detected by $D_k$ in the implementation  for the $j$-th pair, and $\theta_j$ is the 
 parameter of the phase shifter in the implementation.

\section{Closeness of the SHD and its implementation}
\label{sec:clo}
In this section, we explain our formula that estimates the deviation between the SHD and our implementation. 
For this, we use several expected values, including ones of the SHD ${\rm Tr}[\hat{x}_v(\theta)\Lambda_{sr\to v}(\hat{\rho}_2)]$ and of our implementation  $\langle z(\theta) \rangle_{\hat{\rho}_2}$. Also, we employ correlations described by ${\rm Tr}[\hat{x}_{v_k}(\theta_k)\hat{x}_{v_l}(\theta_l)\Lambda^{\otimes N}(\hat{\rho}_{2N})]$ in the SHD and by $\langle z^{(k)}(\theta_k)z^{(l)}(\theta_l) \rangle_{\hat{\rho}_{2N}}$ in our implementation.
For each of these values, we can obtain an upper bound on the difference between the SHD and its implementation.
It is noteworthy that these upper bounds can be statistically evaluated from the data 
obtained from our implementation without any 
assumptions
on the LO pulse, and  the obtained bounds turn out to be very small for a standard input.

The upper bounds on the difference for a single pair in state $\hat \rho_2$ can be expressed in the following form: 
\begin{Thm}\label{th:1}
For any two-mode input state $\hat \rho_2$, the deviation between the expected values, ${\rm Tr}[\hat{x}_v(\theta)\Lambda_{sr\to v}(\hat{\rho}_2)]$ in the SHD and $\langle z(\theta) \rangle_{\hat{\rho}_2}$ in our implementation, is bounded by
\begin{eqnarray}
|{\rm Tr}[\hat{x}_v(\theta)\Lambda_{sr\to v}(\hat{\rho}_2)]-\langle z(\theta) \rangle_{\hat{\rho}_2}|\le 0.525 
\langle d_{\rm hom}\rangle_{\hat{\rho}_2} ,
\label{res:LH_1}
\end{eqnarray}
where $\langle d_{\rm hom}\rangle_{\hat{\rho}_2} $ is 
the
expected value of the quantity 
\begin{align}
d_{\rm hom} &:= \sum_{\theta\in\{0, \pi/4,\pi/2,3\pi/4\}}\frac {f_{\rm hom}(n_1(\theta),n_2(\theta),n_0(\theta))}{4(n_1(\theta)+n_2(\theta)+n_0(\theta)+1)},
\label{def:d_hom}
\end{align}
for the input $\hat \rho_2$ with 
\begin{align}
f_{\rm hom}(n_1,n_2,n_0)
&:=
\delta_{0,n_{0}}\left[ \frac34(n_{1}+n_{2})^2+\frac76(n_{1}+n_{2})+\frac12\right] 
\nonumber \\&\quad\quad
+\frac{(n_{1}-n_{2})^{4}}{6(n_{0}+1)(n_{0}+2)}.
\label{def:f_hom}
\end{align}

Also, for the square of the output, the deviation is bounded as
\begin{align}
&-0.162 \langle d_{\rm hom} \rangle_{\hat{\rho}_2} 
\nonumber\\&\makebox[2cm]{}
\le {\rm Tr}[\hat{x}_v(\theta)^2\Lambda_{sr\to v}(\hat{\rho}_2)]-\langle z(\theta)^2 \rangle_{\hat{\rho}_2} 
\nonumber\\&\makebox[5cm]{}
\le 1.085 \langle d_{\rm hom}\rangle_{\hat{\rho}_2},
\label{res:LH_2}
\\
&\bigl|{\rm Tr}\bigl [( \hat x_v(\pi/4)^2-\hat x_v(3\pi/4)^2) \Lambda_{sr\to v}(\hat \rho)/2\bigr]
\nonumber\\&\makebox[2cm]{}
-(\bigl\langle  z(\pi/4)^2\bigr\rangle_{\hat \rho}-\bigl\langle  z(3\pi/4)^2\bigr\rangle_{\hat \rho_2})/2\bigr|
\nonumber\\&\makebox[5cm]{}
\leq  0.622\langle d_{\rm hom}\rangle_{\hat \rho_2}.
\label{res:LH_2_m}
\end{align}Note that $(\hat x_v(\pi/4)^2-\hat x_v(3\pi/4)^2)/2=(\hat x_v\hat p_v+\hat p_v\hat x_v)/2$.
\end{Thm}
The proof of this theorem is given in Appendix \ref{sec:proof}.
${\hat x}_v({\theta})^{2}$ with particular choices of $\theta=\pi/4, 3\pi/4$ is considered in Eq.~(\ref{res:LH_2_m}), to derive the expected value of $(\hat x_v\hat p_v+\hat p_v\hat x_v)/2$ which is used for evaluating the covariance matrix of the state $\Lambda_{sr\to v}(\hat \rho_2)$, associated with various applications (such as ones exemplified below).

In the experiment, we expect to have $\langle d_{\rm hom} \rangle_{\hat{\rho}_2} \simeq 0$ for any state $\hat{\rho}_2$ when the number of photons in the LO pulse is much larger than that in the signal pulse.
For example, following a similar approximation to estimate $n_1(\theta)-n_2(\theta)$ in the previous section, we have that 
 if the input state $\hat \rho_2$ is a pure state $\ket{\alpha_{ s}}_{ s}\otimes \ket{\beta_{ r}}_{ r}$ with a very strong coherent state $\ket{\beta_{r}}_{r}$ and a very weak coherent state $\ket{\alpha_{s}}_{s}$, we can find that 
  $
{\rm Tr}[
 f_{\rm hom}(\hat n_1(\theta),\hat n_2(\theta),\hat n_0(\theta))\hat \rho_2]$ is approximated to ${}_{s}\bra{\alpha_{ s}}\frac83(\hat{x}_s(\theta)^4\ket{\alpha_{ s}}_{ s}$, which implies $\langle d_{\rm hom} \rangle_{\hat{\rho}_2}$ is 
approximated to $(|\alpha_{ s}|^4+\frac{11}6|\alpha_{ s}|^2+\frac23)/|\beta_{ r}|^2\sim 0$ (see Fig.~\ref{fig:relation}).
Therefore, Theorem 1 implies that our implementation is very close to the SHD, 
and guarantees that our implementation enables us to evaluate the covariance matrix of the state $\Lambda_{sr\to v}(\hat \rho_2)$ with high accuracy without any unverifiable
 assumption for the input state.

Similarly,
 the upper bound of the difference of correlations for $N$ pairs of two input modes in a state $\hat \rho_{2N}$ can be expressed in the following form: 
\begin{Thm}
For $N$ pairs of two input modes in a state $\hat \rho_{2N}$, the deviation between correlations described by ${\rm Tr} [\hat{x}_{v_k}(\theta_k) \hat{x}_{v_l}(\theta_l) \Lambda^{\otimes N} ( \hat{\rho}_{2N})]$ in the SHDs and by $\langle z^{(k)}(\theta_k) z^{(l)}(\theta_l) \rangle_{\hat{\rho}_{2N}}$ in our implementations is bounded by
\begin{align}
&|{\rm Tr} [\hat{x}_{v_k}(\theta_k) \hat{x}_{v_l}(\theta_l) \Lambda^{\otimes N} ( \hat{\rho}_{2N})] - \langle z^{(k)}(\theta_k) z^{(l)}(\theta_l) \rangle_{\hat{\rho}_{2N}} |
\nonumber\\&\makebox[2cm]{}
\le 0.605 \langle d_{\rm hom,hom}^{(k,l)} \rangle_{\hat{\rho}_{2N}},
\label{res:LH_LH}
\end{align}
where $\langle d_{\rm hom,hom}^{(k,l)} \rangle_{\hat{\rho}_{2N}} $ is an expected value of the quantity
\begin{align}
d_{\rm hom,hom}^{(k,l)}
&:=\sum_{\theta,\theta'\in\{0, \pi/4,\pi/2,3\pi/4\}}
\frac{f_{\rm hom}^{(k)}(\theta)g_{\rm hom}^{(l)}(\theta')}{16(N_{\rm hom}^{(k)}(\theta) +1)}
\nonumber\\
&\makebox[1cm]{}
+\sum_{\theta,\theta'\in\{0, \pi/4,\pi/2,3\pi/4\}} \frac{f_{\rm hom}^{(l)}(\theta)g_{\rm hom}^{(k)}(\theta')}{16(N_{\rm hom}^{(l)}(\theta) +1)}
,
\label{eq:d_lm_hom}
\end{align}
for the input $\hat \rho_{2N}$   with 
\begin{align}
&f_{\rm hom}^{(k)}(\theta):=f_{\rm hom}(n_1^{(k)}(\theta),n_2^{(k)}(\theta),n_0^{(k)}(\theta)),
\label{def:f_hom_m}
\\
 &g_{\rm hom}^{(k)}(\theta):=g_{\rm hom}(n_1^{(k)}(\theta),n_2^{(k)}(\theta),n_0^{(k)}(\theta)),
\label{def:g_hom_m}
 \\
& N_{\rm hom}^{(k)}(\theta):=n_1^{(k)}(\theta) + n_2^{(k)}(\theta) +n_0^{(k)}(\theta),
\label{def:n_hom_m}
\\
&g_{\rm hom}(n_1,n_2,n_0):=\frac{1}{2} \delta_{0,n_0} (n_1+n_2+1)+\frac{(n_1-n_2)^2}{2(n_0+1)},
\label{def:g_hom}
\end{align}
and    $f_{\rm hom} (n_1,n_2,n_0)$ being defined by Eq.~(\ref{def:f_hom}).
\end{Thm}
The proof of this theorem is also given 
in Appendix \ref{sec:proof}.

Like before, in the experiment, we expect to have $\langle d_{\rm hom,hom}^{(k,l)} \rangle_{\hat{\rho}_{2N}} \simeq 0$
 for any state $\hat{\rho}_{2N}$, as long as the number of photons in each LO pulse is much larger than that in the paired signal pulse.
For example, if $j$-th input pairs for $j=k,l$ are in a pure state $\ket{\alpha_{ s_j}}_{ s_j}\otimes \ket{\beta_{ r_j}}_{ r_j}$ with a very strong coherent state $\ket{\beta_{ r_j}}_{ r_j}$ and a very weak coherent state $\ket{\alpha_{ s_j}}_{ s_j}$, we can find that ${}_{sr}\bra{\alpha_{s_j},\beta_{r_j}}f_{\rm hom}(\hat n_1(\theta),\hat n_2(\theta),\hat n_0(\theta))\ket{\alpha_{s_j},\beta_{r_j}}_{sr}$ and $
{}_{sr}\bra{\alpha_{s_j},\beta_{r_j}}g_{\rm hom}(\hat n_1(\theta),\hat n_2(\theta),\hat n_0(\theta))\ket{\alpha_{s_j},\beta_{r_j}}_{sr}$ are approximated to ${}_s\bra{\alpha_{s_j}}\frac 83\hat x(\theta)^4\ket{\alpha_{s_j}}_s$ and ${}_s\bra{\alpha_{s_j}}\hat x(\theta)^2\ket{\alpha_{s_j}}_s$, respectively, which concludes $\langle d_{\rm hom,hom}^{(k,l)} \rangle_{\hat{\rho}_{2N}}\sim (|\alpha_{s_k}|^4+\frac{11}6|\alpha_{ s_k}|^2+\frac23)(|\alpha_{ s_l}|^2+\frac12)/|\beta_{ r_k}|^2+(|\alpha_{ s_k}|^2+\frac12)(|\alpha_{ s_l}|^4+\frac{11}6|\alpha_{ s_l}|^2+\frac23)/|\beta_{ r_l}|^2\sim 0$.
Therefore, 
by combining experimental data from our implementation with Theorem 2, we obtain a tight and accurate estimation of the correlation that we would have observed from the SHD.

So far, we have seen how well we can estimate the first and second moments given by the SHDs from the experimental data of our implementation. In different scenarios where we are interested in estimating correlation between outcomes of the {\it ideal} homodyne measurement and those of the SHD, rather than between those of SHDs, we can develop a similar method for the estimation.
In particular, for given two signal pulses $s_1s_2$, one of which, say $s_2$, is sent together with a LO pulse $r_2$, we apply the {\it ideal} homodyne detection to measure a quadrature $\hat{x}_{s_1}(\varphi)$ of the first signal mode $s_1$ and our implementation of the SHD to the pair of the second signal mode $s_2$ and its accompanied LO mode $r_2$.
This allows us to estimate the correlation
${\rm Tr} [\hat x_{s_1}(\varphi) \hat x_{v_2}(\theta) \Lambda_{s_2r_2\to v_2} (\rho_3)]$, where 
the quadrature $\hat x_{v_2}(\theta)$ is
given by 
applying
the ideal SHD to the pair, and
$\hat \rho_3$ represents the state of input three modes. The following theorem shows how good this estimation from our implementation is. 
\begin{Thm}\label{th:3}
For any $3$-mode input state $\hat \rho_3$, 
the relation  
\begin{align}
&|{\rm Tr}[\hat{x}_{s_1}(\varphi) \hat{x}_{v_2}({\theta}) \Lambda_{s_2r_2\to v_2}(\hat{\rho}_3)]-\langle x(\varphi) z^{(2)}(\theta) \rangle_{\hat{\rho}_3}|
\nonumber\\
&\makebox[1cm]{} \le  0.605 (\langle x(0)^2 d_{\rm hom}^{(2)} \rangle_{\hat{\rho}_3} +\langle x(\pi/ 2)^2 d_{\rm hom}^{(2)} \rangle_{\hat{\rho}_3} ),
\label{res:GH_LH}
\end{align}
holds.
$ x(\varphi)$ indicates the outcome of the ideal homodyne measurement with angle $\varphi$ for mode $s_1$.
$d_{\rm hom}^{(2)}$ is defined in the same way as $d_{\rm hom}$:
\begin{align}
    d_{\rm hom}^{(2)}&:=\sum_{\theta\in\{0,\pi/4,\pi/2,3\pi/4\}}
    \frac {f^{(2)}_{\rm hom}(\theta)}{4(N_{\rm hom}^{(2)}(\theta)+1)}.
    \label{def:d_hom_2}
\end{align}
$\langle x(\varphi)^2 d_{\rm hom}^{(2)} \rangle_{\hat{\rho}_3} $ is an expected value of the quantity  $x(\varphi)^2 d_{\rm hom}^{(2)}$ for the input $\hat \rho_3$.
\end{Thm}
The proof of this theorem is given 
in Appendix \ref{sec:proof}.

In Appendices A-C, we derive analogous theorems for cases where it is needed to implement the heterodyne measurement, without assuming anything on the LO pulses.

\section{Applications of the implemented SHD}
\label{sec:app}
Our implementation is useful for various kinds of quantum information processing, and in this section, we consider detection of the entanglement and CVQKD as examples of the applications.

Let us first consider detecting the entanglement by using the SHD.
Suppose that we perform the SHDs on $N$ pairs of a signal pulse and a LO pulse in a state $\hat{\rho}_{2N}$. 
Then, using the statistics of the observed measurement outcomes from our implementation, we can
 estimate the covariance matrix for $\Lambda^{\otimes N} (\hat{\rho}_{2N})$ with accuracy given by Eqs.~(\ref{res:LH_1}), (\ref{res:LH_2}), and (\ref{res:GH_LH}), from which we can judge \cite{HHH09,S00,DGC00,HE06,S06} whether the state $\Lambda^{\otimes N} (\hat{\rho}_{2N})$ has entanglement. If $\Lambda^{\otimes N} (\hat{\rho}_{2N})$ is concluded here to be entangled, the original state $\hat{\rho}_{2N}$ is so because entanglement does not increase under local operations, i.e., the squashing operation $\Lambda^{\otimes N} $.
 As a result, we can confirm the existence of entanglement without any unverifiable assumption for the LO pulses.

Our method can also be applied to CVQKD protocols. 
As an example of such applications, we consider a protocol which is based on the transmission of Gaussian-modulated squeezed states and LO pulses from Alice to Bob, as well as Bob's implementation of the SHDs on the received pulses. Here Alice's preparation of each signal pulse $s_2$ in a Gaussian-modulated squeezed state could have been replaced with preparing an entangled pair $s_1s_2$ in a two-mode squeezed state, followed by the ideal homodyne measurement on the pulse $s_1$. As a result, the security of the protocol above is equivalent to that of a virtual protocol where Alice sends Bob signal pulse $s_2$ in a two-mode squeezed state entangled with local pulse $s_1$, as well as a LO pulse $r_2$, to share a three-mode state $\hat{\rho}_{s_1s_2r_2}$ between Alice's local pulse $s_1$ and Bob's receiving pulses $s_2r_2$, followed by Alice's ideal homodyne measurement on the pulse $s_1$ and Bob's implementation of the SHD on the pulses $s_2r_2$. Therefore, by applying Theorem~\ref{th:1} to pulses $s_2r_2$ and Theorem~\ref{th:3} to three-mode state $\hat{\rho}_{s_1s_2r_2}$, we can estimate the covariance matrix of the state $\Lambda_{s_2r_2\to v_2}(\hat{\rho}_{s_1s_2r_2})$, from which we can estimate an upper bound on the amount of information available to 
Eve~\cite{GC06}.
This estimated information is used to determine the amount of privacy amplification to generate a secret key~\cite{DW04}.

The application for CVQKD protocols 
is not limited only to the specific protocol presented above.
For instance, if Alice's ideal homodyne measurement on her local pulse $s_1$ is replaced with the ideal heterodyne measurement, then the corresponding actual protocol becomes one sending Gaussian-modulated coherent states, rather than squeezed states. 
Bob could also replace the implementation of the SHD on his receiving pulses $s_2r_2$ with a hetrodyne measurement analogous to the SHD, i.e., a ``squashing'' 
heterodyne detection defined in Appendix A.
For this kind of variations of the protocol, to estimate the covariance matrix of the state $\Lambda_{s_2r_2\to v_2}(\hat{\rho}_{s_1s_2r_2})$ 
for proving the security, we establish the estimation formulas analogous to the above Theorems~\ref{th:1}-\ref{th:3}, as presented in Appendix \ref{sec:result}.

\section{Discussion and Conclusion}
The conventional implementation of the homodyne detection forces us to assume that the LO pulse is 
an infinitely strong coherent state. 
Unfortunately, however, it is difficult in practice to verify this assumption due to potential noises or eavesdropping that the LO pulse is subjected to during the transmission.
In order to solve this problems,
 we define a squashing homodyne detection, present an example of its implementation, and show that the difference between them can be evaluated analytically and can be very small. As shown in Appendices~A-C,
we can generalize the squashing homodyne detection to a squashing heterodyne detection, and we present an example of its implementation as well as a theory to estimate the closeness between the two. With our result, we are able to perform fundamental CV information processing tasks, including the detection of the entanglement and generation of the key, without putting any assumption on LO pulses.

We note that our implementation of the SHD assumes to be able to measure the difference between photon numbers output by two photodetectors (for instance, $n_1(\theta)-n_2(\theta)$ in Eq,~(\ref{eq:x_phi})) exactly. Fortunately, this is not a big issue \cite{Yuen:83,Schumaker:84} because 
the accuracy for the difference in the current implementation of the homodyne detection has already been sufficiently better than
the square root of the number of photons detected by each photodetector.
Therefore, our scheme is implementable by using photodetectors which have already been used in the conventional implementation of the homodyne detection.

Our method provides us with a good estimate on the mean and the variance of the output of the squashing operation, which is enough to construct the covariance matrix. On the other hand, we have not discussed the higher order of the moment, and we would make the conjecture that even the {\it probability distribution} itself of the output is very close to that estimated in our implementation for various two-mode inputs, as long as the photon number of the LO pulse is very large.
If our conjecture held, not only the statistics of the output of the implementation, but also the output itself could be regarded as if it were the output of the ideal detector. 
For example, in CVQKD, even if we consider a reverse reconciliation protocol where a part of Bob's measurement outcomes are disclosed to Eve, the security proof may go through with exactly the same manner as the the conventional proof.

\section*{Acknowledgments}
G.K. and M.O. was supported in part by JSPS KAKENHI Grant Numbers 20K03779, 21K03388.
K.T. acknowledges support from JSPS KAKENHI grant JP18H05237 and JST CREST grant JPMJCR 1671. 
K.A. acknowledges the support from Moonshot R\&D, JST Grant No. JPMJMS2061 and from JSPS KAKENHI Grant No. 21H05183 JP.

\appendix

\begin{figure*}[tb]
\begin{center}
\includegraphics[scale=0.4]{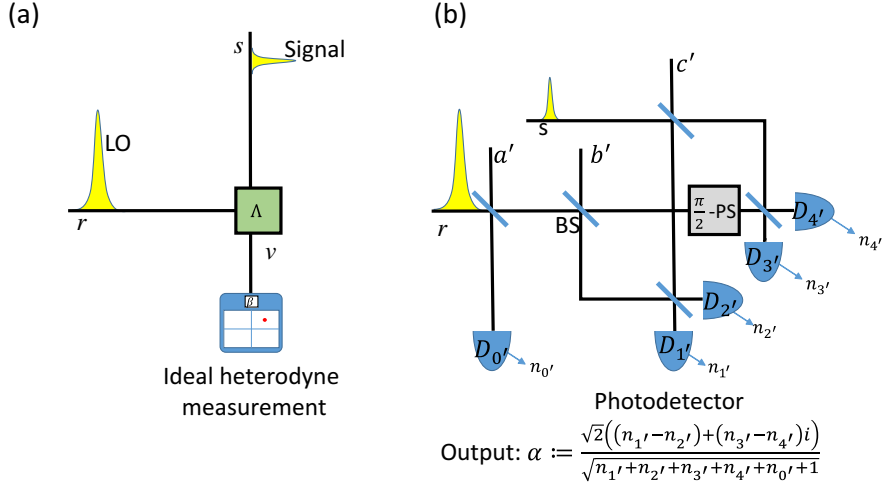}
\end{center}
\caption{
The squashing 
 heterodyne detection. The panel (a) shows the ideal situation. The input state, a pair of the signal pulse and the reference pulse, is squashed into a single mode state with the quantum operation $\Lambda_{sr\to v}$ followed by ideal heterodyne detection. In order to implement this measurement, we employ 50:50 beamsplitters, a $\frac\pi2$-phase shifter, and photodetectors (b). Note that the state in modes  ${a'}$, ${ b'}$, and ${ c'}$ is always the vacuum state, which is useful to clarify the expressions of our proofs in the appendix.
}
\label{fig:heterodyne}
\end{figure*}

\section{Definition of SHeD}
\label{sec:dif_LHeD}
In the main text, we have defined the SHD (Fig.~\ref{fig:implementation}). In this appendix, we define the squashing 
heterodyne detection (SHeD) from the SHD by replacing the ideal homodyne detection used in the SHD with the ideal heterodyne detection.  That is, the  two-mode input state $\hat \rho_2$
 is squashed by the operator $\Lambda_{sr \to v}$ (Eq.~(\ref{eq:lambda})), followed by the ideal heterodyne detection (see Fig.~\ref{fig:heterodyne}(a)).
As a result, the probability density function of the outcome $\beta$ can be expressed by ${}_v\bra{\beta}\Lambda_{sr\to v}(\hat \rho_2)\ket{\beta}_v$, where  $\beta$ is a complex amplitude and $\ket\beta_v$ is a coherent state on mode $ v$ outputted from $\Lambda_{sr\to v}$.

In the following appendices, we use the following two notations in the context of the SHeD:
 $\beta_\theta$ indicates the real number $\Re \beta\cos\theta+\Im \beta\sin\theta$, which is defined by the outcome $\beta$ of the SHeD and the parameter  $\theta\in[0,2\pi)$,  where $\Re X$ ($\Im X$) is the real (imaginary) part of the complex number $X$. 
The operator $\hat F_v(f(\beta) )$ for mode $v$ is defined by  
\begin{eqnarray}
\hat F_v(f(\beta) )&:=&
\int f(\beta)\ket{\beta}\bra{\beta}_v\frac{d^2\beta}\pi,
\label{def:F}
\end{eqnarray}
for any function $f(\beta)$.
These definitions leads to, for instance, $\hat F_v(\beta_\theta)=\hat x_v(\theta)$ and $\hat F_v(\beta_0\beta_{\pi/2})=(\hat x_v \hat p_v+\hat p_v\hat x_v)/2$.

\section{An implementation of SHeD }
\label{sec:imp_LHeD}
In this appendix, we show an example of the implementation of SHeD. Our basic idea is that we first measure the intensity of the LO pulse's, and then perform the conventional implementation of the heterodyne detection.
To be precise (see also Fig.~\ref{fig:heterodyne}(b)), the LO pulse first splits into two pulses by
a 50:50 beamsplitter, one of which is directly observed with a photodetector $D_{0'}$. The other half of the LO pulse and the signal split with  50:50 beamsplitters.
One of the separated LO pulse is subjected to a $\frac\pi2$-phase shift, and after this shift, each pairs of the signal pulse and the LO pulse interfere by a $50:50$ beamsplitter, followed by the detection with $D_{1'}$ and $D_{2'}$ or $D_{3'}$ and $D_{4'}$.      
For later convenience, we express by $n_{k'}$ the number of photons detected by $D_{k'}$.
With this notation, the outcome of our implementation is represented by 
\begin{eqnarray}
\alpha &:=&
 \frac {\sqrt 2\left((n_{1'}-n_{2'})+(n_{3'}-n_{4'})i\right)}{\sqrt{n_{0'}+n_{1'}+n_{2'}+n_{3'}+n_{4'}+1}}.
\label{def:LHeD}
\end{eqnarray}
We note that unlike the implemented SHD, this implementation is uniquely determined and has no characterizing parameter. Similar to the case in the SHD, we define $\left<\alpha_\theta\right>_{\hat \rho_2}$ as the expected value of $\alpha_\theta:=\Re \alpha\cos\theta+\Im \alpha\sin\theta$, which is the the outcome $\alpha$ of the measurement on $\hat \rho_2$ with a parameter $\theta$.

When the input is a $N$-pair state $\hat \rho_{2N}$ and we measure the $j$-th pair with this implementation, all the outcomes in this measurement are described just by adding the subscript $(j)$. For example, we express the outcome of the $j$-th measurement as
\begin{eqnarray}
\alpha^{(j)}&:=&
 \frac {\sqrt 2\left((n_{1'}^{(j)}-n_{2'}^{(j)})+(n_{3'}^{(j)}-n_{4'}^{(j)})i\right)}{\sqrt{n_{0'}^{(j)}+n_{1'}^{(j)}+n_{2'}^{(j)}+n_{3'}^{(j)}+n_{4'}^{(j)}+1}}
,
\end{eqnarray}
where 
$n_{k'}^{(j)}$ is the number of photons detected by  $D_{k'}$ in the SHeD for the $j$-th pair.

\section{Closeness of the ideal measurements  and those implementations of the SHD and the SHeD}
\label{sec:result}
Theorem 1,  Theorem 2, and Theorem 3 in the main text present inequalities that
estimate the deviation between the SHD and our implementation of it. In this appendix, we provide further inequalities to estimate the deviation for the SHeD and the SHD. In the following, we call ``the ideal (implemented) situation'' when SHD or SHeD in the system is ideal (implemented) detection.

We consider the following three cases. 
Case I: the input state is a $2$-mode state $\hat \rho_2$, and the state is measured by the SHD or the SHeD.
Case II: the input state is a $2N$-mode, i.e., $N$ pairs, whose state is $\hat \rho_{2N}$, and each pair is measured by the SHD or the SHeD.
Case III: the input state is a $3$-mode state $\hat \rho_3$, the first mode $s_1$ is measured by the ideal homodyne or heterodyne detection,  and the last 2 modes $s_2r_2$ are measured by the SHD or the SHeD.
For each cases, we compare the ideal and implemented situations by evaluating the deviation between them. Note that, in case III, we consider to measure the first mode by the ideal homodyne or heterodyne detection, which is useful for the analysis of a prepare and measure CVQKD in which the ideal detection is assumed.

Before presenting the inequalities, we summarize notations used below (all the notations except for $\gamma$, $\phi$, $\gamma_\phi$, and $\hat F(f(\gamma))$ have already been defined, but we have summarized them for completeness):
\begin{itemize}
\item $n_0(\theta)$, $n_1(\theta)$, $n_2(\theta)$, $n_{0'}$, $n_{1'}$,$\cdots$, $n_{4'}$:
Photon numbers detected by the photodetectors $D_0$, $D_1$, $D_2$, $D_{0'}$, $D_{1'}$,$\cdots$, $D_{4'}$ in the implemented SHD or SHeD
in the case I.

\item $z(\theta)$:
The outcome of the implemented SHD, which is defined by Eq.~(\ref{eq:x_phi}). We use the angle $\theta$
for the case I.

\item $\alpha$:
A complex number as the outcome of the implemented SHeD for the case I, which is defined by Eq.~(\ref{def:LHeD}).

\item $\beta$:
A complex number as the outcome of the SHeD for the case I.

\item $\hat x_\mu(\theta)$:
The observable defined as $\hat x_\mu \cos\theta+\hat p_\mu\sin \theta$, where $\hat x_\mu$ and $\hat p_\mu$ are the canonical operators for mode $\mu$.

\item $x(\varphi)$:
An outcome of the ideal homodyne detector for the first mode $s_1$ in the case III. Its corresponding observable is $\hat{x}_{s_1}(\varphi)$.

\item $\gamma$:
A complex number as the outcome of the ideal heterodyne detector for the first mode $s_1$ in the case III.

\item $n_0^{(k)}(\theta)$, $n_1^{(k)}(\theta)$, $n_2^{(k)}(\theta)$, 
$z^{(k)}(\theta)$, $n_{0'}^{(k)}$, $n_{1'}^{(k)}$,$\cdots$, $n_{4'}^{(k)}$, $\alpha^{(k)}$, and $\beta^{(k)}$ :
The values $n_0(\theta)$, $n_1(\theta)$, $n_2(\theta)$,  $z(\theta)$, $n_{0'}$, $n_{1'}$,$\cdots$, $n_{2'}$, $\alpha$, and $\beta$
for the $k$-th measurement device in the case II or III.

\item $\alpha_\theta$, $\beta_\theta$, $\gamma_\phi$, $\alpha^{(k)}_{\theta_k}$, $\beta^{(k)}_{\theta_k}$: The real numbers 
$X_\mu:=\Re X\cos\mu+\Im X\sin\mu$ defined from a complex number $X\in\{\alpha,\beta,\gamma,\alpha^{(k)},\beta^{(k)}\}$ and a real number $\mu\in\{\theta,\phi,\theta_k\}$.

\item $\hat F_\mu(f(\beta))$:
An operator on mode $\mu$.
This operator is defined by Eq.~(\ref{def:F}) where $f$ is a function from a complex number to a real number. For example, we will use
 $\hat F_\mu(\beta_\theta)$,
 $\hat F_\mu((\beta_\theta)^2)$,
 $\hat F_\mu(\beta_0\beta_{\pi/2})$, etc.

\item $\left<X\right>_{\hat \sigma}$:
The expected value of the outcome $X$ when the input state is $\hat \sigma\in\{\hat \rho_2,\hat\rho_{2N},\hat \rho_3\}$.
\end{itemize}
Below, we introduce three Theorems, and we will prove them in Appendix \ref{sec:proof}.

In the case of a single pair state $\hat \rho_2$, i.e., in the case I, the upper bound of the deviation can be given for the SHeD as well as for the SHD as follows.
\begin{Thm}
For any input two-mode state $\hat \rho_2$, 
the deviation between the expected value of the SHeD
${\rm Tr}\bigl [ \hat F_v(\beta_\theta) \Lambda_{sr\to v}(\hat \rho_2)\bigr]$
and that of our implementation
$\bigl\langle  \alpha_\theta\bigr\rangle_{\hat \rho_2}$
 is bounded by
\begin{eqnarray}
\bigl|
{\rm Tr}\bigl [ \hat F_v(\beta_\theta) \Lambda_{sr\to v}(\hat \rho_2)\bigr]-\bigl\langle  \alpha_\theta\bigr\rangle_{\hat \rho_2}
\bigr|
&\leq&
0.226\bigl\langle d_{\rm het}\bigr\rangle_{\hat \rho_2},
\label{res:LE_1}
\end{eqnarray}
 where $\bigl\langle d_{\rm het}\bigr\rangle_{\hat \rho_2}$ is the expected value of the quantity
\begin{align}
&
\makebox[1cm]{$d_{\rm het}$} := 
\frac{f_{\rm het}(n_{1'},n_{2'},n_{3'},n_{4'},n_{0'})}{n_{1'}+n_{2'}+n_{3'}+n_{4'}+n_{0'}+1},
\label{def:d_het}
\\
&f_{\rm het}(n_{1'},n_{2'},n_{3'},n_{4'},n_{0'})
\nonumber\\
&\makebox[1cm]{} :=\delta_{0,n_{0'}}\left[\frac72(n_{1'}+n_{2'}+n_{3'}+n_{4'})+2\right]
\nonumber\\
&\makebox[1cm]{}\quad\quad
+\frac{((n_{1'}-n_{2'})^{2}+(n_{3'}-n_{4'})^{2})^{2}}{(n_{0'}+1)(n_{0'}+2)},
\label{def:f_het}
\end{align}
for the input $\hat \rho_2$.

Also, for the square of the output, we can obtain the relation
\begin{align}
-0.084\bigl\langle d_{\rm het}\bigr\rangle_{\hat \rho_2}\leq
\makebox[3cm]{}\nonumber\\
{\rm Tr}\bigl [\hat F_v((\beta_\theta)^2) \Lambda_{sr\to v}(\hat \rho_2)\bigr]-\bigl\langle  (\alpha_\theta)^2\bigr\rangle_{\hat \rho_2}
&\leq 
\bigl\langle d_{\rm het}\bigr\rangle_{\hat \rho_2},
\label{res:LE_2}
\\
\bigl|
{\rm Tr}\bigl [ \hat F_v(\beta_0 \beta_{\pi/2})
 \Lambda_{sr\to v}(\hat \rho_2)\bigr]
-\bigl\langle  \alpha_0\alpha_{\pi/2}\bigr\rangle_{\hat \rho_2}
\bigr|
&\leq
\frac 12\bigl\langle d_{\rm het}\bigr\rangle_{\hat \rho_2}.
\label{res:LE_2_m}
\end{align}
\end{Thm}

These relations imply that  the SHeD and its implementation are close since 
$\bigl\langle d_{\rm het}\bigr\rangle_{\hat \rho_2}\simeq 0$ for typical input states, i.e., the states such that the signal pulse and the LO pulse are not correlated and the number of photons in the LO pulse is much larger than that of the signal pulse.

For $N$-pair state $\hat \rho_{2N}$, i.e., for the case II, we may imagine that some pairs are measured by the SHD, and the other pairs are measured by the SHeD. Even in this case, we can give inequalities which imply the closeness as follows.
\begin{Thm}
For any input $N$-pair state $\hat \rho_{2N}$, the deviation between the correlation of the SHeDs
${\rm Tr}[\hat F_{v_k}(\beta_{\theta_k})\hat F_{v_l}(\beta_{\theta_l})\Lambda^{\otimes N}(\hat \rho_{2N}) ]$
and that of our implementations $\bigl\langle  \alpha_{\theta_k}^{(k)} \alpha_{\theta_l}^{(l)}\bigr\rangle_{\hat \rho_{2N}}
$ is bounded as
\begin{align}
&
\bigl|
       {\rm Tr}[\hat F_{v_k}(\beta_{\theta_k}^{(k)})\hat F_{v_l}(\beta_{\theta_l}^{(l)})\Lambda^{\otimes N}(\hat \rho_{2N}) ]
-\bigl\langle  \alpha_{\theta_k}^{(k)} \alpha_{\theta_l}^{(l)}\bigr\rangle_{\hat \rho_{2N}}
\bigr|
\nonumber\\
 &\makebox[1cm]{}\leq
{0.160}
\bigl\langle d_{\rm het,het}^{(k,l)}\bigr\rangle_{\hat{\rho}_{2N}},
\label{res:LE_LE}
\end{align}
 where $\bigl\langle d_{\rm het,het}^{(k,l)}\bigr\rangle_{\hat{\rho}_{2N}} $ is the expected value of the quantity  
\begin{eqnarray}
d_{\rm het,het}^{(k,l)}&:=&
\frac{f_{\rm het}^{(k)}g_{\rm het}^{(l)}}{N_{\rm het}^{(k)}+1}
+
\frac{f_{\rm het}^{(l)}g_{\rm het}^{(k)}}{N_{\rm het}^{(l)}+1},
\label{eq:d_het_het}
\end{eqnarray}
for the input $\hat \rho_{2N}$. Moreover, 
\begin{align}
f_{\rm het}^{(k)}&:=f_{\rm het}(n_{1'}^{(k)},n_{2'}^{(k)},n_{3'}^{(k)},n_{4'}^{(k)},n_{0'}^{(k)})
\label{def:f_het_m}
\end{align}
 is defined by Eq.~(\ref{def:f_het}), and
 $g_{\rm het}^{(k)}$ and $N_{\rm het}^{(k)}$ are defined as
\begin{align}
&g_{\rm het}^{(k)}:=g_{\rm het}(n_{1'}^{(k)},n_{2'}^{(k)},n_{3'}^{(k)},n_{4'}^{(k)},n_{0'}^{(k)},m_0^{(k)}),
\label{def:g_het_m}
\\
&
g_{\rm het}(n_{1'},n_{2'},n_{3'},n_{4'},n_{0'})
\nonumber\\
&\makebox[1cm]{}:=\delta_{0,n_{0'}}
+\frac{(n_{1'}-n_{2'})^{2}+(n_{3'}-n_{4'})^{2}}{n_{0'}+1},
\label{def:g_het}
\\
&
N_{\rm het}^{(k)}
:= n_{1'}^{(k)}+n_{2'}^{(k)}+n_{3'}^{(k)}+n_{4'}^{(k)}+n_{0'}^{(k)}.
\label{def:N_het_m}
\end{align}

Also, when $k$-th($l$-th) pair is measured by the SHD(SHeD), we can obtain a similar bound: 
\begin{align}
&
\bigl|
{\rm Tr}[\hat x_{v_k}(\theta_k)\hat F_{v_l}(\beta_{\theta_l}^{(l)})\Lambda^{\otimes N}(\hat \rho_{2N}) ]
-\bigl\langle  z^{(k)}(\theta_k) \alpha_{\theta_l}^{(l)}\bigr\rangle_{\hat \rho_{2N}}
\bigr|
\nonumber\\
&\makebox[1cm]{} 
\leq
0.371\bigl\langle d_{{\rm hom},{\rm het}}^{(k,l)}\bigr\rangle_{\hat \rho_{2N}},
\label{res:LH_LE}
\end{align}
where $\bigl\langle d_{{\rm hom},{\rm het}}^{(k,l)}\bigr\rangle_{\hat \rho_N}$ is the expected value of 
\begin{align}
d_{{\rm hom},{\rm het}}^{(k,l)}&:=&
\makebox[-.5cm]{}\sum_{\theta\in\{0, \pi/4,\pi/2,3\pi/4\}}
\frac{f_{\rm hom}^{(k)}(\theta)g_{\rm het}^{(l)}}{4(N_{\rm hom}^{(k)}(\theta)+1)}
+
\frac{f_{\rm het}^{(l)}g_{\rm hom}^{(k)}(\theta)}{4(N_{\rm het}^{(l)}+1)},
\end{align}
 for the input $\hat \rho_{2N}$, the variables $f_{\rm het}^{(l)}$, $g_{\rm het}^{(l)}$, $N_{het}^{(l)}$, 
$f_{\rm hom}^{(k)}(\theta)$, $g_{\rm hom}^{(k)}(\theta)$, and $N_{\rm hom}^{(k)}(\theta)$ are defined by Eqs.~(\ref{def:f_het_m}), (\ref{def:g_het_m}), (\ref{def:N_het_m}), (\ref{def:f_hom_m}) , (\ref{def:g_hom_m}), and (\ref{def:n_hom_m}), respectively.
\end{Thm}
When we estimate the correlation of quadratures for the state $\Lambda^N(\hat \rho_{2N})$ by the implemented SHD or SHeD, Theorem 2 and Theorem 5 can be used to evaluate the upper bound of the estimation error.

In the case of a combination of the ideal homodyne or heterodyne detection and the SHD or the SHeD for three mode state $\hat \rho_3$, i.e., in the case III, we can provide the following three inequalities in addition to the one in Theorem 3, which characterize the closeness.
\begin{Thm}
Suppose that the input state is a $3$-mode state $\hat \rho_3$. When the first mode $s_1$ is measured by the ideal homodyne detection and the last two modes $s_2r_2$ are measured by the SHeD, the following relation holds
\begin{align}
&
\bigl|
{\rm Tr}[  \hat x_{s_1}(\varphi) \hat F_{v_2}(\beta_\theta^{(2)})({\rm id}\otimes\Lambda_{s_2r_2\to v_2})(\hat \rho_3)]-\bigl\langle x(\varphi) \alpha_\theta^{(2)}\bigr\rangle_{\hat\rho_3}
\bigr|
\nonumber\\
&\makebox[1cm]{}
\leq
0.261(\bigl\langle x(0)^2 d_{\rm het}^{(2)}\bigr\rangle_{\hat \rho_3}
+\bigl\langle x(\frac\pi 2)^2 d_{\rm het}^{(2)}\bigr\rangle_{\hat \rho_3}),
\label{res:GH_LE}
\end{align}
where $\langle x(\varphi)^2 d_{\rm het}^{(2)} \rangle_{\hat{\rho}_3} $ is the expected value of 
 $ x(\varphi)^2 d_{\rm het}^{(2)}$ for the input $\hat \rho_3$, which is obtained in the implemention,
 and $d_{\rm het}^{(2)}$ is defined in the same way as Eq.~(\ref{def:d_het}):
 \begin{align}
d_{\rm het}^{(2)} := 
\frac{f_{\rm het}^{(2)}}{N_{\rm het}^{(2)}+1},
 \end{align}

When the first mode is measured by the ideal heterodyne detection and the last two modes are measured by  the SHeD, we have the following relation.
\begin{align}
&
\bigl|{\rm Tr}[ \hat F_{s_1}( \gamma_\phi)\hat F_{v_2}(\beta_\theta^{(2)})({\rm id}\otimes \Lambda_{s_2r_2\to v_2})(\hat \rho_3)]-\bigl\langle \gamma_\phi\alpha_\theta^{(2)}\bigr\rangle_{\hat \rho_3}\bigr|
\nonumber\\
&\makebox[1cm]{} \leq
0.160\bigl\langle \big| \gamma \big|^2d_{\rm het}^{(2)}\bigr\rangle_{\hat \rho_3},
\label{res:GE_LE}
\end{align}
where $\langle  \big| \gamma \big|^2d_{\rm het}^{(2)} \rangle_{\hat{\rho}_3} $ is the expected value of 
 $ \big| \gamma \big|^2d_{\rm het}^{(2)}$ for the input $\hat \rho_3$, which is obtained  in the implementation.

When the first mode is measured by the ideal heterodyne detection and the last two modes are measured by the  SHD, the following relation holds.
\begin{align}
&
\bigl|{\rm Tr}[ \hat F_{s_1}(  \gamma_\phi)\hat x_{v_2}(\theta)\Lambda_{s_2r_2\to v_2}(\hat \rho_3)]-\bigl\langle   \gamma_\phi z^{(2)}(\theta)\bigr\rangle_{\hat \rho_3}\bigr|
\nonumber\\
&\makebox[1cm]{} \leq
0.371\bigl\langle \big|\gamma\big|^2d^{(2)}_{\rm hom}\bigr\rangle_{\hat \rho_3},
\label{res:GE_LH}
\end{align}
where $\langle \big|\gamma\big|^2d_{\rm hom}^{(2)} \rangle_{\hat{\rho}_3} $ is the expected value of $\big|\gamma\big|^2d_{\rm hom}^{(2)}$ for the input $\hat \rho_3$, which is obtained  in the implementation, and $d_{\rm hom}^{(2)}$ is defined by Eq.~(\ref{def:d_hom_2}).
\end{Thm}

\section{Proofs of the theorems}
\label{sec:proof}
Below, we prove all the theorems, from Theorem 1 to Theorem 6. That is, we prove 
Eqs.~(\ref{res:LH_1}),
(\ref{res:LH_2}),
(\ref{res:LH_2_m}),
(\ref{res:LH_LH}),
(\ref{res:GH_LH}),
(\ref{res:LE_1}),
(\ref{res:LE_2}),
(\ref{res:LE_2_m}),
(\ref{res:LE_LE}),
(\ref{res:LH_LE}),
(\ref{res:GE_LE}),
(\ref{res:GH_LE}), and (\ref{res:GE_LH}).
These 13 inequalities can be divided into four sets depending on the similarity of the proofs.
The first one $\mathcal E_1$ consists of Eqs.~(\ref{res:LH_1}) and (\ref{res:LE_1}), and the second one $\mathcal E_2$ is composed of Eqs.~(\ref{res:LH_2}), (\ref{res:LH_2_m}),
 (\ref{res:LE_2}), and (\ref{res:LE_2_m}).
The third set $\mathcal E_3$ is formed by Eqs.~(\ref{res:GH_LH}), (\ref{res:GE_LE}), (\ref{res:GH_LE}), and (\ref{res:GE_LH}), and the final one $\mathcal E_4$ consists of Eqs.~(\ref{res:LH_LH}), (\ref{res:LE_LE}), and (\ref{res:LH_LE}).
The inequalities in the same set can be proved with the same manner, and just a slight change in the parameters suffices. Therefore, in the following, we will first present the proof only a representative inequality of each of the set, and then we show the parameter adjustment that is needed to prove the other inequalities in the same set.

\begin{figure}[tb]
\begin{center}
\includegraphics[scale=0.4]{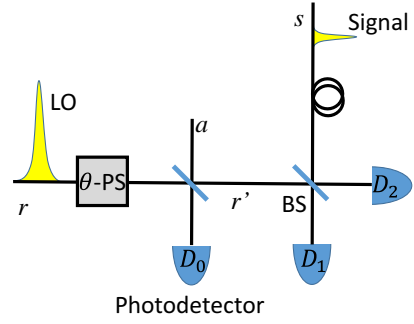}
\end{center}
\caption{The ancillary mode $a$ in the implemented SHD. The expressions used in the proof become clearer by noting that this mode always contains the vacuum only.
}
\label{fig:implementation_d}
\end{figure}
Before showing the proofs, we define several notations.
\begin{itemize}
\item $\{  s,r,r',v,a,a',b',c',0,1,2,0',1',2',3',4',
s_j,r_j,$ $r'_j,v_j,a_j,a'_j,b'_j,c'_j,0_j,1_j,2_j,0'_j,1'_j,2'_j,3'_j,4'_j
\}=:\mathcal M$: The set of identifiers of the modes for the SHD in the case of $\theta=0$ or the SHeD.
${ s}$  and ${r}$ means that the input modes are the signal pulse and the LO pulse, respectively. Mode ${v}$ is the output mode of the operator $\Lambda_{sr\to v}$ for the SHD or the  SHeD.
Mode ${r'}$ is the one of the modes split from mode ${r}$ by a beam splitter, which is not observed by the photodetector $D_{0(0')}$ in the implemented SHD(SHeD). 
${ a}$, ${ a}'$, ${ b}'$, ${ c}'$ are the modes combined with the modes of the signal or the LO pulse   at beam splitters (see Fig.~\ref{fig:heterodyne}(b) and \ref{fig:implementation_d}). 
 $0$,$\cdots$, $0'$,$\cdots$ refer to the modes that are measured by photodetectors $D_0,\cdots,D_{0'},\cdots$, respectively in the implemented SHD or SHeD.
 The subscript $j$ is used to identify the $j$-th measurement device when we have multiple measurement devices in the setup. 
\item $\hat n_{\mu\in \mathcal M}$: The photon number operator for mode $\mu$.
\item $ \hat a_{\mu\in \mathcal M}$: The annihilation  operator for mode $\mu$, i.e., $[\hat a_\mu,\hat a_\mu^\dagger]=1$, $[\hat a_\mu^\dagger,\hat a_\nu^\dagger]=[\hat a_\mu,\hat a_\nu]=0$, and $\hat n_\mu=\hat a_\mu^\dagger\hat a_\mu$ for any $\mu,\nu\in\mathcal M$.
\item $\ket{m_1,m_2,\cdots}_{\mu_1\mu_2\cdots \subseteq \mathcal M}$: The photon number state for modes $\mu_1\mu_2\cdots$, i.e., $\hat a_\mu\ket {m}_\mu=\sqrt {m} \ket {m-1}_\mu$ and $\hat a_\mu^\dagger\ket{m-1}_\mu=\sqrt{m}\ket{m}_\mu$.
\item $\Lambda^\dagger_{\mu\to \nu_1\nu_2}$ for $\mu,\nu_1,\nu_2\in\mathcal M$: The conjugate of the map $\Lambda_{\nu_1,\nu_2\to \mu}$, i.e. 
for any operator $\hat X$, we define the conjugate of the map as $\Lambda^\dagger_{\mu\to \nu_1\nu_2}(\hat X):=\sum_{m=0}^\infty M_m^\dagger\hat X M_m$ for
$M_m:=\sum_{n=0}^{m}\ket{n}_{ \mu}{}_{ \nu_1\nu_2}\bra{ n,m-n}$.
\end{itemize}
Note that, from the definitions, we have that the three modes $ sra$ are mutually independent, i.e., $[\hat a_\nu^{\dagger},\hat a_\mu]=0$ for $\nu\neq\mu\in\{  s,r,a\}$. In the same way,
 modes 
$ sr'0$, 
$ 120$,  
$ srb'c'a'$,  
$ sr'b'c'{0'}$, and 
$  1'2'3'4'0'$ are  mutually independent, respectively. The above definition also implies that the vacuum state for modes 
$sra$ is equal to the one for $ sr'0$ or 
$ 120$, i.e. $\ket{0,0,0}_{ sra}=\ket{0,0,0}_{sr'0}=\ket{0,0,0}_{ 120}$. Similarly, the relation
 $\ket{0,0,0,0,0}_{ srb'c'a'}=\ket{0,0,0,0,0}_{ sr'b'c'{0'}}=\ket{0,0,0,0,0}_{ 1'2'3'4'0'}$ also holds.

For ease of the proof, we assume that the $\pi$ phase shift is applied only when the input light from the above is reflected to the right at the beamsplitters in Figs. \ref{fig:heterodyne} and \ref{fig:implementation_d}. This leads to the following relationships:
\begin{align}
\hat a_1&
=\frac 1{\sqrt 2}\hat a_{ s}+\frac12\hat a_{ r}-\frac12\hat a_{ a}
=\frac 1{\sqrt 2}\hat a_{ s}+\frac 1{\sqrt 2}\hat a_{ r'},
\label{rel:a_1}
\\
\hat a_2&=-\frac 1{\sqrt 2}\hat a_{ s}+\frac12\hat a_{ r}-\frac12\hat a_{ a}
=-\frac 1{\sqrt 2}\hat a_{ s}+\frac 1{\sqrt 2}\hat a_{ r'},
\label{rel:a_2}
\\
\hat a_{ r}&=\frac 1{\sqrt 2} \hat a_{ r'}+\frac 1{\sqrt 2}\hat a_{0},
\label{rel:a_r}
\\
\hat a_{1'}&
=\frac 1{ 2}\hat a_{ s}+\frac1{2}\hat a_{ r'}+\frac12\hat a_{ b'}+\frac12\hat a_{c'},
\label{rel:a_1'}
\\
\hat a_{2'}&
=-\frac 1{ 2}\hat a_{ s}+\frac1{2}\hat a_{ r'}+\frac12\hat a_{ b'}-\frac12\hat a_{ c'},
\label{rel:a_2'}
\\
\hat a_{3'}&
=+\frac 1{ 2}\hat a_{ s}+\frac i{2}\hat a_{ r'}-\frac i2\hat a_{ b'}-\frac12\hat a_{ c'},
\label{rel:a_3'}
\\
\hat a_{4'}&
=-\frac 1{ 2}\hat a_{ s}+\frac i{2}\hat a_{ r'}-\frac i2\hat a_{ b'}+\frac12\hat a_{ c'},
\label{rel:a_4'}
\\
\hat a_{ r'}&=\frac 1{\sqrt 2} \hat a_{ r}-\frac 1{\sqrt 2}\hat a_{a'},
\label{rel:a_r'_}
\\
\hat a_{ r}&=\frac 1{\sqrt 2} \hat a_{r'}+\frac 1{\sqrt 2}\hat a_{0'}.
\label{rel:a_r'}
\end{align}
Note that the coefficients of $\hat a_r$ in Eqs.~(\ref{rel:a_1})$-$(\ref{rel:a_r}) are due to the particular choice of the parameter $\theta=0$ in the SHD.

\subsection{Proof of Eq.~(\ref{res:LH_1})}
As a representative of the first set $\mathcal E_1$, we prove Eq.~(\ref{res:LH_1}). We start with providing a sufficient condition of the inequality in Eq.~(\ref{res:LH_1}), which is an operator inequality on the two-mode $ rs$:
\begin{align}
\pm \bigl(
\Lambda_{v\to sr}^\dagger(\hat x_v(\theta))
- {}_{ a}\big\langle0\bigr|\hat z_{\rm hom}(\theta)\bigl|0\big\rangle_{ a}
\bigr)
&\leq 
c_1\frac{\hat f_{\rm hom}}{\hat n_{ s}+\hat n_{ r}+\hat I}
,
\label{eq:proof_1_1}
\end{align}
where
\begin{align}
   c_1&:=\sqrt{\frac85}(\sqrt2-1)= 0.52494\cdots,
\\
\hat z_{\rm hom}(\theta)&:=e^{-i\theta\hat n_{ r}}\frac{\hat n_1-\hat n_2
}{\sqrt{2(\hat n_{ 1}+\hat n_{ 2}+\hat n_{ 0}+\hat I)}}e^{i\theta\hat n_{ r}},
\label{def:x_l_hom}
\end{align}
\begin{align}
&\hat f_{\rm hom}
\nonumber\\
&\makebox[0.5cm]{}:=\makebox[-0.5cm]{} \sum_{\theta\in\{0,\pi/4,\pi/2,3\pi/4\}}\frac 14 e^{-i\theta\hat n_{r}}
{}_{a}\bra{0}
  f_{\rm hom}(\hat n_1,\hat n_2,\hat n_0)
\ket{0}_{ a}
e^{i\theta\hat n_{r}},
\label{def:hat_f_hom}
\end{align}
and $f_{\rm hom}(n_1,n_2,n_0)$ is defined by Eq.~(\ref{def:f_hom}).
The inequality in Eq.~(\ref{res:LH_1}) is reconstructed from the inequality in Eq.~(\ref{eq:proof_1_1}) by evaluating the expected value of operators in both sides for the input state $\hat \rho_2 $ and maximizing the sign in the left-hand side.
Note that the division and the square root used above are well-defined since the operator $\hat n_{ s}+\hat n_{ r}+\hat I$ is commutes with  $\hat f_{\rm hom}$, and the operator  $\hat n_1+\hat n_2+\hat n_0+\hat I$ are positive and commutes with  $\hat n_1-\hat n_2$.

The proof of Eq.~(\ref{eq:proof_1_1}) can be decomposed into the proof of the non-negativity of the following two operators,
\begin{align}
&
\frac{\hat f_{\rm hom}}{\hat n_{ s}+\hat n_{ r}+\hat I}-
\frac{\hat n_{ s}^2+\hat n_{ s}+\frac 12\hat I}{\hat n_{ s}+\hat n_{ r}+\hat I},
\label{eq:proof_1_2}
\\
&
c_1\frac{
\hat n_{ s}^2+\hat n_{ s}+\frac 12\hat I.
}{\hat n_{ s}+\hat n_{ r}+\hat I}
\mp \bigl(
\Lambda_{v\to sr}^\dagger(\hat x_v)
-{}_{ a}\big\langle0\bigr|\hat z_{\rm hom}(0)\bigl|0\big\rangle_{ a}
\bigr).
\label{eq:proof_1_3}
\end{align}
If those operators are non-negative, then Eq.~(\ref{eq:proof_1_1}) is proved. This is so because $c_1\times (\ref{eq:proof_1_2})+e^{-i \hat n_{ r}\theta}(\ref{eq:proof_1_3})e^{i \hat n_{ r}\theta}$
is equal to the right hand side minus the left hand side of the inequality in Eq.~(\ref{eq:proof_1_1}), where  we have used the relation $e^{i \hat n_{ r}\theta}\Lambda_{v\to sr}^\dagger(\hat X)e^{-i \hat n_{ r}\theta}=
\Lambda_{v\to sr}^\dagger(e^{-i \hat n_{ v}\theta}\hat X e^{i \hat n_{ v}\theta})$ holds for any operator $\hat X$. 

We first prove the non-negativity of the operator in Eq.~(\ref{eq:proof_1_2}). For this, we evaluate 
its matrix elements with the photon number basis  $\{\bigl|n,m\big\rangle_{ sr}\}_{n,m=0,1,\cdots}$:
\begin{widetext}
\begin{eqnarray}
&&(n+m+1)\times{}_{ sr}\bra{n,m}
\frac{\hat f_{\rm hom}}{\hat n_{ s}+\hat n_{ r}+\hat I}-
\frac{\hat n_{ s}^2+\hat n_{ s}+\frac 12\hat I}{\hat n_{ s}+\hat n_{ r}+\hat I}
\ket{n',m'}_{ sr}
\nonumber\\
&=&
\sum_{\theta\in\{0,\pi/4,\pi/2,3\pi/4\}}\frac14e^{-i\theta(m'-m)}
\left( 
\left(\frac34(n+m)^2+\frac76(n+m)+\frac12\right)
{}_{ sra}\bra{n,m,0}\delta_{0, \hat n_0}\ket{n',m',0}_{ sra}\right.
\nonumber\\
&&{}\left.+
{}_{ sra}\bra{n,m,0}\frac{(\hat a_{ s}\hat a_{ r'}^\dagger+\hat a_{ s}^\dagger\hat a_{ r'})^4 }{6(\hat n_0+1)(\hat n_0+2)}\ket{n',m',0}_{ sra}\right)
-
\delta_{n,n'}\delta_{m,m'}\left(n^2+n+\frac12\right)
\nonumber\\
&=&
\delta_{n,n'}\delta_{m,m'}
\left( 
\left(\frac34(n+m)^2+\frac76(n+m)+\frac12\right)
{}_{ sra}\bra{n,m,0}\delta_{0, \hat n_0}\ket{n,m,0}_{ sra}
-
\left(n^2+n+\frac12\right)
\right.\nonumber\\
&&{}\left.+
{}_{ sra}\bra{n,m,0}
\frac{
(6\hat n_{ s}^2+6\hat n_{ s}+3)\hat n_{ r'}^2+(6\hat n_{ s}^2-4\hat n_{ s}-2)\hat n_{ r'}+(3\hat n_{ s}^2-2\hat n_{ s})}
{6(\hat n_0+1)(\hat n_0+2)}\ket{n,m,0}_{ sra}\right)
\nonumber\\
&=&
\delta_{n,n'}\delta_{m,m'}\left(
2^{-m}\left(\frac34(n+m)^2+\frac76(n+m)+\frac12\right)
-\left(n^2+n+\frac12\right)\right.
\nonumber\\
&&{}\left.+
\sum_{k=0}^m\frac{m!\left((6n^2+6n+3)(m-k)(m-k-1)+(12n^2+2n+1)(m-k)+(3n^2-2n)\right)
}{6 (k+2)!(m-k)!2^m}
\right)
\nonumber\\
&=&
\delta_{n,n'}\delta_{m,m'}\left(\frac{1-2^{-m}}{3(m+1)}+\frac{m(9m^2+17m+6)}{12(m+1)}2^{-m}
+\frac{m(4+(3m^2+8m+3)2^{-m})}{6(m+1)(m+2)}n\right.
\nonumber\\&&\left.{}
+\frac{2(2m+5)n^2}{(m+1)(m+2)}(1-(1+m+\frac12m(m-1))2^{-m})
+\frac{m(4(m-1)^2+15(m-1)+2)}{4(m+1)(m+2)2^{m}}n^2\right).
\label{eq:diagonal_a1}
\end{eqnarray}
\end{widetext}
From this evaluation, we find that the operator in Eq.~(\ref{eq:proof_1_2}) is diagonal for the photon number basis of modes $sr$, and the last expression clarifies that all the diagonal elements are non-negative, i.e., all the 5 terms in the last expression are trivially non-negative for $n,m\in\mathbb Z_{\geq 0}$.
Therefore, the operator in Eq.~(\ref{eq:proof_1_2}) is non-negative, given that Eq.~(\ref{eq:diagonal_a1}) holds.

Each of the equalities in Eq.~(\ref{eq:diagonal_a1}) can be derived by noting the following points. In the first equality, we rewrite the annihilation and creation operators for modes $120$ by using those for modes $ sr'0$, i.e. we employ the relations in Eqs.~(\ref{rel:a_1}) and (\ref{rel:a_2}). 
In the second equality, we expand the term $(\hat a_{ s}\hat a_{ r'}^\dagger+\hat a_{ s}^\dagger\hat a_{ r'})^4$ and find that 
the off-diagonal terms vanish due to averaging with respect to $\theta$.
In the third equality, we exploit the relation
$\ket{m,0}_{ ra}=\frac{1}{\sqrt{m!2^{m}}}(\hat a_{ r'}^\dagger+\hat a_0^\dagger)^{m}\ket{0,0}_{ r'0}$, which comes from Eq.~(\ref{rel:a_r}), to expand the term $(\hat a_{ r'}^\dagger+\hat a_0^\dagger)^{m}$.
In the last equality, we take the summation over $k$, i.e., the relation $\sum_{k=0}^m\frac{m!}{(k+2)!(m-k-2)!2^m}=1-(1+m)2^{-m}$ etc. is used.

Next, we move on to showing the non-negativity of the operator in Eq.~(\ref{eq:proof_1_3}), which can be proved from the non-negativity of the small matrices. 
By definition, the operator can be expressed as 
\begin{eqnarray}
&&\sum_{n,m=0}^\infty
\hat \Delta^{(1)}_{n,m}
+
\sum_{(n,m)\in\Omega^{(1)}}
\hat W_{\pm,n,m}^{(1)},
\label{eq:submat_a1}
\end{eqnarray}
where
\begin{eqnarray}
\hat \Delta^{(1)}_{n,m}&:=&
(\delta_{n,m}+\delta_{n,0})h_2(n,m)
\hat P_{n,n,m},
\label{eq:proof_1_4_1}
\\
\Omega^{(1)}&:=&\{(n,m)|n,m\in \{1,2,\cdots\}\;\land \;n\leq m\},
\label{eq:proof_1_4_2}
\\
\hat W_{\pm, n,m}^{(1)}&:=&
 h_2(n-1,m)\hat P_{n-1,n-1,m}
\mp  h_1(n,m)^* \hat P_{n-1,n,m}
\nonumber\\&&
\mp  h_1(n,m) \hat P_{n,n-1,m}
+h_2(n,m) \hat P_{n,n,m},
\label{eq:proof_1_4_3}
\\
\hat P_{n,n',m}&:=&
\ket{n,m-n}_{ sr}
{}_{ sr}
\bra{n',m-n'},
\label{def:P_1}
\\
 h_1(n,m)&:=&
\frac{1}{2}\sqrt{n}(1-\sqrt{1-\frac{n}{m+1}}),
\label{eq:proof_1_4_6}
\end{eqnarray}
\begin{eqnarray}
 h_2(n,m)&:=&\frac{c_1}2\frac{n^2+n+\frac{1}{2}}{m+1}.
\label{eq:proof_1_4_5}
\end{eqnarray}
Here, we have used the relations
\begin{align}
\Lambda_{v\to sr}^\dagger(
\hat x_v
)
&=\frac12 \Lambda_{v\to sr}^\dagger(\hat a_{ v}+\hat a_{ v}^\dagger)
\nonumber\\
&=\sum_{n,m=1}^\infty\frac{\sqrt n}2(\ket{n,m-1}_{ sr}{}_{ sr}\bra{n-1,m}
\nonumber\\&\quad\quad\quad
+\ket{n-1,m}_{ sr}{}_{ sr}\bra{n,m-1}),
\label{def:x_v_hom_derive}
\\
{}_{ a}\big\langle0\bigr|\hat z_{\rm hom}(0)\bigl|0\big\rangle_{ a}\!\!&
=\frac{\hat a_{ s}\hat a_{ r}^\dagger+\hat a_{ s}^\dagger\hat a_{ r}}{2\sqrt{\hat n_{ s}+\hat n_{ r}+\hat I}}
\nonumber\\
&=\!\!\!\!\sum_{m,n=1}^\infty\!\!\!\sqrt{\frac{m}{m+n}}\frac{\sqrt n}{2}
(\ket{n,m-1}_{ sr}{}_{ sr}\bra{n-1,m}
\nonumber\\&\quad\quad\quad
+\ket{n-1,m}_{ sr}{}_{ sr}\bra{n,m-1}).
\label{def:x_l_hom_derive}
\end{align}
The first relation in Eq.~(\ref{def:x_l_hom_derive}) is justified by converting the annihilation and creation operators in $\hat z_{\rm hom}(0)$, i.e., $\hat a_{1}^\dagger \hat a_{1}=\hat n_{1}$,  for modes $ 120$ into those for modes $ sra$. In doing so, we used the relations in Eqs.~(\ref{rel:a_1}) and (\ref{rel:a_2}). 

The sufficient condition of the non-negativity of the expression in Eq.~(\ref{eq:submat_a1}) is  
that all the terms in the expression are non-negative, that is, $\hat \Delta_{n,m}^{(1)}\geq0$  for $n,m\in\mathbb Z_{\geq 0}$ and
 $\hat W_{\pm ,n,m}^{(1)}\geq0$ for $(n,m)\in\Omega^{(1)}$, and below, we prove each of the non-negativity.

The non-negativity of $\hat \Delta_{n,m}^{(1)}$ can be confirmed as follows. In the photon number basis of modes $ rs$, this operator is diagonal from the definition in Eq.~(\ref{eq:proof_1_4_1}), and moreover, its diagonal elements, which are defined by Eq.~(\ref{eq:proof_1_4_5}), are trivially non-negative for $n,m\in\mathbb Z_{\geq 0}$. 

The non-negativity of $\hat W_{\pm ,n,m}^{(1)}$ is guaranteed from the non-negativity of the $2\times2$ matrix
\begin{eqnarray}
\left(
\begin{array}{cc}
h_2(n-1,m)&\mp h_1(n,m)^*\\
\mp h_1(n,m)&h_2(n,m)\end{array}
\right),
\label{eq:principal_submatrix}
\end{eqnarray}
for $(n,m)\in\Omega^{(1)}$ since 
$\hat W_{\pm ,n,m}^{(1)}$ has all zero elements in the photon number basis of modes $ rs$ except for a principal submatrix of this form, which can be checked from the definition in Eq.~(\ref{eq:proof_1_4_3}).
Therefore, it is enough to check that its trace and determinant of the small matrix are non-negative.
The non-negativity of the trace is trivial from the definition in Eq.~(\ref{eq:proof_1_4_5}).
On the other hand, the non-negativity of the determinant can be confirmed by the following inequality.
\begin{widetext}
\begin{eqnarray}
&&
4\frac{(m+1)^2}{n(n+1)(\sqrt{n+1}-1)^2}
 (h_2(m,n)h_2(m,n-1)-|h_1(m,n)|^2)
\nonumber\\
&=&
\frac{1}{(n+1)(\sqrt{n+1}-1)^2}
\left(
c_1^2\frac{n^4+\frac14}n-(m+1)(\sqrt{m+1}-\sqrt{m-n+1})^2
\right)
\nonumber\\
&\geq&
c_1^{2}
\frac{n^4+\frac14}{n(n+1)(\sqrt{n+1}-1)^2}
-1
\geq
\frac5{8(\sqrt{2}-1)^2}
c_1^{2}-1
=0.
\label{eq:proof_1_5}
\end{eqnarray}
\end{widetext}
In the first equality, we employ the definitions in Eqs.~(\ref{eq:proof_1_4_6}) and (\ref{eq:proof_1_4_5}).
 The second relation comes from 
\begin{align}
(m+1)(\sqrt{m+1}-\sqrt{m-n+1})^2
&\leq(n+1)(\sqrt{n+1}-1)^2,
\label{eq:basic_relation_1}
\end{align}
for $1\leq n\leq m$, which is satisfied by any $(n,m)\in\Omega^{(1)}$. The last inequality is due to  the constraint $1\leq n$.

Now that we have shown the non-negativity of the operators in Eqs.~(\ref{eq:proof_1_2}) and (\ref{eq:proof_1_3}), following that Eq.~(\ref{res:LH_1}) is proved.

\subsection{Proof of Eq.~(\ref{res:LE_1})}
 Eq.~(\ref{res:LE_1}) can be proved in the same manner as Eq.~(\ref{res:LH_1}). For this, it is
enough to show the following inequality
\begin{align}
&\pm 
(\Lambda^\dagger_{v\to sr}\bigl(\hat F_v(\beta_\theta)\bigr)-
{}_{ a'b'c'}\big\langle0,0,0\bigr|
\hat z_{\rm het}(\theta))
\bigl|0,0,0\big\rangle_{ a'b'c'})
\nonumber\\
&\quad\quad\quad\leq
c_{1,1}\frac{\hat f_{\rm het}}{\hat{n}_s+\hat{n}_r+\hat I},
\label{def:Delta_het}
\end{align}
where 
\begin{align}
& c_{1,1}:=\frac{1}{\sqrt{30}}(\sqrt{5}-1)=0.22567\cdots,\\
&\hat f_{\rm het}:=
{}_{ a'b'c'}\big\langle0,0,0\bigr|
  f_{\rm het}(\hat n_{1'},\hat n_{2'},\hat n_{3'},\hat n_{4'},\hat n_{0'})
\bigl|0,0,0\big\rangle_{a'b'c'},
\label{def:hat_f_het}
\\
&\hat z_{\rm het}(\theta):=
\sqrt2\frac{(\hat n_{1'}-\hat n_{2'})\cos \theta
\quad+(\hat n_{3'}-\hat n_{4'})\sin \theta}
{\sqrt{\hat n_{ 1'}+\hat n_{ 2'}+\hat n_{ 3'}+\hat n_{ 4'}+\hat n_{ 0'}+\hat I}},
\label{def:x_l_het}
\end{align}
and $f_{\rm het}(n_{1'},n_{2'},n_{3'},n_{4'},n_{0'})$ is defined by Eq.~(\ref{def:f_het}). Eq.~(\ref{res:LE_1}) can be reproduced from Eq.~(\ref{def:Delta_het}) by using the input state $\hat \rho_2 $.

The sufficient condition for the relation in Eq.~(\ref{def:Delta_het}) to hold is the non-negativity of the two operators
\begin{align}
&
\frac{\hat f_{\rm het}}{\hat{n}_s+\hat{n}_r+\hat I}-\frac{\hat n_{s}^2+3\hat n_{ s}+2\hat I}{\hat{n}_s+\hat{n}_r+\hat I},
\label{eq:proof_1_1_1}
\\
&c_{1,1}\frac{\hat n_{ s}^2+3\hat n_{ s}+2\hat I}{\hat{n}_s+\hat{n}_r+\hat I}
\mp 
\bigl(
\Lambda_{v\to sr}^\dagger(\hat x_v)-
{}_{ a}\big\langle0\bigr|\hat z_{\rm hom}(0)\bigl|0\big\rangle_{ a}
\bigr),
\label{eq:proof_1_1_2}
\end{align}
since the non-negativity of the operator $c_{1,1}\times (\ref{eq:proof_1_1_1})+e^{-i \hat n_{ r}\theta}(\ref{eq:proof_1_1_2})e^{i \hat n_{ r}\theta}$ leads to the relation 
(\ref{def:Delta_het}), where we have used the relations $\hat F_v(\beta_\theta)=\hat x_v(\theta)$ and
\begin{align}
&e^{i\theta\hat n_{ r}}{}_{ a'b'c'}\big\langle0,0,0\bigr|
\hat z_{\rm het}(\theta)\bigl|0,0,0\big\rangle_{ a'b'c'}e^{-i\theta \hat n_{ r}}
\nonumber\\
&=
e^{i\theta \hat n_{ r}}\frac{e^{i\theta}\hat a_{ s}^\dagger \hat a_{ r}+e^{-i\theta}\hat a_{ s} \hat a_{ r}^\dagger}{2\sqrt{\hat n_{ s}+\hat n_{ r}+\hat I}}e^{-i\theta \hat n_{ r}}
=
{}_{ a}\big\langle0\bigr|\hat z_{\rm hom}(0)\bigl|0\big\rangle_{ a}.
\label{def:x_l_het_derive}
\end{align}
Note that in order to derive the relation in Eq.~(\ref{def:x_l_het_derive}),  we rewrite the annihilation and creation operators in $\hat z_{\rm het}(0)$ for modes $1'2'3'4'0'$ into those for modes $ srb'c'a'$, i.e., we use the relations in Eqs.~(\ref{rel:a_1'}),  (\ref{rel:a_2'}),  (\ref{rel:a_3'}), (\ref{rel:a_4'}), and (\ref{rel:a_r'_}). Below, we show the non-negativity of the operators in Eqs.~(\ref{eq:proof_1_1_1}) and (\ref{eq:proof_1_1_2}).

First, the non-negativity of the operator in Eq.~(\ref{eq:proof_1_1_1}) can be shown by the expression
\begin{widetext}
\begin{eqnarray}
&&(n+m+1){}_{ sr} \bra{n,m}
\frac{\hat f_{\rm het}}{\hat{n}_s+\hat{n}_r+\hat I}-\frac{\hat n_{ s}^2+3\hat n_{ s}+2\hat I}{\hat{n}_s+\hat{n}_r+\hat I}
\ket{n',m'}_{ sr}
\nonumber\\
&=&
\left( (\frac72(n+m)+2)
{}_{ sra'}\bra{n,m,0}\delta_{0,\hat n_{0'}}\ket{n',m',0}_{sra'}\right.
\nonumber\\
&&{}\left.+
{}_{ sra'}\bra{n,m,0}\frac{
(\hat n_{ s}^2+3\hat n_{ s}+2)\hat n_{ r'}(\hat n_{ r'}-1)
+(\hat 4n_{ s}^2+4\hat n_{ s}+1)\hat n_{ r'}+
\hat 2n_{ s}^2-\hat n_{ s}
 }{(\hat n_{0'}+1)(\hat n_{0'}+2)}\ket{n',m',0}_{ sra'}\right)
\nonumber\\
&&{}-
\delta_{n,n'}\delta_{m,m'}\left(n^2+3n+2\right)
\nonumber\\
&=&
\delta_{n,n'}\delta_{m,m'}\left(
2^{-m}\left(\frac72(n+m)+2\right)\right.
\nonumber\\
&&{}+
\sum_{k=0}^m\frac{m!\left((n^2+3n+2)(m-k)(m-k-1)+(4n^2+4n+1)(m-k)+(2n^2-n)\right)
}{6 (k+2)!(m-k)!2^m}
\nonumber\\
&&{}-
\left.\left(n^2+3n+2\right)\right)
\nonumber\\
&=&
\delta_{n,n'}\delta_{m,m'}\left(
\frac2{m+1}(1-2^{-m})+\frac{m(3m+1)}{2(m+1)}2^{-m}\right.
+\frac{m(8(m-1)^2+27(m-1)+2)}{2(m+1)(m+2)}2^{-m}n
\nonumber\\&&{}
+\frac{m(3m-1)(m+3)}{(m+1)(m+2)}2^{-m}n(n-1)
\left.{}
+\frac{4(2m+3)+8(m+3)n}{(m+1)(m+2)}
(1-(1+m+\frac12m(m-1))2^{-m})n\right).
\label{eq:503}
\end{eqnarray}
Here, in the first equality, we transform the annihilation and creation operators for modes $ 1'2'3'4'0'$ into those for modes $ sr'b'c'0'$, i.e., we use the relations in Eqs.~(\ref{rel:a_1'}),  (\ref{rel:a_2'}),  (\ref{rel:a_3'}), and  (\ref{rel:a_4'}). 
We can show the other two equalities in the same manner as that in the case of Eq.~(\ref{eq:diagonal_a1}). Note that we can easily check that the five terms in the last expression of the diagonal elements are non-negative for any $n,m\in\mathbb Z_{\geq 0}$.
\end{widetext}

Next, we show the non-negativity of the operators in Eq.~(\ref{eq:proof_1_1_2}). For this, note that it can be expressed by the form of 
(\ref{eq:submat_a1})
where we replace the definition $h_2(n,m)$ in Eq.~(\ref{eq:proof_1_4_5}) with 
\begin{eqnarray}
\frac{c_{1,1}}2\frac{n^2+3n+2}{m+1},
\label{def:mod_h_1_1}
\end{eqnarray}
but we use the other definitions in Eqs.~(\ref{eq:proof_1_4_1})$-$(\ref{eq:proof_1_4_6}) as they are. The modification of the definition of $h_2(n,m)$ causes the change in the definitions of $\hat \Delta_{n,m}^{(1)}$ and $\hat W_{\pm ,n,m}^{(1)}$ implicitly.
Therefore, in order to prove the non-negativity of the operator in Eq,~(\ref{eq:proof_1_1_2}), it suffices if we can show the non-negativity of the modified operator  $\hat \Delta_{n,m}^{(1)}$ for $n,m\in\mathbb Z_{\geq 0}$, 
as well as the non-negativity of the trace and the determinant of the modified $2\times 2$ matrix in Eq.~(\ref{eq:principal_submatrix}), which is the principal submatrix of $\hat W_{\pm,n,m}^{(1)}$, for $(n,m)\in\Omega^{(1)}$. 
The non-negativity of the modified $\hat \Delta_{n,m}^{(1)}$ and the trace of the matrix can be checked from the explicit expression of $h_2(n,m)$ in Eq.~(\ref{def:mod_h_1_1}).
The non-negativity of  the determinant of the modified matrix can be shown by using the following inequality.
\begin{widetext}
\begin{eqnarray}
&&
\frac{4(m+1)^2}{n(n+1)(\sqrt{n+1}-1)^2}
(
h_2(m,n)h_2(m,n-1)-|h_1(m,n)|^2)
\nonumber\\
&=&
\frac{1}{(n+1)(\sqrt{n+1}-1)^2}
\Bigl(c_{1,1}^2(n+1)^2(n+2)
-(m+1)(\sqrt{m+1}-\sqrt{m-n+1})^2
\Bigr)
\nonumber\\&\geq&
\frac{n^2+3n+2}{(\sqrt{n+1}-1)^2}c_{1,1}^{2}
-1
\geq
\frac{30}{(\sqrt{5}-1)^2}c_{1,1}^{2}-1
=0.
\end{eqnarray}
\end{widetext}
The second relation is due to Eq.~(\ref{eq:basic_relation_1}), and the third relation comes from the fact that the left hand side is minimized at $n=4$ for $ n\in\mathbb Z_{\geq 1}$.

Now, we have finished showing the non-negativity of the operators in Eqs.~(\ref{eq:proof_1_1_1}) and (\ref{eq:proof_1_1_2}), leading to the proof of  Eq.~(\ref{res:LE_1}).

\subsection{Proof of  the second inequality of Eq.~(\ref{res:LH_2})}
As a representative  of the second set $\mathcal E_2$, i.e., the set of 
Eqs.~(\ref{res:LH_2}), (\ref{res:LH_2_m}),
 (\ref{res:LE_2}), and (\ref{res:LE_2_m}),
we select the second inequality in Eq.~(\ref{res:LH_2}). This relation holds, if the following inequality holds,
\begin{align}
\left(
\Lambda_{v\to sr}^\dagger(\hat x_v^2(\theta))
- {}_{ a}\big\langle0\bigr|\hat z_{\rm hom}(\theta)^2\bigl|0\big\rangle_{ a}
\right)
&\leq 
c_2\frac{\hat f_{\rm hom}}{\hat n_{ s}+\hat n_{ r}+\hat I},
\label{eq:proof_2_1}
\end{align}
 where  
\begin{eqnarray}
c_2&:=&\frac{12+\sqrt{1145-624\sqrt 2}}{26}=1.08472\cdots,
\end{eqnarray}
and, $\hat z_{\rm hom}(\theta)$ and $\hat f_{\rm hom}$ are defined by Eq.~(\ref{def:x_l_hom}) and Eq.~(\ref{def:hat_f_hom}), respectively.
The second inequality of Eq.~(\ref{res:LH_2}) can be rebuilt from Eq.~(\ref{eq:proof_2_1}) in the same manner as the derivation of  Eq.~(\ref{res:LH_1}) from Eq.~(\ref{eq:proof_1_1}).

Like the case of Eq.~(\ref{eq:proof_1_1}), the correctness of Eq.~(\ref{eq:proof_2_1}) can be demonstrated from the non-negativity of the operators in Eq.~(\ref{eq:proof_1_2}), which we have already shown, and that of
\begin{align}
c_2\frac{\hat n_{ s}^2+\hat n_{ s}+\frac 12\hat I}{\hat n_{ s}+\hat n_{ r}+\hat I}
-
\Lambda_{v\to sr}^\dagger(\hat x_v^2(0))
+ {}_{ a}\big\langle0\bigr|\hat z_{\rm hom}(0)^2\bigl|0\big\rangle_{ a}.
\label{eq:proof_2_3}
\end{align}
To show the non-negativity of the latter operator, like the case in Eqs.~(\ref{def:x_v_hom_derive}) and (\ref{def:x_l_hom_derive}), we rewrite the last two operators in Eq.~(\ref{eq:proof_2_3}) as
\begin{widetext}
\begin{align}
\Lambda_{v\to sr}^\dagger(\hat x_v^2(0))
&=\frac14\Lambda_{v\to sr}^\dagger((\hat a_{ v}+\hat a_{ v}^\dagger)^2)
\nonumber\\
&=
\sum_{n,m=2}^\infty
\frac{\sqrt{ n(n-1)}}4(\ket{n,m-2}_{ sr}{}_{ sr}\bra{n-2,m}
+\ket{n-2,m}_{ sr}{}_{ sr}\bra{n,m-2})
+\sum_{n,m=0}^\infty(\frac12n +\frac14)\ket{n,m}_{ sr}{}_{ sr}\bra{n,m}),
\label{def:x2_v_hom_derive}
\nonumber\\
\\
 {}_{ a}\big\langle0\bigr|\hat z_{\rm hom}(0)^2\bigl|0\big\rangle_{ a}
&=
\frac{\hat a_{ s}^{\dagger 2}\hat a_{ r}^2+\hat a_{ s}^{ 2}\hat a_{ r}^{\dagger 2}+2\hat n_{ s} \hat n_{ r}+2\hat n_{ s} +\hat n_{ r}}{4(\hat n_{ s}+\hat n_{ r}+\hat I)}
\nonumber\\
&=
\sum_{n,m=2}^\infty
\frac{\sqrt {n(n-1)}}{4}
\frac{\sqrt{m(m-1)}}{m+n-1}
\left(
\ket{n,m-2}_{ sr}{}_{ sr}\bra{n-2,m}
+\ket{n-2,m}_{ sr}{}_{ sr}\bra{n,m-2}
\right)\nonumber\\&{}
\makebox[1cm]{}+
\sum_{n,m=0}^\infty
\left(\frac12n+\frac14-\frac{2n^2+n+1}{4(m+n+1)}\right)
\ket{n,m}_{ sr}{}_{ sr}\bra{n,m}.
\label{def:x2_l_hom_derive}
\end{align}
\end{widetext}
With the help of these expressions, the operator in Eq.~(\ref{eq:proof_2_3}) can be rewritten as 
\begin{eqnarray}
&&
\sum_{n,m=0}^\infty
\hat \Delta^{(2)}_{n,m} 
+
\sum_{(n,m)\in\Omega^{(2)}}
\hat W^{(2)}_{n,m},
\label{eq:submat_a2}
\end{eqnarray}
where we have made the following definitions
\begin{eqnarray}
\hat \Delta^{(2)}_{n,m}&=&
(\delta_{n,m}+\delta_{n,m-1}+\delta_{n,0}+\delta_{n,1})h_4(n,m)\hat P_{n,n,m},
\label{eq:proof_sub_3_1}
\nonumber\\
\\
\Omega^{(2)}&:=&\{(n,m)|n,m\in\mathbb Z_{\geq 1}\;\land \;n<m\},
\label{eq:proof_sub_3_2}
\\
\hat W^{(2)}_{n,m}&:=&
h_4(n-1,m)\hat P_{n-1,n-1,m}
\nonumber\\&&{}
-  h_3(n,m)^*\hat P_{n-1,n+1,m}
\nonumber
\end{eqnarray}
\begin{eqnarray}
&&
{}-  h_3(n,m)\hat P_{n+1,n-1,m}
\nonumber\\&&{}
+h_4(n+1,m)\hat P_{n+1,n+1,m},
\label{eq:proof_sub_3_3}
\\
h_3(n,m)&:=&
\frac{\sqrt{n(n+1)}}{4}(1-\frac{\sqrt{(m-n)(m-n+1)}}{m+1}),
\nonumber\\
\label{eq:proof_sub_3_5}
\\
h_4(n,m)&:=&\frac{(2c_2-1)(2n^{2}+2n+1)+n}{8(m+1)},
\label{eq:proof_sub_3_4}
\end{eqnarray}
and $\hat P_{n,n',m}$ is defined by Eq.~(\ref{def:P_1}).

In order to show the non-negativity of the operator in Eq.~(\ref{eq:proof_2_3}), it suffices to show the non-negativity of $\hat \Delta_{n,m}^{(2)}$ ($\hat W^{(2)}_{n,m}$) for $n,m\in\mathbb Z_{\geq 0}$($(n,m)\in\Omega^{(2)}$).
The non-negativity of  $\hat \Delta_{n,m}^{(2)}$ for  $n,m\in\mathbb Z_{\geq 0}$
 is trivial from the fact $c_2> 1$ and the definitions in Eqs.~(\ref{eq:proof_sub_3_1}) and (\ref{eq:proof_sub_3_4}).
As for the non-negativity of the operator $\hat W^{(2)}_{n,m}$, it can be demonstrated if we can show the non-negativity of the principal submatrix
\begin{eqnarray}
\left(
\begin{array}{cc}
h_4(n-1,m)&h_3(n,m)^*\\
 h_3(n,m)&h_4(n+1,m)\end{array}
\right),
\label{eq:principal_submatrix_2}
\end{eqnarray}
\newpage
in the photon number basis of modes $ rs$ since the other matrix elements in the basis are zero (see Eq.~(\ref{eq:proof_sub_3_3})).
The non-negativity of this matrix can be confirmed from the facts that the definition in Eq.~(\ref{eq:proof_sub_3_4}) leads to the non-negativity of the trace of it, and the non-negativity of the determinant of it can be checked as follows:
\begin{widetext}
\begin{eqnarray}
&&
16(m+1)^2(h_4(m,n-1)h_4(m,n+1)
-|h_3(m,n)|^2)
\nonumber\\
&=&
16(m+1)^2h_4(m,n-1)h_4(m,n+1)
-n(n+1)(m+1-\sqrt{(m-n)(m-n+1)})^2
\nonumber\\
&\geq&
16(m+1)^2h_4(m,n-1)h_4(m,n+1)
-n(n+1)(n+2-\sqrt{2})^2
\nonumber\\
&=&
(n-1)\left(
4c_2(c_2-1)(n-1)^3
+\frac{1}{24}(524c_2(c_2-1)+44(c_2-1)+25)(n-1)^2
\right. \nonumber\\&&{}\left.
+\frac{7}{2}(10c_2(c_2-1)+2(c_2-1)+1)(n-1)
+\frac{1}{24}(292c_2(c_2-1)+172(c_2-1)+59)
\right)
\geq
 0.
\end{eqnarray}
\end{widetext}
The second relation comes from 
\begin{eqnarray}
0&\leq&
m+1-\sqrt{(m-n)(m-n+1))}
\leq
n+2-\sqrt{2},
\label{rel:lb_2}
\nonumber\\
\end{eqnarray}
for $1\leq n\leq m-1$, which is satisfied by any $(n,m)\in\Omega^{(2)}$. For the derivation of the fourth expression from the third one,
we use the relation $\sqrt 2=-\frac{13}{12}c_2^2+c_2+\frac{77}{48}$ in order to simplify the expression. The last inequality comes from $c_2>1$.

Therefore, the operator in Eq.~(\ref{eq:proof_2_3}) is shown to be non-negative, and therefore, the second inequality of Eq.~(\ref{res:LH_2}) is proved.

\subsection{Proof of the first inequality  of Eq.~(\ref{res:LH_2}), and Eqs.~(\ref{res:LH_2_m}), (\ref{res:LE_2}), and (\ref{res:LE_2_m})}
The first inequality of Eq.~(\ref{res:LH_2}) and
Eqs.~(\ref{res:LH_2_m}), (\ref{res:LE_2}), and (\ref{res:LE_2_m}), i.e., the rest of the relations in the second set $\mathcal E_2$, can be proved in exactly the same way as the second inequality of  Eq.~(\ref{res:LH_2}). Therefore, we will only give key expressions that enable us to construct the entire proof.

The operator relations that are sufficient for proving the four respective relations are
\begin{widetext}
\begin{align}
-
\left(
\Lambda_{v\to sr}^\dagger(\hat x_v^2(\theta))
- {}_{ a}\big\langle0\bigr|\hat z_{\rm hom}(\theta)^2\bigl|0\big\rangle_{ a}
\right)
&\leq 
\frac{c_{2,1}\hat f_{\rm hom}}{\hat n_{ s}+\hat n_{ r}+\hat I},
\label{eq:proof_2_1_1}
\\
\pm
\frac12\left(
\Lambda_{v\to sr}^\dagger(\hat x_v^2(\pi/4)-\hat x_v^2(3\pi/4))
- {}_{ a}\big\langle0\bigr|(\hat z_{\rm hom}(\pi/4)^2-\hat z_{\rm hom}(3\pi/4)^2)\bigl|0\big\rangle_{ a}
\right)
&\leq 
\frac{c_{2,2}\hat f_{\rm hom}}{\hat n_{s}+\hat n_{r}+\hat I},
\label{eq:proof_2_2_1}
\\
b
\left(
\Lambda_{v\to sr}^\dagger\bigl(\hat F_v(\beta_\theta^2)\bigr)
-{}_{a'b'c'}\big\langle0,0,0\bigr|
\hat z_{\rm het}(\theta)^2
\bigl|0,0,0\big\rangle_{a'b'c'}
\right)
&\leq 
\frac{c_{2,3}^{(b)}\hat f_{\rm het}}{\hat n_{ s}+\hat n_{ r}+\hat I},
\label{eq:proof_2_3_1}
\\
\pm 
\left(
\Lambda_{v\to sr}^\dagger\bigl(\hat F_v(\beta_{0}\beta_{\pi/2})\bigr)
-{}_{a'b'c'}\big\langle0,0,0\bigr|
\hat z_{\rm het}(0)\hat z_{\rm het}(\pi/2)
\bigl|0,0,0\big\rangle_{a'b'c'}
\right)
&\leq 
\frac12\frac{\hat f_{\rm het}}{\hat n_{ s}+\hat n_{ r}+\hat I},
\label{eq:proof_2_4_1}
\end{align}
\end{widetext}
where $b\in\{+,-\}$,
\begin{eqnarray}
c_{2,1}&:=&\frac{\sqrt{1145-624\sqrt 2}-12}{26}= 0.16164\cdots,
\end{eqnarray}
\begin{eqnarray}
c_{2,2}&:=&\sqrt{\frac{2}{13}}(3-\sqrt2)= 0.62199\cdots,
\\
c_{2,3}^{(+)}&:=&1, 
\end{eqnarray}
\begin{eqnarray}
c_{2,3}^{(-)}&:=&\frac{\sqrt{5(653-288\sqrt 2)}-25}{120}= 0.08374\cdots,
\end{eqnarray}
and $\hat f_{\rm hom}$,  $\hat f_{\rm het}$, $\hat {z}_{\rm hom}(\theta)$, and $\hat {z}_{\rm het}(\theta)$ are defined by Eqs.~(\ref{def:hat_f_hom}),
 (\ref{def:hat_f_het}), (\ref{def:x_l_hom}), and (\ref{def:x_l_het}),
 respectively.

We can evaluate the operators in Eqs.~(\ref{eq:proof_2_1_1})$-$(\ref{eq:proof_2_4_1}) as
\begin{align}
&
\frac12\Lambda_{v\to sr}^\dagger(\hat x_v^2(\pi/4)-\hat x_v^2(3\pi/4))
=\hat Y_I,
\\
&
\frac12{}_{ a}\big\langle0\bigr|(\hat z_{\rm hom}(\pi/4)^2-\hat z_{\rm hom}(3\pi/4)^2)\bigl|0\big\rangle_{ a}
=
\hat Y_R,
\\
&e^{i\theta\hat n_{r}}
\Lambda_{v\to sr}^\dagger
\bigl(
\hat F_v(\beta_\theta^2)
\bigr)
e^{-i\theta\hat n_{ r}}
=
\Lambda_{v\to sr}^\dagger(\hat x_v(0)^2)+\frac14 \hat I,
\end{align}
\begin{align}
&e^{i\theta\hat n_{ r}}
{}_{a'b'c'}\big\langle0,0,0\bigr|
\hat z_{\rm het}(\theta)^2
\bigl|0,0,0\big\rangle_{a'b'c'}
e^{-i\theta\hat n_{ r}}
\nonumber\\&
=
{}_{ a}\big\langle0\bigr|\hat z_{\rm hom}(0)^2\bigl|0\big\rangle_{ a}
+
\sum_{n,m=0}^\infty
\frac{m+2n}{4(m+n+1)}\ket{n,m}_{ sr}{}_{ sr}\bra{n,m},
\\
&\Lambda_{v\to sr}^\dagger
\bigl(
\hat F_v(\beta_{0}\beta_{\pi/2})
\bigr)
=\hat Y_I,
\\
&{}_{a'b'c'}\big\langle0,0,0\bigr|
\hat z_{\rm het}(0)\hat z_{\rm het}(\pi/2)
\bigl|0,0,0\big\rangle_{a'b'c'}
=
\hat Y_R\,,
\end{align}
where
\begin{widetext}
\begin{align}
\hat Y_I
&:=\sum_{n,m=2}^\infty\frac{\sqrt {n(n-1)}}4i
(\ket{n,m-2}_{sr}{}_{sr}\bra{n-2,m}
-\ket{n-2,m}_{sr}{}_{sr}\bra{n,m-2}),
\\
\hat Y_R
&:=
\sum_{n,m=2}^\infty
\frac{\sqrt {n(n-1)}}{4}
\frac{\sqrt{m(m-1)}}{m+n
+
1}
i\left(
\ket{n,m-2}_{ sr}{}_{ sr}\bra{n-2,m}
-\ket{n-2,m}_{ sr}{}_{ sr}\bra{n,m-2}
\right).
\end{align}
Using these expressions, the relations in Eqs.~(\ref{eq:proof_2_1_1})$-$(\ref{eq:proof_2_4_1}) can respectively be derived from the non-negativity of the following operators
\begin{eqnarray}
&&
c_{2,1}\frac{\hat n_{ s}^2+\hat n_{ s}+\frac 12\hat I}{\hat n_{ s}+\hat n_{ r}+\hat I}
+
\Lambda_{v\to sr}^\dagger(\hat x_v(0)^2)
-
{}_{ a}\big\langle0\bigr|\hat z_{\rm hom}(0)^2\bigl|0\big\rangle_{ a},
\label{eq:proof_2_1_2}
\\
&&
c_{2,2}\frac{\hat n_{ s}^2+\hat n_{ s}+\frac 12\hat I}{\hat n_{ s}+\hat n_{ r}+\hat I}
\mp
\left(
\hat Y_I-\hat Y_R
\right),
\label{eq:proof_2_2_2}
\\
&&
c_{2,3}^{(b)}\frac{\hat n_{ s}^2+3\hat n_{ s}+2\hat I}{\hat n_{ s}+\hat n_{ r}+\hat I}
-
b\left(
\Lambda_{v\to sr}^\dagger(\hat x_v(0)^2)
+\frac14 \hat I-
{}_{ a}\big\langle0\bigr|\hat z_{\rm hom}(0)^2\bigl|0\big\rangle_{ a}
-\sum_{n,m=0}^\infty
\frac{m+2n}{4(m+n+1)}\ket{n,m}_{ sr}{}_{ sr}\bra{n,m}\right),
\label{eq:proof_2_3_2}
\\
&&\frac12\frac{\hat n_{ s}^2+3\hat n_{ s}+2\hat I}{\hat n_{ s}+\hat n_{ r}+\hat I}
\mp
\left(
\hat Y_I-\hat Y_R
\right)\,.
\label{eq:proof_2_4_2}
\end{eqnarray}
Here, we have also use the non-negativity of 
 the operator in Eq.~(\ref{eq:proof_1_2}) or Eq.~(\ref{eq:proof_1_1_1}), which was already shown.
 
\end{widetext}

Any of the operators in Eqs.~(\ref{eq:proof_2_1_2})$-$(\ref{eq:proof_2_4_2}) can be expressed with the form of Eq.~(\ref{eq:submat_a2}), in which we use the definitions in Eqs.~(\ref{eq:proof_sub_3_1})$-$(\ref{eq:proof_sub_3_3}), except for 
the definitions of $h_3(n,m)$ and $h_4(n,m)$, i.e.,
Eqs.~(\ref{eq:proof_sub_3_5}) and (\ref{eq:proof_sub_3_4}). Depending on the respective operators in Eqs.~(\ref{eq:proof_2_1_2})$-$(\ref{eq:proof_2_4_2}), the definition of $h_3(n,m)$ is changed into
\begin{eqnarray}
&&-
\frac{\sqrt{n(n+1)}}{4}(1-\frac{\sqrt{(m-n)(m-n+1)}}{m+1}),
\label{def:h_4_1}
\\
&&\pm i
\frac{\sqrt{n(n+1)}}{4}(1-\frac{\sqrt{(m-n)(m-n+1)}}{m+1}),
\label{def:h_4_2}
\\
&&b
\frac{\sqrt{n(n+1)}}{4}(1-\frac{\sqrt{(m-n)(m-n+1)}}{m+1}),
\label{def:h_4_3}
\\
&&\pm i
\frac{\sqrt{n(n+1)}}{4}(1-\frac{\sqrt{(m-n)(m-n+1)}}{m+1}),
\label{def:h_4_4}
\end{eqnarray}
  and the definition of   $h_4(n,m)$ is modified  to 
\begin{eqnarray}
&&\frac{(2c_{2,1}+1)(2n^{2}+2n+1)-n}{8(m+1)},
\label{def:h_3_2_1}
\\
&&\frac{c_{2,2}}2\frac{n^{2}+n+\frac12}{m+1},
\label{def:h_3_2_2}
\\
&&
\frac{2c_{2,3}^{(b)}(n+1)(n+2)-b(n^{2}+1)}{4(m+1)},
\label{def:h_3_2_3}
\\
&&
\frac{1}4\frac{n^{2}+3n+2}{m+1}.
\label{def:h_3_2_4}
\end{eqnarray}
The modification of the definitions of $h_3(n,m)$ and  $h_4(n,m)$ affects the definitions of $\hat \Delta_{n,m}^{(2)}$ and $\hat W_{n,m}^{(2)}$, implicitly.
Therefore, the non-negativity of the operators in Eqs.~(\ref{eq:proof_2_1_2})$-$(\ref{eq:proof_2_4_2}) are guaranteed from the non-negativity of the modified $\hat \Delta_{n,m}^{(2)}$ for $n,m\in\mathbb Z_{\geq 0}$
and the non-negativity of the trace and the determinant of the modified $2\times 2$ matrix in Eq.~(\ref{eq:principal_submatrix_2}), which is the principal submatrix of $\hat W_{n,m}^{(2)}$, for $(n,m)\in\Omega^{(2)}$.
For any of the cases, the non-negativity of the modified $\hat \Delta_{n,m}^{(2)}$ and the trace of the matrix can be checked from the explicit expressions of the modified $h_4(n,m)$ in Eqs.~(\ref{def:h_3_2_1})$-$(\ref{def:h_3_2_4}).
By using the relation in Eq.~(\ref{rel:lb_2}), a lower bound of the the determinant of the modified matrix for any of the cases can be expressed as
\begin{eqnarray}
&&16(m+1)^2(h_4(n-1,m)h_4(n+1,m)
-|h_3(n,m)|^2)
\nonumber\\
&\geq&
16(m+1)^2h_4(n-1,m)h_4(n+1,m)
\nonumber\\&&{}
-n(n+1)(n+2-\sqrt{2})^2\,.
\end{eqnarray}
This is so because the value of $|h_3(n,m)|^2$ remain the same in all the cases as we can see from the modified definitions of $h_3(n,m)$ in Eqs.~(\ref{def:h_4_1})$-$(\ref{def:h_4_4}). In each of the cases, the right hand side can be respectively evaluated as
\begin{widetext}
\begin{eqnarray}
&&(n-1)
\left(4(c_{2,1}+1)c_{2,1}(n-2)^3
+\frac1{24}(812c_{2,1}^2+768(c-\frac18)+77)(n-2)^2
\right.\nonumber\\ &&{}\left.
+\frac1{12}(1088c_{2,1}^2+960(c_{2,1}-\frac18)+59)(n-2)
+\frac1{49}(3577(c_{2,1}-\frac17)^2+3815(c_{2,1}-\frac17)+31)\right),
\\
\nonumber\\
&&(n-1)
\left(
4(c_{2,2}^2-\frac14)(n-2)^3
+\frac1{24}(812(c_{2,2}^2-\frac14)+3)(n-2)^2
+\frac{272}3(c_{2,2}^2-\frac14)(n-2)
+\frac1{3}(219(c_{2,2}^2-\frac13)+13)\right),
\nonumber\\
\\
\nonumber\\
&&
(7+2\sqrt 2)(n-1)^3+3(17+4\sqrt2)(n-1)^2
+11(9+2\sqrt2)(n-1)+35+12\sqrt2,
\\
\nonumber\\
&&(n-2)\left(
4c_{2,3}^{(-)}(c_{2,3}^{(-)}+1)(n-3)^2(n-1)
+\frac1{12}(480(c_{2,3}^{(-)}-\frac1{12})^2+556(c_{2,3}^{(-)}-\frac1{12})+27)
(n-3)(n-1)
\right.
\nonumber\\&&{}\left.
+\frac16(648(c_{2,3}^{(-)}-\frac1{12})^2+714(c_{2,3}^{(-)}-\frac1{12})+13)
\left(n-3+\frac{(12c_{2,3}^{(-)}-1)(20c_{2,3}^{(-)}+19)}{(c_{2,3}^{(-)}+1)(108c_{2,3}^{(-)}-7)}\right)\right),
\\
\nonumber\\
&&n(1+n)((2\sqrt2+1)n+4\sqrt2)\,.
\end{eqnarray}
\end{widetext}
From top to bottom, they correspond to 
 the case of Eq.~(\ref{eq:proof_2_1_2}), i.e.,  $h_4(n,m)$ is equal to Eq.~(\ref{def:h_3_2_1}),
 the case of Eq.~(\ref{eq:proof_2_2_2}), i.e.,  $h_4(n,m)$ is equal to Eq.~(\ref{def:h_3_2_2}),
 the case of Eq.~(\ref{eq:proof_2_3_2}) with $b=+$, i.e., $h_4(n,m)$ and $b$ are equal to Eq.~(\ref{def:h_3_2_3}) and $+$,
 the case of Eq.~(\ref{eq:proof_2_3_2}) with $b=-$, i.e., $h_4(n,m)$ and $b$ are equal to Eq.~(\ref{def:h_3_2_3}) and $-$,
 the case of Eq.~(\ref{eq:proof_2_4_2}), i.e.,  $h_4(n,m)$ is equal to Eq.~(\ref{def:h_3_2_4}).
The non-negativity of these values for $(n,m)\in\Omega^{(2)}$, i.e. $n\in\mathbb Z_{\geq 1}$,
can be confirmed from the above expressions as well as the facts $c_{2,1}>\frac17$, $c_{2,2}^2>\frac13$, $c_{2,3}^{(-)}>\frac{1}{12}$, and $1>\frac{(12c_{2,3}^{(-)}-1)(20c_{2,3}^{(-)}+19)}{(c_{2,3}^{(-)}+1)(108c_{2,3}^{(-)}-7)}>0$.

These are key expressions for confirming the correctness of the first inequality in Eq.~(\ref{res:LH_2}) and
Eqs.~(\ref{res:LH_2_m}), (\ref{res:LE_2}), and (\ref{res:LE_2_m}).

\subsection{Proof of Eq.~(\ref{res:GH_LH})}
As a representative of the third set $\mathcal E_3$ , i.e., the set in Eqs.~(\ref{res:GH_LH}), (\ref{res:GE_LE}), (\ref{res:GH_LE}), and (\ref{res:GE_LH}), we select  Eq.~(\ref{res:GH_LH}) and prove this.

Eq.~(\ref{res:GH_LH}) is a relation for the three-mode states $\hat \rho_3$. The first mode $s_1$ is measured by the ideal homodyne detection, and the other two modes $s_2r_2$ are measured by the SHD.
Using this notation, a sufficient condition for the inequality in Eq.~(\ref{res:GH_LH}) to hold can be expressed as followings:
\begin{eqnarray}
&&
\pm 
\hat x_{s_1}(\phi)
\left(
\Lambda_{r_2\to s_2r_2}(\hat x_{r_2}(\theta))-{}_{a_1}\bra{0}\hat z^{(2)}_{\rm hom}(\theta)\ket{0}_{a_1}
\right)
\nonumber\\
&\leq &{}
c_3  (\hat n_{s_1}+\frac12 \hat I)\frac{\hat f^{(2)}_{\rm hom}
}{\hat n_{ s_2}+\hat n_{ r_2}+\hat I},
\label{eq:proof_3_1}
\end{eqnarray}
where     
\begin{eqnarray}
c_3:=\sqrt{\frac{32}{15}}(\sqrt2-1)= 0.60499\cdots,
\label{def:c_3}
\end{eqnarray}
and the operators $\hat f_{\rm hom}^{(j)}$ and $\hat z^{(j)}_{\rm hom}(\theta)$ are defined in the similar way as 
$\hat f_{\rm hom}$ in Eq.~(\ref{def:hat_f_hom})  and $\hat z_{\rm hom}(\theta)$ in Eq.~(\ref{def:x_l_hom}): 
\begin{align}
&\hat f_{\rm hom}^{(j)}:=
\nonumber\\
&
\sum_{\theta\in\{\pi/4,\pi/2,3\pi/4\}}\!\!\!\frac 14 e^{-i\theta\hat n_{ r_j}}
{}_{ a_j}\bra{0}
  f_{\rm hom}(\hat n_{1_j},\hat n_{2_j},\hat n_{0_j})
\ket{0}_{ a_j}
e^{i\theta\hat n_{ r_j}},
\label{def:hat_f_hom_j}
\\
&\hat z^{(j)}_{\rm hom}(\theta)
:=e^{-i\theta\hat n_{ r_j}}\frac{\hat n_{1_j}-\hat n_{2_j}
}{\sqrt{2(\hat n_{ 1_j}+\hat n_{ 2_j}+\hat n_{ 0_j}+\hat I)}}e^{i\theta\hat n_{ r_j}}.
\label{def:hat_x_l_hom_j}
\end{align}

The inequality in Eq.~(\ref{res:GH_LH}) can be reproduced by using input state $\hat \rho_3$ like we did in the previous proofs.

As was the case in Eq.~(\ref{eq:proof_1_1}), Eq.~(\ref{eq:proof_3_1}) is derived from the non-negativity of 
\begin{align}
&&\frac{\hat f^{(j)}_{\rm hom}}{\hat{n}_{s_j}+\hat{n}_{r_j}+\hat I}
-\frac{(\hat n_{ s_j})^2+\hat n_{ s_j}+\frac12 \hat I}{\hat{n}_{s_j}+\hat{n}_{r_j}+\hat I}=:\hat{ \mathcal F}_{\rm hom}^{(j)},
\label{eq:proof_4_2_0}
\end{align}
which is the operator in Eq.~(\ref{eq:proof_1_2}) for the $j$-the pair in $\hat\rho_{2N}$, and
\begin{align}
&
c_3 (\hat n_{s_1}+\frac12 \hat I)
\frac{
\hat n_{ s_2}^2+\hat n_{ s_2}+\frac 12 \hat I.
}{\hat n_{ s_2}+\hat n_{ r_2}+\hat I}
\nonumber\\
&\makebox[1cm]{}\mp 
\hat x_{s_1}
(
\Lambda_{r_2\to s_2r_2}(\hat x_{r_2}(\theta))-{}_{a_1}\bra{0}\hat z^{(2)}_{\rm hom}(\theta)\ket{0}_{a_1})\,.
\label{eq:proof_3_3}
\nonumber\\
\end{align}
This is so because 
$c_3\times (\hat n_{s_1}+\frac12 \hat I)\cdot \hat{ \mathcal F}_{\rm hom}^{(2)}+e^{i\phi \hat n_{s_1}-i\theta\hat n_{ r_2}}
(\ref{eq:proof_3_3})
e^{-i\phi \hat n_{s_1}+i\theta\hat n_{ r_2}}$
is equal to the right hand side minus the left hand side of the inequality in Eq.~(\ref{eq:proof_3_1}). 

The non-negativity of Eq.~(\ref{eq:proof_4_2_0}) is provided from that of Eq.~(\ref{eq:proof_1_2}) just by replacing the variables. As for the operator in Eq.~(\ref{eq:proof_3_3}), it can be rewritten as 
\begin{eqnarray}
&&
\sum_{
\begin{array}{c}
u,n,m\in\mathbb Z_{\geq 0}
\\
s.t. \quad n\leq m
\end{array}
}
\hat \Delta^{(3)}_{u,n,m}
+
\sum_{(u,n,m,t)\in\Omega^{(3)}}
\hat W_{\pm,u,n,m,t}^{(3)},
\label{eq:submat_a3}
\nonumber\\
\end{eqnarray}
where
\begin{widetext}
\begin{eqnarray}
\hat \Delta^{(3)}_{u,n,m}&:=&(4-(\delta_{n,m}+\delta_{n,0}-2)(\delta_{u,0}-2))
 h_6(u,n,m)\hat P_{u,u,n,n,m},
\label{eq:def_Delta_3}
\\
\Omega^{(3)}
&:=&\{(u,n,m,t)|u,m,n\in\mathbb Z_{\geq 1}
\land\; n\leq m\;\land\; t\in\{0,1\}\},
\label{eq:def_Omega_3}
\\
\hat W_{\pm,u,n,m,t}^{(3)}
&:=&
h_6(u+t-1,n-1,m)\hat P_{u+t-1,u+t-1,n-1,n-1,m}
\mp h_5(u,n,m)^*\hat P_{u+t-1,u-t,n-1,n,m}
\nonumber\\&&{}
\mp h_5(u,n,m)\hat P_{u-t,u+t-1,n,n-1,m}
+h_6(u-t,n,m)\hat P_{u-t,u-t,n,n,m},
\label{eq:def_W_3}
\end{eqnarray}
\end{widetext}
and
\begin{eqnarray}
\hat P_{u,u',n,n',m}
&:=&\bigl|u,n,m-n\bigr\rangle
\bigl\langle u',n',m-n'\bigr|,
\label{eq:def_P_3}
\\
h_5(u,n,m)&:=&\frac{\sqrt{un}}{4}(1-\sqrt{1-\frac{n}{m+1}}),
\label{eq:def_h6_3}
\\
h_6(u,n,m)&:=&
\frac{c_3}{4}(u+\frac12)\frac{n^2+n+\frac{1}{2}}{m+1}.
\label{eq:def_h5_3}
\end{eqnarray}
As a sufficient condition for  the non-negativity of the operator in Eq.~(\ref{eq:submat_a3}),
we will show the non-negativity of  $\hat \Delta_{u,n,m}^{(3)}$ ($\hat W^{(3)}_{\pm,u,n,m,t}$) for $u,n,m\in\mathbb Z_{\geq 0}$ such that $n\leq m$($(u,n,m,t)\in\Omega^{(3)}$).
The non-negativity of  $\hat \Delta_{u,n,m}^{(3)}$
 is trivial from $c_3> 0$ and the definitions in Eqs.~(\ref{eq:def_Delta_3}) and (\ref{eq:def_h5_3}). In order to guarantee the non-negativity of the operator $\hat W^{(3)}_{\pm,u,n,m,t}$, it suffices to check the non-negativity of the principal submatrix
\begin{eqnarray}
\left(
\begin{array}{cc}
h_6(u+t-1,n-1,m)&h_5(u,n,m)^*\\
 h_5(u,n,m)&h_6(u-t,n,m)\end{array}
\right),
\label{eq:principal_submatrix_3}
\end{eqnarray}
in the photon number basis of the three input modes since the other matrix elements in the basis are zero (see Eq.~(\ref{eq:def_W_3})). The non-negativity of the trace of this matrix can be confirmed from the definition in Eq.~(\ref{eq:def_h5_3}). As for the non-negativity of the determinant, it can be checked as follows:
\begin{widetext}
\begin{eqnarray}
&&
\frac{16(m+1)^2}{un(n+1)(\sqrt{n+1}-1)^2}
(h_6(u-t,m,n)h_6(u+t-1,m,n-1)
-|h_5(u,m,n)|^2)
\nonumber\\
&=&
\frac {1}{(n+1)(\sqrt{n+1}-1)^2}
\left(
c_3^2\frac {u^2-\frac14}u\cdot\frac{n^4+\frac14}{n}
-(m+1)(\sqrt{m+1}-\sqrt{m-n+1})^2
\right)
\nonumber\\
&\geq&
c_3^2
\frac {(u^2-\frac14)(n^4+\frac14)}{un(n+1)(\sqrt{n+1}-1)^2}
-1
\geq
\frac{15}{32(\sqrt{2}-1)^2}c_3^2-1
=0.
\end{eqnarray}
\end{widetext}
The second relation is due to Eq.~(\ref{eq:basic_relation_1}),
and the left hand side of the third relation is minimized in the case of $u=1$ and $n=1$ when $(u,n,m,t)$ is in $\Omega^{(3)}$.

Therefore, the operator in Eq.~(\ref{eq:proof_3_3}) is non-negative, and therefore, Eq.~(\ref{res:GH_LH}) is proved.

\subsection{Proof of Eqs.~(\ref{res:GH_LE}), (\ref{res:GE_LE}), and (\ref{res:GE_LH})}

Eqs.~(\ref{res:GH_LE}), (\ref{res:GE_LE}), and (\ref{res:GE_LH}), i.e., the rest of relations in the third  set $\mathcal E_3$, can be proved with minor modifications to the proof of Eq.~(\ref{res:GH_LH}) in the previous subsection.
Below, we provide only the key equations, which are essential for the proof.

To guarantee Eqs.~(\ref{res:GH_LE}), (\ref{res:GE_LE}), and (\ref{res:GE_LH}), it is sufficient to respectively show
\begin{widetext}
\begin{eqnarray}
\pm 
\hat x_{s_1}(\phi)
(
\Lambda^\dagger_{v_2\to s_2r_2}\bigl(\hat F_{v_2}(\beta_\theta^{(2)})\bigr)-
{}_{ a'_2b'_2c'_2}\big\langle0,0,0\bigr|
\hat z_{\rm het}^{(2)}(\theta))
\bigl|0,0,0\big\rangle_{ a'_2b'_2c'_2}
)
&\leq &
c_{3,1}  (\hat n_{s_1}+\frac12 \hat I)\frac{\hat f_{\rm het}^{(2)}
}{\hat n_{ s_2}+\hat n_{ r_2}+\hat I},
\label{eq:proof_3_1_1}
\\
\pm 
\hat x_{s_1}(\phi)
(
\Lambda^\dagger_{v_2\to s_2r_2}\bigl(\hat F_{v_2}(\beta_\theta^{(2)})\bigr)-
{}_{ a'_2b'_2c'_2}\big\langle0,0,0\bigr|
\hat z_{\rm het}^{(2)}(\theta))
\bigl|0,0,0\big\rangle_{ a'_2b'_2c'_2}
)
&\leq &
c_{3,2}  (\hat n_{s_1}+\hat I)\frac{\hat f_{\rm het}^{(2)}
}{\hat n_{ s_2}+\hat n_{ r_2}+\hat I},
\label{eq:proof_3_2_1}
\\
\pm 
\hat x_{s_1}(\phi)
(\Lambda^\dagger_{v_2\to s_2r_2}(\hat x_{v_2}(\theta))
-
{}_{ a_2}\big\langle0\bigr|
\hat z_{\rm hom}^{(2)}(\theta))
\bigl|0\big\rangle_{ a_2}
)
&\leq &
c_{3,3}  (\hat n_{s_1}+\hat I)\frac{\hat f_{\rm hom}^{(2)}
}{\hat n_{ s}+\hat n_{ r}+\hat I},
\label{eq:proof_3_3_1}
\end{eqnarray}
\end{widetext}
where
\begin{eqnarray}
c_{3,1}&:=&\frac{\sqrt{2}}{3\sqrt 5}(\sqrt5-1)=0.26058\cdots,
\\
c_{3,2}&:=&\frac{1}{2\sqrt {15}}(\sqrt5-1)=0.15957\cdots,
\label{def:c_3_2}
\\
c_{3,3}&:=&\frac{2}{\sqrt 5}(\sqrt{2}-1)= 0.37048\cdots,
\label{def:c_3_3}
\end{eqnarray}
and, $\hat f_{\rm hom}^{(2)}$,  $\hat {z}_{\rm hom}^{(2)}(\theta)$, $\hat f_{\rm het}^{(2)}$, and $\hat {z}_{\rm het}^{(2)}(\theta)$
 are defined by Eqs.~(\ref{def:hat_f_hom}), (\ref{def:hat_x_l_hom_j}), (\ref{def:hat_f_het}), and (\ref{def:hat_x_l_het_j}),
 respectively. Moreover, the operators $\hat f_{\rm het}^{(j)}$ and $\hat z^{(j)}_{\rm het}(\theta)$ are defined in the similar way as 
$\hat f_{\rm hom}$ in Eq.~(\ref{def:hat_f_het})  and $\hat z_{\rm het}(\theta)$ in Eq.~(\ref{def:x_l_het}): 
\begin{align}
&\hat f_{\rm het}^{(j)}:=
\nonumber\\
&\quad {}_{ a'_jb'_jc'_j}\big\langle0,0,0\bigr|
  f_{\rm het}(\hat n_{1'_j},\hat n_{2'_j},\hat n_{3'_j},\hat n_{4'_j},\hat n_{0'_j})
\bigl|0,0,0\big\rangle_{a'_jb'_jc'_j},
\label{def:hat_f_het_j}
\\
&\hat z_{\rm het}^{(j)}(\theta):=
\sqrt2\frac{(\hat n_{1'_j}-\hat n_{2'_j})\cos \theta
\quad+(\hat n_{3'_j}-\hat n_{4'_j})\sin \theta}
{\sqrt{\hat n_{1'_j}+\hat n_{ 2'_j}+\hat n_{ 3'_j}+\hat n_{ 4'_j}+\hat n_{ 0'_j}+\hat I}}.
\label{def:hat_x_l_het_j}
\end{align}

Since the operators in Eq.~(\ref{eq:proof_4_2_0}) and
\begin{align}
&\frac{\hat f^{(j)}_{\rm het}}{\hat n_{ s_j}+\hat n_{ r_j}+\hat I}
-
\frac{(\hat n_{ s_j})^2+3\hat n_{ s_j}+2\hat I}{\hat n_{ s_j}+\hat n_{ r_j}+\hat I}
=:
\hat{\mathcal F}_{\rm het}^{(j)}
\label{eq:proof_4_1_1}
\end{align}
 are non-negative, the inequalities in Eq.~(\ref{eq:proof_3_1_1}),
 (\ref{eq:proof_3_2_1}), and
 (\ref{eq:proof_3_3_1}) can be respectively derived from the non-negativity of the operators 
\begin{align}
&
c_{3,1}
 (\hat n_{s_1}+\frac12\hat I) \frac{(\hat n_{ s_2}^2+3\hat n_{ s_2}+2\hat I)
}{\hat n_{ s_2}+\hat n_{ r_2}+\hat I}
\nonumber\\
&\quad\quad
\mp
\hat x_{s_1}
(\Lambda_{v_2\to s_2r_2}^\dagger(\hat x_{v_2})-
{}_{ a_2}\big\langle0\bigr|\hat z^{(2)}_{\rm hom}(0)\bigl|0\big\rangle_{ a_2}),
\label{eq:proof_3_1_2}
\\
&
c_{3,2}
 (\hat n_{s_1}+\hat I) \frac{(\hat n_{ s_2}^2+3\hat n_{ s_2}+\hat I)
 }{\hat n_{ s_2}+\hat n_{ r_2}+\hat I}
\nonumber\\
&\quad\quad
\mp
\hat x_{s_1}
(\Lambda_{v_2\to s_2r_2}^\dagger(\hat x_{v_2})-
{}_{ a_2}\big\langle0\bigr|\hat z^{(2)}_{\rm hom}(0)\bigl|0\big\rangle_{ a_2}),
\label{eq:proof_3_2_2}
\\
&
c_{3,3}  (\hat n_{s_1}+\hat I)\frac{(\hat n_{ s_2}^2+\hat n_{ s_2}+\frac12\hat I)
}{\hat n_{ s_2}+\hat n_{ r_2}+\hat I}
\nonumber\\
&\quad\quad
 \pm 
\hat x_{s_1}
(\Lambda_{v_2\to s_2r_2}^\dagger(\hat x_{v_2})-
{}_{ a_2}\big\langle0\bigr|\hat z^{(2)}_{\rm hom}(0)\bigl|0\big\rangle_{ a_2})\,.
\label{eq:proof_3_3_2}
\end{align}
Here, we used the fact that  
\begin{eqnarray}
\int\frac{d^2\gamma}{\pi}|\gamma^2|\ket\gamma\bra{\gamma}_{s_1}&=&\hat n_{s_1}+\hat I,
\end{eqnarray}
and the non-negativity of $\mathcal F_{\rm hom}^{(j)}$ in Eq.~(\ref{eq:proof_4_2_0}) and  $\mathcal F_{\rm het}^{(j)}$ in Eq.~(\ref{eq:proof_4_1_1}),
which can be derived from the non-negativity of Eq.~(\ref{eq:proof_1_1_1}) just by replacing the variables.

The three operators in Eqs.~(\ref{eq:proof_3_1_2})$-$(\ref{eq:proof_3_3_2}) can be written in the form in Eq.~(\ref{eq:submat_a3}). In doing so, the definitions of all the terms in Eqs.~(\ref{eq:def_Delta_3})$-$(\ref{eq:def_h6_3}) remain the same except for the definition of $h_6(u,n,m)$ for each of the Eqs.~(\ref{eq:proof_3_1_2})$-$(\ref{eq:proof_3_3_2}), i.e., Eq.~(\ref{eq:def_h5_3}), which is respectively replaced with
\begin{eqnarray}
\frac{c_{3,1}}4(u+\frac12)\frac{n^{2}+3n+2}{m+1},
\label{def_h_5_m1}
\\
\frac{c_{3,2}}4(u+1)\frac{n^{2}+3n+2}{m+1},
\label{def_h_5_m2}
\\
\frac{c_{3,3}}4(u+1)\frac{n^{2}+n+\frac12}{m+1}\,.
\label{def_h_5_m3}
\end{eqnarray}
This modification implicitly affects the definitions of $\hat \Delta_{u,n,m}^{(3)}$ and $\hat W_{\pm, u,n,m,t}^{(3)}$. As a result, we can prove the non-negativity of the operators in Eqs.~(\ref{eq:proof_3_1_2}), (\ref{eq:proof_3_2_2}), and  (\ref{eq:proof_3_3_2}) in the same way as the case of Eq.~(\ref{eq:proof_3_3}), i.e., it is sufficient to check the non-negativity of 
 the modified operators $\hat \Delta_{u,n,m}^{(3)}$ and $\hat W_{\pm,u,n,m,t}^{(3)}$ for appropriate parameters $u,n,m,t$. 
   The non-negativity of the modified operator $\hat W_{\pm,u,n,m,t}^{(3)}$ is obtained from
that of   the modified $2\times 2$ matrix in Eq.~(\ref{eq:principal_submatrix_3}), i.e., the non-negativity of 
the trace and the determinant of it.
 The non-negativity of the modified operator $\hat \Delta_{u,n,m}^{(3)}$ and 
    the trace of the matrix for appropriate parameters $u,n,m,t$ is guaranteed from the non-negativity of the modified function $h_6(u,n,m)$ for $u,n,m\in\mathbb Z_{\geq 0}$ such that $n\leq m$.
    This can be checked from the definitions in Eqs.~(\ref{def_h_5_m1}), (\ref{def_h_5_m2}), and (\ref{def_h_5_m3}) directly.
  The non-negativity of the determinant can be confirmed by the following expression:
\begin{widetext}
\begin{eqnarray}
&&
\frac{16(m+1)^2}{un(n+1)(\sqrt{n+1}-1)^2}
(h_6(u-t,m,n)h_6(u+t-1,m,n-1)
-|h_5(u,m,n)|^2)
\nonumber\\
&\geq&
\frac{16(m+1)^2}{un(n+1)(\sqrt{n+1}-1)^2}h_6(u,m,n)h_6(u-1,m,n-1)-1
\geq 0.
\end{eqnarray}
\end{widetext}
Here, the first inequality is due to the relation in Eq.~(\ref{eq:basic_relation_1}). The last expression can be evaluated as 
\begin{eqnarray}
c_{3,1}^2\frac{
(u^2-\frac14)
}{u}\cdot
\frac{
(n+1)(n+2)
}{(\sqrt{n+1}-1)^2}-1&\geq0,
\\
c_{3,2}^2(u+1)\frac{
(n+1)(n+2)
}{(\sqrt{n+1}-1)^2}-1&\geq0,
\\
c_{3,3}^2(u+1)\frac{
(n^4+\frac14)
}{n(n+1)(\sqrt{n+1}-1)^2}-1&\geq0,
\end{eqnarray}respectively.
Note that these values are minimized at the point $u=1$, $n=4$ for the first two values and at the point $u=1$, $n=1$ for the last one when $(u,n,m,t)\in\Omega^{(3)}$.

This ends the derivation of Eqs.~(\ref{res:GH_LE}), (\ref{res:GE_LE}), and (\ref{res:GE_LH}).

\subsection{Proof of Eq.~(\ref{res:LH_LH})}

As a representative of the fourth set $\mathcal E_4$ , i.e., the set of Eqs.~(\ref{res:LH_LH}), (\ref{res:LE_LE}), and (\ref{res:LH_LE}), we select Eq.~(\ref{res:LH_LH}), which we will prove in this subsection.

In the case of Eq.~(\ref{res:LH_LH}), $N$ pairs of the signal pulse and LO pulse are input. Especially, $k$-th and $l$-th pair is observed by the SHD. The operator $\hat g_{\rm hom}^{(j)}$ is defined by: 
\begin{widetext}
\begin{eqnarray}
\hat g_{\rm hom}^{(j)}&:=& \sum_{\theta\in\{\pi/4,\pi/2,3\pi/4\}}\frac 14 e^{-i\theta\hat n_{ r_j}}
{}_{ a_j}\bra{0}
  g_{\rm hom}(\hat n_{1_j},\hat n_{2_j},\hat n_{0_j})
\ket{0}_{ a_j}
e^{i\theta\hat n_{ r_j}},
\label{def:hat_g_hom_j}
\end{eqnarray}
where  $g_{\rm hom}(n_1,n_2,n_0)$ is defined by Eq.~(\ref{def:g_hom}).

The inequality in Eq.~(\ref{res:LH_LH}) can be derived from the following operator relation: 
\begin{align}
&\pm 
(
\Lambda^\dagger_{v_k\to s_kr_k}(\hat x_{v_k}(\theta_k))
\Lambda^\dagger_{v_l\to s_lr_l}(\hat x_{v_l}(\theta_l))
-
{}_{ a_k}\big\langle0\bigr|
\hat z_{\rm hom}^{(k)}(\theta_k))
\bigl|0\big\rangle_{ a_k}
{}_{ a_l}\big\langle0\bigr|
\hat z_{\rm hom}^{(l)}(\theta_l))
\bigl|0\big\rangle_{ a_l}
)
\nonumber\\
&
\makebox[1cm]{}\leq 
c_3
(\frac{\hat f_{\rm hom}^{(k)}
}{\hat n_{ s_k}+\hat n_{ r_k}+\hat I}\hat g^{(l)}_{\rm hom}
+
\hat g^{(k)}_{\rm hom}\frac{\hat f_{\rm hom}^{(l)}
}{\hat n_{ s_l}+\hat n_{ r_l}+\hat I}
),
\label{eq:proof_4_1}
\end{align}
\end{widetext}
where $c_3$ is defined by Eq.~(\ref{def:c_3}).
The inequality in Eq.~(\ref{res:LH_LH}) is rebuilt by using  $\hat \rho_{2N}$.

Eq.~(\ref{eq:proof_4_1}) can be derived from the non-negativity of the following three operators: $\mathcal F_{\rm hom}^{(j)}$~(\ref{eq:proof_4_2_0}), 
\begin{eqnarray}
&&
\hat g^{(j)}_{\rm hom}
-\left(\hat n_{ s_j}+\frac12 \hat I\right)=:\hat{\mathcal G}_{\rm hom}^{(j)},
\label{eq:proof_4_2_1}
\end{eqnarray}
 and 
\begin{widetext}
\begin{align}
&c_3
\left(
\frac{(\hat n_{ s_k})^2+\hat n_{ s_k}+\frac12 \hat I}{\hat n_{ s_k}+\hat n_{ r_k}+\hat I}
\left(\hat n_{ s_l}+\frac12 \hat I\right)
+
\left(\hat n_{ s_k}+\frac12 \hat I\right)
\frac{(\hat n_{ s_l})^2+\hat n_{ s_l}+\frac12 \hat I}{\hat n_{ s_l}+\hat n_{ r_1}+\hat I}
\right)
\nonumber\\
&
\makebox[1cm]{}
\mp 
\bigl(
\Lambda^\dagger_{v_k\to s_kr_k}(\hat x_{v_k}(0))
\Lambda^\dagger_{v_l\to s_lr_l}(\hat x_{v_l}(0))
-
{}_{ a_k}\big\langle0\bigr|
\hat z_{\rm hom}^{(k)}(0))
\bigl|0\big\rangle_{ a_k}
{}_{ a_l}\big\langle0\bigr|
\hat z_{\rm hom}^{(l)}(0))
\bigl|0\big\rangle_{ a_l}
\bigl),
\label{eq:proof_4_3}
\nonumber\\
\end{align}
\end{widetext}
since the difference between the both sides of the relation in Eq.~(\ref{eq:proof_4_1}) is equal to $c_3(
\frac{\hat f^{(k)}_{\rm hom}}{\hat{n}_{s_k}+\hat{n}_{ r_k}+\hat I}\hat{\mathcal G}^{(l)}_{\rm hom}+
\hat{\mathcal F}^{(k)}_{\rm hom}\left(\hat n_{ s_l}+\frac12\hat I\right)+
\hat{\mathcal G}^{(k)}_{\rm hom}
\frac{\hat f^{(l)}_{\rm hom}}{\hat{n}_{ s_l}+\hat{n}_{ r_l}+\hat I}
+\left(\hat n_{ s_k}+\frac12 \hat I\right)
\hat{\mathcal F}^{(l)}_{\rm hom}
)+
e^{-i\theta_k\hat n_{ s_k}-i\theta_l\hat n_{ r_l}}
(\ref{eq:proof_4_3})e^{i\theta_k\hat n_{ s_l}+i\theta_l\hat n_{ r_l}}
$. Here, trivial relations $\frac{\hat f^{(j)}_{\rm hom}}{\hat{n}_{ s_j}+\hat{n}_{ r_j}+\hat I}\geq 0$ and $\hat n_{ s_j}\geq 0$ are also employed.

From the matrix element of Eq.~(\ref{eq:proof_4_2_1}), which can be evaluated as
\begin{widetext} 
\begin{eqnarray}
&&{}_{ s_jr_j}\bra{n,m}
\hat g_{\rm hom}^{(j)}-(\hat n_{ s_j}+ \frac12 \hat I)
\ket{n',m'}_{ s_jr_j}
\nonumber\\
&=&
\sum_{\theta\in\{0,\pi/4,\pi/2,3\pi/4\}}\frac14e^{-i\theta(m-m')}
\left( 
\frac12\left(n+m+1\right)
{}_{ s_jr_ja_j}\bra{n,m,0}\delta_{\hat n_{0_j},0}\ket{n',m',0}_{ s_jr_ja_j}\right.
\nonumber\\
&&{}\left.+
{}_{ s_jr_ja_j}\bra{n,m,0}\frac{(\hat a_{ s_j}\hat a^{\dagger}_{ r'_j}+\hat a^{\dagger}_{ s_j}\hat a_{ r'_j})^2 }{2(\hat n_{0_j}+1)}
\ket{n',m',0}_{ s_jr_ja_j}
\right)
-
\delta_{n,n'}\delta_{m,m'}\left(n+\frac12\right)
\nonumber\\
&=&
\delta_{n,n'}\delta_{m,m'}
\left( 
\frac12\left(n+m+1\right)
{}_{ s_jr_ja_j}\bra{n,m,0}\delta_{\hat n_{0_j},0}\ket{n,m,0}_{ s_jr_ja_j}
+
{}_{ s_jr_ja_j}\bra{n,m,0}\frac{2\hat n_{ s_j}\hat n_{ r'_j}+\hat n_{ s_j}+\hat n_{ r'_j} }{2(\hat n_{0_j}+1)}
\ket{n,m,0}_{ s_jr_ja_j}
-
\left(n+\frac12\right)\right)
\nonumber\\
&=&
\delta_{n,n'}\delta_{m,m'}\left(
2^{-m-1}\left(n+m+1\right)
+
\sum_{k=0}^m\frac{m!\left((2n+1)(m-k)+n\right)
}{(k+1)!(m-k)!2^{m+1}}
-\left(n+\frac12\right)\right)
\nonumber\\
&=&
\delta_{n,n'}\delta_{m,m'}\left(
\frac n{m+1}(1-(1+m/2)2^{-m})+m2^{-m-1}
\right),
\label{eq:501}
\end{eqnarray}
\end{widetext}
we can see that the operator in Eq.~(\ref{eq:proof_4_2_1}) is diagonal for the photon number basis $\{\ket{n,m}^{(j)}_{sr}\}_{n,m=0,1,\cdots}$. 
The equalities in Eq.~(\ref{eq:501}) are assured for the same reasons as in Eq.~(\ref{eq:diagonal_a1}). Since all the 2 terms in the diagonal elements on the last expression are non-negative for $n,m\in\mathbb Z_{\geq 0}$,
we have that Eq.~(\ref{eq:proof_4_2_1}) is non-negative.

Next, we show the non-negativity of the operator in Eq.~(\ref{eq:proof_4_3}).
By definition, the operator can be rewritten as 
\begin{eqnarray}
&&\sum_{
\begin{array}{c}
n,m,u,w\in\mathbb Z_{\geq 0}
\\
s.t. \; n\leq m \;\land\; u\leq w
\end{array}
}\!\!\!\!\hat \Delta^{(4)}_{n,m,u,w}
+\!\!\!\!\!\!\sum_{(n,m,u,w,t)\in\Omega^{(4)}}
W_{\pm ,n,m,u,w,t}^{(4)},
\nonumber\\
\label{eq:proof_4_4_0}
\end{eqnarray}
where all the notations are defined as follows:
\begin{widetext}
\begin{eqnarray}
\hat\Delta^{(4)}_{n,m,u,w}&:=&(4-(\delta_{n,m}+\delta_{n,0}-2)(\delta_{u,w}+\delta_{u,0}-2))
 h_8(n,m,u,w)\hat P_{n,n,m,u,u,w},
\label{def:Delta_4}
\\
 \Omega^{(4)}&:=&\{(n,m,u,w,t)|n,m,u,w\in\mathbb Z_{\geq 0}
 \land\; n\leq m\;\land\; u\leq w \;\land\; t\in \{0,1\}\},
\label{def:Omega_4}
\\
\hat W^{(4)}_{\pm,n, m,u,w,t}&:=& h_8(n-1,m,u+t-1,w)\hat P_{n-1,n-1,m,u+t-1,u+t-1,w}
\mp h_7(n,m,u,w)^*\hat P_{n-1,n,m,u+t-1,u-t,w}
\nonumber\\&&{}
\mp h_7(n,m,u,w)\hat P_{n,n-1,m,u-t,u+t-1,w}
+h_8(n,m,u-t,w)\hat P_{n,n,m,u-t,u-t,w},
\label{eq:proof_4_4}
\\
\hat P_{n,n',m,u,u',w}
&:=&
\ket{n,m-n}_{ s,r}^{(k)}{}_{\; s,r}^{(k)}\bra{n',m-n'}\otimes
\ket{u,w-u}_{ s,r}^{(l)}{}_{\; s,r}^{(l)}\bra{u',w-u'},
\label{def:P_4}
\\
h_7(n,m,u,w)
&:=&\frac{\sqrt{nu}}{4}
\Big(1-\sqrt{(1-\frac{n}{m+1})(1-\frac{u}{w+1})}\Big),
\label{def:h_8}
\\
h_8(n,m,u,w)
&:=&
\frac {c_3}4
\bigl(
\frac{n^2+n+\frac{1}{2}}{m+1}
(u+\frac12)
+
(n+\frac12)
\frac{u^2+u+\frac{1}{2}}{w+1}
\bigr).
\label{def:h_7}
\end{eqnarray}
\end{widetext}
As a sufficient condition for the non-negativity of the operator in Eq.~(\ref{eq:proof_4_4_0}),
we will show the non-negativity of  $\hat \Delta_{n,m,u,w}^{(4)}$ ($\hat W^{(4)}_{\pm,n,m,u,w,t}$) for $n,m,u,w\in\mathbb Z_{\geq 0}$ such that $n\leq m$ and $u\leq w$($(n,m,u,w,t)\in\Omega^{(4)}$).
The non-negativity of  $\hat \Delta_{n,m,u,w}^{(4)}$ for $n,m,u,w\in\mathbb Z_{\geq 0}$ such that $n\leq m$ and $u\leq w$
 is trivial from the fact $c_3> 0$ and the definitions in Eqs.~(\ref{def:Delta_4}) and (\ref{def:h_7}).
In order to confirm the non-negativity of the  operator $\hat W^{(4)}_{\pm,n,m,u,w,t}$, it is enough to check the non-negativity of the principal submatrix
\begin{align}
\left(
\begin{array}{cc}
h_8(n-1,m,u+t-1,w)&\mp h_7(n,m,u,w)^*\\
\mp  h_7(n,m,u,w)&h_8(n,m,u-t,w)\end{array}
\right),
\label{eq:principal_submatrix_4}
\end{align}
in the photon number basis of the four input modes since the other matrix elements in the basis are zero (see Eq.~(\ref{eq:proof_4_4})).
The non-negativity of this matrix can be confirmed from the facts that 
the definition in Eq.~(\ref{def:h_7}) guarantees the non-negativity of the trace of it, and the non-negativity of the determinant can be checked by the following inequality
\begin{widetext}
\begin{eqnarray}
&&
|h_7(n,m,u,w)|^{-2}({h_8(n,m,u-t,w)h_8(n-1,m,u-1+t,w)}-{|h_7(n,m,u,w)|^2}
)
\nonumber\\
&=&
\frac {c_3^2}{nu}
\prod_{s\in\{1,-1\}}
\frac{
(n^2+sn+\frac{1}{2})
(u+(-1)^t\frac{s}2)n^{-1}\frac n{m+1}
+
(n+\frac s2)
(u^2+(-1)^t s u+\frac{1}{2})u^{-1}\frac u{w+1}
}{1-\sqrt{(1-\frac n{m+1})(1-\frac u{w+1})}}
-1
\nonumber\\
&\geq&
\frac{c_3^2}{nu}
\min\left[
\frac{(n^4+\frac14)(u^2-\frac14)}{(n+1)(\sqrt{n+1}-1)^2}
,\frac{(u^4+\frac14)(n^2-\frac14)}{(u+1)(\sqrt{u+1}-1)^2}
\right]-1
\geq
c_3^2\frac{15}{32(\sqrt{2}-1)^2}
-1=0,
\label{eq:tmp_00}
\end{eqnarray}
\end{widetext}
for ${(n,m,u,w,t)\in \Omega^{(4)}}$. In the second relation, we apply the following inequality:
\begin{eqnarray}
&&
\prod_{b\in\{1,-1\}}\frac{\alpha_b\frac n{m+1}+\beta_b\frac u{w+1}}
{1-\sqrt{(1-\frac n{m+1})(1-\frac u{w+1})}}
\nonumber\\&\geq&
\min[\frac{\alpha_1\alpha_{-1}n^2}{(n+1)(\sqrt{n+1}-1)^{2}},\frac{\beta_1\beta_{-1}w^2}{(w+1)(\sqrt{w+1}-1)^{2}}],
\label{eq:tmp_1}
\nonumber\\
\end{eqnarray}
 under the constraints  $\alpha_1,\alpha_{-1},\beta_1,\beta_{-1}>0$, $0<n\leq m$, and $0<u\leq w$, which will be proved in the next subsection. In order to apply this inequality, we assign values as follows:  $\alpha_b:={( n^2+b n+\frac 12)(u+(-1)^t \frac b2)}n^{-1}$, and $\beta_b :={(n+ \frac b2)( u^2+(-1)^tb u+\frac12)}u^{-1}$. 
The third expression in Eq.~(\ref{eq:tmp_00}) is minimized when $n=1$ and $u=1$.

Therefore, we have derived the non-negativity of the  operator in Eq.~(\ref{eq:proof_4_4}), and we have shown the correctness of the relation in Eq.~(\ref{res:LH_LH}).

\subsection{Proof of Eq.~(\ref{eq:tmp_1})}
In this subsection, we prove Eq.~(\ref{eq:tmp_1}). Under the constraints  $\alpha_1,\alpha_{-1},\beta_1,\beta_{-1}>0$, $1>x_0:=\frac n{n+1}\geq x:=\frac n{m+1}>0$, and $1>y_0:=\frac u{u+1}\geq y:=\frac u{w+1}>0$,
the relation in Eq.~(\ref{eq:tmp_1}) can be shown as follows:
\begin{eqnarray}
&&
\prod_{b\in\{1,-1\}}\frac{\alpha_bx+\beta_by}{1-\sqrt{(1-x)(1-y)}}
\nonumber\\
&\geq&
\left(
\frac{\sqrt{\alpha_1\alpha_{-1}}x+\sqrt{\beta_1\beta_{-1}}y}{1-\sqrt{(1-x)(1-y)}}\right)^{2}
\nonumber
\end{eqnarray}
\begin{eqnarray}
&\geq&
\min[\frac{\alpha_1\alpha_{-1}x^2}{(1-\sqrt{1-x})^2},\frac{\beta_1\beta_{-1}y^2}{(1-\sqrt{1-y})^2}]
\nonumber
\\
&\geq&
\min[\frac{\alpha_1\alpha_{-1}x_0^2}{(1-\sqrt{1-x_0})^{2}},\frac{\beta_1\beta_{-1}y_0^2}{(1-\sqrt{1-y_0})^{2}}].
\end{eqnarray}
The first inequality comes from  
$\alpha_1\beta_{-1}+\alpha_{-1}\beta_1\geq 2\sqrt{\alpha_1\alpha_{-1}\beta_1\beta_{-1}}$ for any positive variables $\alpha_1$, $\alpha_{-1}$, $\beta_1$, and $\beta_{-1}$.
The last inequality comes from the fact that the function $\frac{x}{1-\sqrt{1-x}}$ is a monotonically decreasing and positive function when $0<x<1$. Therefore, it is sufficient to prove the second inequality.

The second inequality is derived from the fact that 
\begin{eqnarray}
&&
\left(\frac{z_1(1-w_1^2)+z_2 (1-w_2^2)}{1-w_1w_2}\right)^2
\nonumber\\
&\geq&
\min[z_1^2(1+w_1)^2,z_2^2 (1+w_2)^2]
\label{eq:tmp_2}
\end{eqnarray}
holds when  $1> w_{b\in\{1,2\}}\geq0$ and $z_{b\in\{1,2\}}>0$, since 
the second inequality is obtained by  
substituting $\sqrt{1-x}$, $\sqrt{1-y}$,
$\sqrt{\alpha_1\alpha_{-1}}$,
 and 
$\sqrt{\beta_1\beta_{-1}}$
with $w_1$, $w_2$, $z_1$, and $z_2$, respectively.
In the rest of this subsection, we give the proof of  Eq.~(\ref{eq:tmp_2}). For this, we consider three regions for a given $z_{b\in\{1,2\}}>0$;
\begin{widetext}
\begin{eqnarray}
R_1&:=&\{(w_1',w_2')| z_1(1-w_1^{\prime 2})+z_2 (1-w_2^{\prime 2})\geq z_1(1+w_1')(1-w_1'w_2')
 \},
\nonumber\\
R_2&:=&\{(w_1',w_2')|  1-w_2'\geq 0\;\land \; z_2( w_2'+1) \geq (w_1'+1)w_1' z_1
 \},
\nonumber\\
R_3&:=&\{(w_1',w_2')| 1\geq w_1'\geq 0\;\land\; 1\geq w_2'\geq 0\;\land\;z_1^2w_1'\leq z_2^2w_2' \}.
\end{eqnarray}
\end{widetext}
The relation $R_1\supseteq R_2$ holds since the inequality identifying $R_1$ can be rewrite 
$(1-w_2')(z_2(w_2'+1)-(w_1'+1)w_1'z_1)\geq 0$.
The relation $R_2\supseteq R_3$ also holds since the region $R_2$ is a convex set and all the vertices of the polytope $R_3$ is included in $R_2$, i.e., $(0,0),(0,1),(z_1^{-2}z^2_2,1)\in R_2$ in the case of $z_1\geq z_2>0$ and 
$(0,0),(0,1),(1,1),(1,z_1^{2}z^{-2}_2)\in R_2$ in the case of $z_2\geq z_1>0$.
Therefore, $R_1\supseteq R_3$ holds.
This relation assures that
when 
 $1\geq w_{b\in\{1,2\}}\geq  0$,   
$z_1^2w_1\leq z^2_2w_2$, and $z_{b\in\{1,2\}}\geq 0$, i.e., $(w_1,w_2)\in R_3$, 
\begin{eqnarray}
z_1(1-w_1^2)+z_2 (1-w_2^2)
&\geq&
z_1(1+w_1)(1-w_1w_2),
\nonumber\\
\end{eqnarray}
holds, i.e. $(w_1,w_2)\in R_1$. Since $w_1w_2<1$ and $z_1(1+w_1)>0$,
this relation leads to
\begin{eqnarray}
\left(\frac{z_1(1-w_1^2)+z_2 (1-w_2^2)}{1-w_1w_2}\right)^2
&\geq&
z_1^2(1+w_1)^2.
\label{eq:tmp_101}
\end{eqnarray}

By replacing $w_b$ and $z_b$ with $w_{3-b}$ and $z_{3-b}$, respectively, we can also obtain the relation
\begin{eqnarray}
\left(\frac{z_1(1-w_1^2)+z_2 (1-w_2^2)}{1-w_1w_2}\right)^2
&\geq&
z_2^2(1+w_2)^2,
\label{eq:tmp_102}
\end{eqnarray}
 when 
 $1> w_{b\in\{1,2\}}\geq  0$,   
$z_1^2w_1\geq z^2_2w_2$, and $z_{b\in\{1,2\}}\geq 0$.

As a result, when $1> w_{b\in\{1,2\}}\geq  0$ and $z_{b\in\{1,2\}}\geq 0$,
the relation in Eqs.~(\ref{eq:tmp_101}) or (\ref{eq:tmp_102}) hold. This is equivalent to the relation in Eq.~(\ref{eq:tmp_2}).

\subsection{Proof of Eqs.~(\ref{res:LE_LE}) and (\ref{res:LH_LE})}
Eqs.~(\ref{res:LE_LE}) and (\ref{res:LH_LE}), i.e., the rest of the relations in the fourth set $\mathcal E_4$, 
 can be proved with minor modifications of the proof of Eq.~(\ref{res:LH_LH}). All the notations used here are the same as those in the proof of Eq.~(\ref{res:LH_LH}).

\newpage
For the modification, we employ additional operators that will be proved to be non-negative:
\begin{eqnarray}
&&{\hat g^{(j)}_{\rm het}}
-({\hat n_{ s_j}+\hat I})
=:\hat{\mathcal G}_{\rm het}^{(j)},
\label{eq:proof_4_1_2}
\end{eqnarray}
where the operator
 $\hat g_{\rm het}^{(j)}$
is defined from  
$g_{\rm het}(n_{1'},n_{2'},n_{3'},n_{4'},n_{0'})$ in Eq.~(\ref{def:g_het}) as:
\begin{widetext}
\begin{eqnarray}
\hat g_{\rm het}^{(j)}&:=&
{}_{ a'_jb'_jc'_j}
\bra{0,0,0}
  g_{\rm het}(\hat n_{1'_j},\hat n_{2'_j},\hat n_{3'_j},\hat n_{4'_j},\hat n_{0'_j})
\ket{0,0,0}_{a'_jb'_jc'_j}.
\end{eqnarray}

The matrix element of Eq.~(\ref{eq:proof_4_1_2})
can be  evaluated as 
\begin{eqnarray}
&&{}_{ s_jr_j}\bra{ n,m}
\hat g_{\rm het}^{(j)}-(\hat n_{ s_j}+\hat I)
\ket{n',m'}_{ s_jr_j}
\nonumber\\
&=&
\left( 
{}_{ s_jr_ja'_j}\bra{n,m,0}\delta_{\hat n_{0'_j},0}\ket{n',m',0}_{ s_jr_ja'_j}
+
{}_{ s_jr_ja'_j}\bra{n,m,0}\frac{(\hat n_{ s_j}+1)\hat n_{ r'_j}+\hat n_{ s_j} }{\hat n_{0'_j}+1}
\ket{n',m',0}_{ s_jr_ja'_j}
\right)
\nonumber\\&&{}
-
\delta_{n,n'}\delta_{m,m'}\left(n+1\right)
\nonumber\\
&=&
\delta_{n,n'}\delta_{m,m'}\left(
2^{-m}
+
\sum_{k=0}^m\frac{m!\left((n+1)(m-k)+n\right)
}{(k+1)!(m-k)!2^{m}}
-\left(n+1\right)\right)
=
\delta_{n,n'}\delta_{m,m'}\left(
\frac n{m+1}(2-(2+m)2^{-m})
\right).
\label{eq:502}
\end{eqnarray}
\end{widetext}

From this expression, we can find that the operator in Eq.~(\ref{eq:proof_4_1_2}) is diagonal for the photon number basis of the input modes.
The equalities in Eq.~(\ref{eq:502}) are justified with the same reasons as in Eq.~(\ref{eq:503}).
From the last expression, we have that the diagonal elements are positive for $n,m\in\mathbb Z_{\geq 0}$, and the non-negativity of  Eq.~(\ref{eq:proof_4_1_2}) is guaranteed.

In order to confirm Eqs.~(\ref{res:LE_LE}) and (\ref{res:LH_LE}), it is sufficient to respectively show
\begin{widetext}
\begin{align}
&\pm 
\bigl(
\Lambda^\dagger_{v_k\to s_kr_k}\bigl(\hat F_{v_k}(\beta_{\theta_k})\bigr)
\Lambda^\dagger_{v_l\to s_lr_l}\bigl(\hat F_{v_l}(\beta_{\theta_l})\bigr)
-
{}_{ a'_kb'_kc'_k}\big\langle0,0,0\bigr|
\hat z_{\rm het}^{(k)}(\theta_k))
\bigl|0,0,0\big\rangle_{ a'_kb'_kc'_k}
{}_{ a'_lb'_lc'_l}\big\langle0,0,0\bigr|
\hat z_{\rm het}^{(l)}(\theta_l))
\bigl|0,0,0\big\rangle_{ a'_lb'_lc'_l}
)
\nonumber\\
&\makebox[3cm]{}\leq {}
{c_{3,2}}
(\frac{\hat f_{\rm het}^{(k)}
}{\hat n_{ s_k}+\hat n_{ r_k}+\hat I}
\hat g^{(l)}_{\rm het}+
\hat g^{(k)}_{\rm het}
\frac{\hat f_{\rm het}^{(l)}
}{\hat n_{ s_l}+\hat n_{ r_l}+\hat I}
),
\label{eq:tmp201}
\\
&\pm 
\bigl(
\Lambda^\dagger_{v_k\to s_kr_k}(\hat x_{v_k}(\theta_k\bigr))
\Lambda^\dagger_{v_l\to s_lr_l}\bigl(\hat F_{v_l}(\beta_{\theta_l})\bigr)
-
{}_{ a_k}\big\langle0\bigr|
\hat z_{\rm hom}^{(k)}(\theta_k))
\bigl|0\big\rangle_{ a_k}
{}_{ a'_lb'_lc'_l}\big\langle0,0,0\bigr|
\hat z_{\rm het}^{(l)}(\theta_l))
\bigl|0,0,0\big\rangle_{ a'_lb'_lc'_l}
)
\nonumber\\
&\makebox[3cm]{}\leq {}
c_{3,3}
(\frac{\hat f_{\rm hom}^{(k)}
}{\hat n_{ s_k}+\hat n_{ r_k}+\hat I}
\hat g^{(l)}_{\rm het}+
\hat g^{(k)}_{\rm hom}\frac{\hat f_{\rm het}^{(l)}
}{\hat n_{ s_l}+\hat n_{ r_l}+\hat I}
)\,.
\label{eq:tmp202}
\end{align}
\end{widetext}
Here,
 $\hat f_{\rm hom}^{(j)}$, $\hat g_{\rm hom}^{(j)}$,$\hat f_{\rm het}^{(j)}$,  $\hat z_{\rm hom}^{(j)}(\theta)$,  $\hat z_{\rm het}^{(j)}(\theta)$,  $c_{3,2}$, and $c_{3,3}$ are defined by
Eqs.~(\ref{def:hat_f_hom_j}),
(\ref{def:hat_g_hom_j}),
(\ref{def:hat_f_het_j}),
(\ref{def:hat_x_l_hom_j}),
(\ref{def:hat_x_l_het_j}),
 (\ref{def:c_3_2}), and (\ref{def:c_3_3}), respectively. 

 Since we already checked the non-negativity of  
the operators in Eqs.~(\ref{eq:proof_4_2_0}),
(\ref{eq:proof_4_2_1}),
(\ref{eq:proof_4_1_1}), and
(\ref{eq:proof_4_1_2}),
Eqs.~(\ref{eq:tmp201}) and (\ref{eq:tmp202}) can be derived from the non-negativity of the operators
\begin{widetext}
\begin{align}
&
{c_{3,2}}
(
\frac{(\hat n_{ s}^{(k)})^2+3\hat n_{ s}^{(k)}+2 \hat I
}{\hat n_{ s}^{(k)}+\hat n_{ r}^{(k)}+\hat I}
(\hat n_{ s}^{(l)}+\hat I)
+(\hat n_{ s}^{(k)}+\hat I)
\frac{(\hat n_{ s}^{(l)})^2+3\hat n_{ s}^{(l)}+2 \hat I
}{\hat n_{ s}^{(l)}+\hat n_{ r}^{(l)}+\hat I}
)
\nonumber\\
&\makebox[3cm]{}
\mp 
\bigl(
\Lambda^\dagger_{v_k\to s_kr_k}(\hat x_{v_k}(0))
\Lambda^\dagger_{v_l\to s_lr_l}(\hat x_{v_l}(0))
-
{}_{ a_k}\big\langle0\bigr|
\hat z_{\rm hom}^{(k)}(0))
\bigl|0\big\rangle_{ a_k}
{}_{ a_l}\big\langle0\bigr|
\hat z_{\rm hom}^{(l)}(0))
\bigl|0\big\rangle_{ a_l}
\bigl),
\label{eq:tmp_203}
\\
&
c_{3,3}
(
\frac{(\hat n_{ s}^{(k)})^2+\hat n_{ s}^{(k)}+\frac12 \hat I
}{\hat n_{ s}^{(k)}+\hat n_{ r}^{(k)}+\hat I}
(\hat n_{ s}^{(l)}+\hat I)
+
(\hat n_{ s}^{(k)}+\frac12 \hat I)\frac{(\hat n_{ s}^{(l)})^2+3\hat n_{ s}^{(l)}+2 \hat I
}{\hat n_{ s}^{(l)}+\hat n_{ r}^{(l)}+\hat I}
)
\nonumber\\
&\makebox[3cm]{}
\mp 
\bigl(
\Lambda^\dagger_{v_k\to s_kr_k}(\hat x_{v_k}(0))
\Lambda^\dagger_{v_l\to s_lr_l}(\hat x_{v_l}(0))
-
{}_{ a_k}\big\langle0\bigr|
\hat z_{\rm hom}^{(k)}(0))
\bigl|0\big\rangle_{ a_k}
{}_{ a_l}\big\langle0\bigr|
\hat z_{\rm hom}^{(l)}(0))
\bigl|0\big\rangle_{ a_l}
\bigl),
\label{eq:tmp_204}
\end{align}
\end{widetext}
respectively, as was the case of the justification of Eq.~(\ref{eq:proof_4_1}). Here, we have used the relations $\hat F_v(\beta_\theta)=\hat x_v(\theta)$ and Eq.~(\ref{def:x_l_het_derive}).
The above two operators can be written in the form in Eq.~(\ref{eq:proof_4_4_0}) where all the definitions in Eqs.~(\ref{def:Delta_4})$-$(\ref{def:h_8}) remain the same, but the definition of the function $h_8(n,m,u,w)$ is modified from Eq.~(\ref{def:h_7}) into the following two values
\begin{eqnarray}
&&
\frac{{c_{3,2}}}4(
\frac{n^{2}+3n+2}{m+1}(u+1)
+(n+1)\frac{u^{2}+3u+2}{w+1}
),
\label{eq:tmp208}
\nonumber\\
\\
&&
\frac{c_{3,3}}4(
\frac{n^{2}+n+\frac12}{m+1}(u+1)
+(n+\frac12)\frac{u^{2}+3u+2}{w+1}
),
\nonumber\\
\label{eq:tmp209}
\end{eqnarray}
respectively.
 This modification implicitly affects the definitions of $\hat \Delta_{n,m,u,w}^{(4)}$ and $\hat W_{\pm, n,m,u,w,t}^{(4)}$.
 As a result, we can prove the non-negativity of the above operators in Eqs.~(\ref{eq:tmp_203}) and (\ref{eq:tmp_204}) in the same manner as the case of Eq.~(\ref{eq:proof_4_3}).
  That is, it is sufficient to check the non-negativity of 
 the modified operators $\hat \Delta_{n,m,u,w}^{(4)}$ and $\hat W_{\pm,n,m,u,w,t}^{(4)}$ for appropriate parameters $n,m,u,w$. 
   The non-negativity of the modified operator $\hat W_{\pm,n,m,u,w,t}^{(4)}$ is obtained from
that of   the modified $2\times 2$ matrix in Eq.~(\ref{eq:principal_submatrix_4}), i.e., the non-negativity of 
the trace and the determinant of it.
 The non-negativity of the modified operator $\hat \Delta_{n,m,u,w}^{(4)}$ and 
    the trace of the matrix for appropriate parameters $n,m,u,w$ is guaranteed from the non-negativity of the modified function $h_8(n,m,u,w)$ for $n,m,u,w\in\mathbb Z_{\geq 0}$.
    This can be checked from the definitions in Eqs.~(\ref{eq:tmp208}) and (\ref{eq:tmp209}).
Lower bounds of the determinant times $|h_7(n,m,u,w)|^{-2}$ can be derived as 
\begin{widetext}
\begin{eqnarray}
&&c_{3,2}^2
\min\left[
\frac{(n^2+3n+2)(u+1)}{(\sqrt{n+1}-1)^2},
\frac{(u^2+3u+2)(n+1)}{(\sqrt{u+1}-1)^2}
\right]-1
\geq
c_{3,2}^2\frac{60}{(\sqrt{5}-1)^2}
-1=0,
\label{eq:tmp501}
\\
&&
c_{3,3}^2
\min\left[
\frac{(n^4+\frac14)(u+1)}{n(n+1)(\sqrt{n+1}-1)^2},
\frac{(u^3+3u+2)(n^2-\frac14)}{n(\sqrt{u+1}-1)^2}
\right]-1
\geq
c_{3,3}^3\frac{5}{4(\sqrt{2}-1)^2}
-1=0,
\label{eq:tmp502}
\end{eqnarray}
\end{widetext}
respectively. 
In the first case, the left-most lower bound in Eq.~(\ref{eq:tmp501}) is justified by the inequality in Eq.~(\ref{eq:tmp_1}), where 
${(n+1)( n+b +1)(u+(-1)^t \frac b2+\frac12)}n^{-1}$
 and 
${(n+ \frac b2+\frac12)(u+1)( u+(-1)^tb+1)}u^{-1}$
 are substituted as the parameters 
$\alpha_b$ and $\beta_b$, respectively.
This left-most bound is minimized at the point  $u=1$ and $n=4$ when $(n,m,u,w,t)\in \Omega^{(4)}$ to obtain the second lower bound.
In the second case, the left-most lower bound in Eq.~(\ref{eq:tmp502}) is confirmed by the inequality in Eq.~(\ref{eq:tmp_1}), in which
${( n^2+b n+\frac12)(u+ (-1)^t\frac b2+\frac12)}n^{-1}$
 and 
${(n+ \frac b2)(u+1)( u+(-1)^tb+1)}u^{-1}$
are substituted as the parameters 
$\alpha_b$ and $\beta_b$, respectively.
The left-most bound is minimized  at the point  $u=1$ and $n=1$ when $(n,m,u,w,t)\in \Omega^{(4)}$.

As a result, Eqs.~(\ref{res:LE_LE}) and (\ref{res:LH_LE}) are justified.

\bibliography{ref_LO}

\begin{thebibliography}{36}%
\makeatletter
\providecommand \@ifxundefined [1]{%
 \@ifx{#1\undefined}
}%
\providecommand \@ifnum [1]{%
 \ifnum #1\expandafter \@firstoftwo
 \else \expandafter \@secondoftwo
 \fi
}%
\providecommand \@ifx [1]{%
 \ifx #1\expandafter \@firstoftwo
 \else \expandafter \@secondoftwo
 \fi
}%
\providecommand \natexlab [1]{#1}%
\providecommand \enquote  [1]{``#1''}%
\providecommand \bibnamefont  [1]{#1}%
\providecommand \bibfnamefont [1]{#1}%
\providecommand \citenamefont [1]{#1}%
\providecommand \href@noop [0]{\@secondoftwo}%
\providecommand \href [0]{\begingroup \@sanitize@url \@href}%
\providecommand \@href[1]{\@@startlink{#1}\@@href}%
\providecommand \@@href[1]{\endgroup#1\@@endlink}%
\providecommand \@sanitize@url [0]{\catcode `\\12\catcode `\$12\catcode
  `\&12\catcode `\#12\catcode `\^12\catcode `\_12\catcode `\%12\relax}%
\providecommand \@@startlink[1]{}%
\providecommand \@@endlink[0]{}%
\providecommand \url  [0]{\begingroup\@sanitize@url \@url }%
\providecommand \@url [1]{\endgroup\@href {#1}{\urlprefix }}%
\providecommand \urlprefix  [0]{URL }%
\providecommand \Eprint [0]{\href }%
\providecommand \doibase [0]{http://dx.doi.org/}%
\providecommand \selectlanguage [0]{\@gobble}%
\providecommand \bibinfo  [0]{\@secondoftwo}%
\providecommand \bibfield  [0]{\@secondoftwo}%
\providecommand \translation [1]{[#1]}%
\providecommand \BibitemOpen [0]{}%
\providecommand \bibitemStop [0]{}%
\providecommand \bibitemNoStop [0]{.\EOS\space}%
\providecommand \EOS [0]{\spacefactor3000\relax}%
\providecommand \BibitemShut  [1]{\csname bibitem#1\endcsname}%
\let\auto@bib@innerbib\@empty
\bibitem [{\citenamefont {Lvovsky}\ and\ \citenamefont {Raymer}(2009)}]{LR09}%
  \BibitemOpen
  \bibfield  {author} {\bibinfo {author} {\bibfnamefont {A.~I.}\ \bibnamefont
  {Lvovsky}}\ and\ \bibinfo {author} {\bibfnamefont {M.~G.}\ \bibnamefont
  {Raymer}},\ }\href {\doibase 10.1103/RevModPhys.81.299} {\bibfield  {journal}
  {\bibinfo  {journal} {Rev. Mod. Phys.}\ }\textbf {\bibinfo {volume} {81}},\
  \bibinfo {pages} {299} (\bibinfo {year} {2009})}\BibitemShut {NoStop}%
\bibitem [{\citenamefont {Ou}\ \emph {et~al.}(1992)\citenamefont {Ou},
  \citenamefont {Pereira}, \citenamefont {Kimble},\ and\ \citenamefont
  {Peng}}]{OPK92}%
  \BibitemOpen
  \bibfield  {author} {\bibinfo {author} {\bibfnamefont {Z.~Y.}\ \bibnamefont
  {Ou}}, \bibinfo {author} {\bibfnamefont {S.~F.}\ \bibnamefont {Pereira}},
  \bibinfo {author} {\bibfnamefont {H.~J.}\ \bibnamefont {Kimble}}, \ and\
  \bibinfo {author} {\bibfnamefont {K.~C.}\ \bibnamefont {Peng}},\ }\href@noop
  {} {\bibfield  {journal} {\bibinfo  {journal} {Phys. Rev. Lett.}\ }\textbf
  {\bibinfo {volume} {68}},\ \bibinfo {pages} {3663} (\bibinfo {year}
  {1992})}\BibitemShut {NoStop}%
\bibitem [{\citenamefont {Boyer}\ \emph {et~al.}(2008)\citenamefont {Boyer},
  \citenamefont {Marino}, \citenamefont {Pooser},\ and\ \citenamefont
  {Lett}}]{BMP08}%
  \BibitemOpen
  \bibfield  {author} {\bibinfo {author} {\bibfnamefont {V.}~\bibnamefont
  {Boyer}}, \bibinfo {author} {\bibfnamefont {A.~M.}\ \bibnamefont {Marino}},
  \bibinfo {author} {\bibfnamefont {R.~C.}\ \bibnamefont {Pooser}}, \ and\
  \bibinfo {author} {\bibfnamefont {P.~D.}\ \bibnamefont {Lett}},\ }\href@noop
  {} {\bibfield  {journal} {\bibinfo  {journal} {Science}\ }\textbf {\bibinfo
  {volume} {321}},\ \bibinfo {pages} {544} (\bibinfo {year}
  {2008})}\BibitemShut {NoStop}%
\bibitem [{\citenamefont {Janousek}\ \emph {et~al.}(2009)\citenamefont
  {Janousek}, \citenamefont {Wagner}, \citenamefont {Morizur}, \citenamefont
  {Treps}, \citenamefont {Lam}, \citenamefont {Harb},\ and\ \citenamefont
  {Bachor}}]{JWM09}%
  \BibitemOpen
  \bibfield  {author} {\bibinfo {author} {\bibfnamefont {J.}~\bibnamefont
  {Janousek}}, \bibinfo {author} {\bibfnamefont {K.}~\bibnamefont {Wagner}},
  \bibinfo {author} {\bibfnamefont {J.-F.}\ \bibnamefont {Morizur}}, \bibinfo
  {author} {\bibfnamefont {N.}~\bibnamefont {Treps}}, \bibinfo {author}
  {\bibfnamefont {P.~K.}\ \bibnamefont {Lam}}, \bibinfo {author} {\bibfnamefont
  {C.~C.}\ \bibnamefont {Harb}}, \ and\ \bibinfo {author} {\bibfnamefont
  {H.-A.}\ \bibnamefont {Bachor}},\ }\href@noop {} {\bibfield  {journal}
  {\bibinfo  {journal} {Nat. Photon.}\ }\textbf {\bibinfo {volume} {3}},\
  \bibinfo {pages} {399} (\bibinfo {year} {2009})}\BibitemShut {NoStop}%
\bibitem [{\citenamefont {Palomaki}\ \emph {et~al.}(2013)\citenamefont
  {Palomaki}, \citenamefont {Teufel}, \citenamefont {Simmonds},\ and\
  \citenamefont {Lehnert}}]{PTS13}%
  \BibitemOpen
  \bibfield  {author} {\bibinfo {author} {\bibfnamefont {T.~A.}\ \bibnamefont
  {Palomaki}}, \bibinfo {author} {\bibfnamefont {J.~D.}\ \bibnamefont
  {Teufel}}, \bibinfo {author} {\bibfnamefont {R.~W.}\ \bibnamefont
  {Simmonds}}, \ and\ \bibinfo {author} {\bibfnamefont {K.~W.}\ \bibnamefont
  {Lehnert}},\ }\href@noop {} {\bibfield  {journal} {\bibinfo  {journal}
  {Science}\ }\textbf {\bibinfo {volume} {342}},\ \bibinfo {pages} {710}
  (\bibinfo {year} {2013})}\BibitemShut {NoStop}%
\bibitem [{\citenamefont {Roslund}\ \emph {et~al.}(2014)\citenamefont
  {Roslund}, \citenamefont {de~Ara\'ujo}, \citenamefont {Jiang}, \citenamefont
  {Fabre},\ and\ \citenamefont {Treps}}]{RAJ14}%
  \BibitemOpen
  \bibfield  {author} {\bibinfo {author} {\bibfnamefont {J.}~\bibnamefont
  {Roslund}}, \bibinfo {author} {\bibfnamefont {R.~M.}\ \bibnamefont
  {de~Ara\'ujo}}, \bibinfo {author} {\bibfnamefont {S.}~\bibnamefont {Jiang}},
  \bibinfo {author} {\bibfnamefont {C.}~\bibnamefont {Fabre}}, \ and\ \bibinfo
  {author} {\bibfnamefont {N.}~\bibnamefont {Treps}},\ }\href@noop {}
  {\bibfield  {journal} {\bibinfo  {journal} {Nat. Photon.}\ }\textbf {\bibinfo
  {volume} {8}},\ \bibinfo {pages} {109} (\bibinfo {year} {2014})}\BibitemShut
  {NoStop}%
\bibitem [{\citenamefont {Masada}\ \emph {et~al.}(2015)\citenamefont {Masada},
  \citenamefont {Miyata}, \citenamefont {Politi}, \citenamefont {Hashimoto},
  \citenamefont {O'Brien},\ and\ \citenamefont {Furusawa}}]{MMP15}%
  \BibitemOpen
  \bibfield  {author} {\bibinfo {author} {\bibfnamefont {G.}~\bibnamefont
  {Masada}}, \bibinfo {author} {\bibfnamefont {K.}~\bibnamefont {Miyata}},
  \bibinfo {author} {\bibfnamefont {A.}~\bibnamefont {Politi}}, \bibinfo
  {author} {\bibfnamefont {T.}~\bibnamefont {Hashimoto}}, \bibinfo {author}
  {\bibfnamefont {J.~L.}\ \bibnamefont {O'Brien}}, \ and\ \bibinfo {author}
  {\bibfnamefont {A.}~\bibnamefont {Furusawa}},\ }\href@noop {} {\bibfield
  {journal} {\bibinfo  {journal} {Nat. Photon.}\ }\textbf {\bibinfo {volume}
  {9}},\ \bibinfo {pages} {316} (\bibinfo {year} {2015})}\BibitemShut {NoStop}%
\bibitem [{\citenamefont {Marino}\ \emph {et~al.}(2009)\citenamefont {Marino},
  \citenamefont {Pooser}, \citenamefont {Boyer},\ and\ \citenamefont
  {Lett}}]{MPB09}%
  \BibitemOpen
  \bibfield  {author} {\bibinfo {author} {\bibfnamefont {A.~M.}\ \bibnamefont
  {Marino}}, \bibinfo {author} {\bibfnamefont {R.~C.}\ \bibnamefont {Pooser}},
  \bibinfo {author} {\bibfnamefont {V.}~\bibnamefont {Boyer}}, \ and\ \bibinfo
  {author} {\bibfnamefont {P.~D.}\ \bibnamefont {Lett}},\ }\href@noop {}
  {\bibfield  {journal} {\bibinfo  {journal} {Nature}\ }\textbf {\bibinfo
  {volume} {457}},\ \bibinfo {pages} {859} (\bibinfo {year}
  {2009})}\BibitemShut {NoStop}%
\bibitem [{\citenamefont {Barbosa}\ \emph {et~al.}(2010)\citenamefont
  {Barbosa}, \citenamefont {Coelho}, \citenamefont {de~Faria}, \citenamefont
  {Cassemiro}, \citenamefont {Villar}, \citenamefont {Nussenzveig},\ and\
  \citenamefont {Martinelli}}]{BCF10}%
  \BibitemOpen
  \bibfield  {author} {\bibinfo {author} {\bibfnamefont {F.~A.~S.}\
  \bibnamefont {Barbosa}}, \bibinfo {author} {\bibfnamefont {A.~S.}\
  \bibnamefont {Coelho}}, \bibinfo {author} {\bibfnamefont {A.~J.}\
  \bibnamefont {de~Faria}}, \bibinfo {author} {\bibfnamefont {K.~N.}\
  \bibnamefont {Cassemiro}}, \bibinfo {author} {\bibfnamefont {A.~S.}\
  \bibnamefont {Villar}}, \bibinfo {author} {\bibfnamefont {P.}~\bibnamefont
  {Nussenzveig}}, \ and\ \bibinfo {author} {\bibfnamefont {M.}~\bibnamefont
  {Martinelli}},\ }\href@noop {} {\bibfield  {journal} {\bibinfo  {journal}
  {Nat. Photon.}\ }\textbf {\bibinfo {volume} {4}},\ \bibinfo {pages} {858}
  (\bibinfo {year} {2010})}\BibitemShut {NoStop}%
\bibitem [{\citenamefont {Jensen}\ \emph {et~al.}(2011)\citenamefont {Jensen},
  \citenamefont {Wasilewski}, \citenamefont {Krauter}, \citenamefont
  {Fernholz}, \citenamefont {Nielsen}, \citenamefont {Owari}, \citenamefont
  {Plenio}, \citenamefont {Serafini}, \citenamefont {Wolf},\ and\ \citenamefont
  {Polzik}}]{JWK11}%
  \BibitemOpen
  \bibfield  {author} {\bibinfo {author} {\bibfnamefont {K.}~\bibnamefont
  {Jensen}}, \bibinfo {author} {\bibfnamefont {W.}~\bibnamefont {Wasilewski}},
  \bibinfo {author} {\bibfnamefont {H.}~\bibnamefont {Krauter}}, \bibinfo
  {author} {\bibfnamefont {T.}~\bibnamefont {Fernholz}}, \bibinfo {author}
  {\bibfnamefont {B.~M.}\ \bibnamefont {Nielsen}}, \bibinfo {author}
  {\bibfnamefont {M.}~\bibnamefont {Owari}}, \bibinfo {author} {\bibfnamefont
  {M.~B.}\ \bibnamefont {Plenio}}, \bibinfo {author} {\bibfnamefont
  {A.}~\bibnamefont {Serafini}}, \bibinfo {author} {\bibfnamefont {M.~M.}\
  \bibnamefont {Wolf}}, \ and\ \bibinfo {author} {\bibfnamefont {E.~S.}\
  \bibnamefont {Polzik}},\ }\href@noop {} {\bibfield  {journal} {\bibinfo
  {journal} {Nat. Phys.}\ }\textbf {\bibinfo {volume} {7}},\ \bibinfo {pages}
  {13} (\bibinfo {year} {2011})}\BibitemShut {NoStop}%
\bibitem [{\citenamefont {Furusawa}\ \emph {et~al.}(1998)\citenamefont
  {Furusawa}, \citenamefont {rensen}, \citenamefont {Braunstein}, \citenamefont
  {Fuchs}, \citenamefont {Kimble},\ and\ \citenamefont {Polzik}}]{FSB98}%
  \BibitemOpen
  \bibfield  {author} {\bibinfo {author} {\bibfnamefont {A.}~\bibnamefont
  {Furusawa}}, \bibinfo {author} {\bibfnamefont {J.~L.~S.}\ \bibnamefont
  {rensen}}, \bibinfo {author} {\bibfnamefont {S.~L.}\ \bibnamefont
  {Braunstein}}, \bibinfo {author} {\bibfnamefont {C.~A.}\ \bibnamefont
  {Fuchs}}, \bibinfo {author} {\bibfnamefont {H.~J.}\ \bibnamefont {Kimble}}, \
  and\ \bibinfo {author} {\bibfnamefont {E.~S.}\ \bibnamefont {Polzik}},\
  }\href@noop {} {\bibfield  {journal} {\bibinfo  {journal} {Science}\ }\textbf
  {\bibinfo {volume} {282}},\ \bibinfo {pages} {706} (\bibinfo {year}
  {1998})}\BibitemShut {NoStop}%
\bibitem [{\citenamefont {Yokoyama}\ \emph {et~al.}(2013)\citenamefont
  {Yokoyama}, \citenamefont {Ukai}, \citenamefont {Armstrong}, \citenamefont
  {Sornphiphatphong}, \citenamefont {Kaji}, \citenamefont {Suzuki},
  \citenamefont {ichi Yoshikawa}, \citenamefont {Yonezawa}, \citenamefont
  {Menicucci},\ and\ \citenamefont {Furusawa}}]{YUS13}%
  \BibitemOpen
  \bibfield  {author} {\bibinfo {author} {\bibfnamefont {S.}~\bibnamefont
  {Yokoyama}}, \bibinfo {author} {\bibfnamefont {R.}~\bibnamefont {Ukai}},
  \bibinfo {author} {\bibfnamefont {S.~C.}\ \bibnamefont {Armstrong}}, \bibinfo
  {author} {\bibfnamefont {C.}~\bibnamefont {Sornphiphatphong}}, \bibinfo
  {author} {\bibfnamefont {T.}~\bibnamefont {Kaji}}, \bibinfo {author}
  {\bibfnamefont {S.}~\bibnamefont {Suzuki}}, \bibinfo {author} {\bibfnamefont
  {J.}~\bibnamefont {ichi Yoshikawa}}, \bibinfo {author} {\bibfnamefont
  {H.}~\bibnamefont {Yonezawa}}, \bibinfo {author} {\bibfnamefont {N.~C.}\
  \bibnamefont {Menicucci}}, \ and\ \bibinfo {author} {\bibfnamefont
  {A.}~\bibnamefont {Furusawa}},\ }\href@noop {} {\bibfield  {journal}
  {\bibinfo  {journal} {Nat. Photon.}\ }\textbf {\bibinfo {volume} {7}},\
  \bibinfo {pages} {982} (\bibinfo {year} {2013})}\BibitemShut {NoStop}%
\bibitem [{\citenamefont {Su}\ \emph {et~al.}(2013)\citenamefont {Su},
  \citenamefont {Hao}, \citenamefont {Deng}, \citenamefont {Ma}, \citenamefont
  {Wang}, \citenamefont {Jia}, \citenamefont {Xie},\ and\ \citenamefont
  {Peng}}]{SHD13}%
  \BibitemOpen
  \bibfield  {author} {\bibinfo {author} {\bibfnamefont {X.}~\bibnamefont
  {Su}}, \bibinfo {author} {\bibfnamefont {S.}~\bibnamefont {Hao}}, \bibinfo
  {author} {\bibfnamefont {X.}~\bibnamefont {Deng}}, \bibinfo {author}
  {\bibfnamefont {L.}~\bibnamefont {Ma}}, \bibinfo {author} {\bibfnamefont
  {M.}~\bibnamefont {Wang}}, \bibinfo {author} {\bibfnamefont {X.}~\bibnamefont
  {Jia}}, \bibinfo {author} {\bibfnamefont {C.}~\bibnamefont {Xie}}, \ and\
  \bibinfo {author} {\bibfnamefont {K.}~\bibnamefont {Peng}},\ }\href@noop {}
  {\bibfield  {journal} {\bibinfo  {journal} {Nat. Commun.}\ }\textbf {\bibinfo
  {volume} {4}},\ \bibinfo {pages} {2828} (\bibinfo {year} {2013})}\BibitemShut
  {NoStop}%
\bibitem [{\citenamefont {Hage}\ \emph {et~al.}(2008)\citenamefont {Hage},
  \citenamefont {Samblowski}, \citenamefont {DiGuglielmo}, \citenamefont
  {Franzen}, \citenamefont {Fiur\'as\v{c}ek},\ and\ \citenamefont
  {Schnabel}}]{HSD08}%
  \BibitemOpen
  \bibfield  {author} {\bibinfo {author} {\bibfnamefont {B.}~\bibnamefont
  {Hage}}, \bibinfo {author} {\bibfnamefont {A.}~\bibnamefont {Samblowski}},
  \bibinfo {author} {\bibfnamefont {J.}~\bibnamefont {DiGuglielmo}}, \bibinfo
  {author} {\bibfnamefont {A.}~\bibnamefont {Franzen}}, \bibinfo {author}
  {\bibfnamefont {J.}~\bibnamefont {Fiur\'as\v{c}ek}}, \ and\ \bibinfo {author}
  {\bibfnamefont {R.}~\bibnamefont {Schnabel}},\ }\href@noop {} {\bibfield
  {journal} {\bibinfo  {journal} {Nat. Phys.}\ }\textbf {\bibinfo {volume}
  {4}},\ \bibinfo {pages} {915} (\bibinfo {year} {2008})}\BibitemShut {NoStop}%
\bibitem [{\citenamefont {Gross}\ \emph {et~al.}(2011)\citenamefont {Gross},
  \citenamefont {Strobel}, \citenamefont {Nicklas}, \citenamefont {Zibold},
  \citenamefont {Bar-Gill}, \citenamefont {Kurizki},\ and\ \citenamefont
  {Oberthaler}}]{GSN11}%
  \BibitemOpen
  \bibfield  {author} {\bibinfo {author} {\bibfnamefont {C.}~\bibnamefont
  {Gross}}, \bibinfo {author} {\bibfnamefont {H.}~\bibnamefont {Strobel}},
  \bibinfo {author} {\bibfnamefont {E.}~\bibnamefont {Nicklas}}, \bibinfo
  {author} {\bibfnamefont {T.}~\bibnamefont {Zibold}}, \bibinfo {author}
  {\bibfnamefont {N.}~\bibnamefont {Bar-Gill}}, \bibinfo {author}
  {\bibfnamefont {G.}~\bibnamefont {Kurizki}}, \ and\ \bibinfo {author}
  {\bibfnamefont {M.~K.}\ \bibnamefont {Oberthaler}},\ }\href@noop {}
  {\bibfield  {journal} {\bibinfo  {journal} {Nature}\ }\textbf {\bibinfo
  {volume} {480}},\ \bibinfo {pages} {219} (\bibinfo {year}
  {2011})}\BibitemShut {NoStop}%
\bibitem [{\citenamefont {Zavatta}\ \emph {et~al.}(2011)\citenamefont
  {Zavatta}, \citenamefont {Fiur\'as\v{c}ek},\ and\ \citenamefont
  {Bellini}}]{ZFB11}%
  \BibitemOpen
  \bibfield  {author} {\bibinfo {author} {\bibfnamefont {A.}~\bibnamefont
  {Zavatta}}, \bibinfo {author} {\bibfnamefont {J.}~\bibnamefont
  {Fiur\'as\v{c}ek}}, \ and\ \bibinfo {author} {\bibfnamefont {M.}~\bibnamefont
  {Bellini}},\ }\href@noop {} {\bibfield  {journal} {\bibinfo  {journal} {Nat.
  Photon.}\ }\textbf {\bibinfo {volume} {5}},\ \bibinfo {pages} {52} (\bibinfo
  {year} {2011})}\BibitemShut {NoStop}%
\bibitem [{\citenamefont {Chrzanowski}\ \emph {et~al.}(2014)\citenamefont
  {Chrzanowski}, \citenamefont {Walk}, \citenamefont {Assad}, \citenamefont
  {Janousek}, \citenamefont {Hosseini}, \citenamefont {Ralph}, \citenamefont
  {Symul},\ and\ \citenamefont {Lam}}]{CWA14}%
  \BibitemOpen
  \bibfield  {author} {\bibinfo {author} {\bibfnamefont {H.~M.}\ \bibnamefont
  {Chrzanowski}}, \bibinfo {author} {\bibfnamefont {N.}~\bibnamefont {Walk}},
  \bibinfo {author} {\bibfnamefont {S.~M.}\ \bibnamefont {Assad}}, \bibinfo
  {author} {\bibfnamefont {J.}~\bibnamefont {Janousek}}, \bibinfo {author}
  {\bibfnamefont {S.}~\bibnamefont {Hosseini}}, \bibinfo {author}
  {\bibfnamefont {T.~C.}\ \bibnamefont {Ralph}}, \bibinfo {author}
  {\bibfnamefont {T.}~\bibnamefont {Symul}}, \ and\ \bibinfo {author}
  {\bibfnamefont {P.~K.}\ \bibnamefont {Lam}},\ }\href@noop {} {\bibfield
  {journal} {\bibinfo  {journal} {Nat. Photon.}\ }\textbf {\bibinfo {volume}
  {8}},\ \bibinfo {pages} {333} (\bibinfo {year} {2014})}\BibitemShut {NoStop}%
\bibitem [{\citenamefont {Ulanov}\ \emph {et~al.}(2015)\citenamefont {Ulanov},
  \citenamefont {Fedorov}, \citenamefont {Pushkina}, \citenamefont {Kurochkin},
  \citenamefont {Ralph},\ and\ \citenamefont {Lvovsky}}]{UFP15}%
  \BibitemOpen
  \bibfield  {author} {\bibinfo {author} {\bibfnamefont {A.~E.}\ \bibnamefont
  {Ulanov}}, \bibinfo {author} {\bibfnamefont {I.~A.}\ \bibnamefont {Fedorov}},
  \bibinfo {author} {\bibfnamefont {A.~A.}\ \bibnamefont {Pushkina}}, \bibinfo
  {author} {\bibfnamefont {Y.~V.}\ \bibnamefont {Kurochkin}}, \bibinfo {author}
  {\bibfnamefont {T.~C.}\ \bibnamefont {Ralph}}, \ and\ \bibinfo {author}
  {\bibfnamefont {A.~I.}\ \bibnamefont {Lvovsky}},\ }\href@noop {} {\bibfield
  {journal} {\bibinfo  {journal} {Nat. Photon.}\ }\textbf {\bibinfo {volume}
  {9}},\ \bibinfo {pages} {764} (\bibinfo {year} {2015})}\BibitemShut {NoStop}%
\bibitem [{\citenamefont {Grosshans}\ \emph {et~al.}(2003)\citenamefont
  {Grosshans}, \citenamefont {Assche}, \citenamefont {Wenger}, \citenamefont
  {Brouri}, \citenamefont {Cerf},\ and\ \citenamefont {Grangier}}]{GAW03}%
  \BibitemOpen
  \bibfield  {author} {\bibinfo {author} {\bibfnamefont {F.}~\bibnamefont
  {Grosshans}}, \bibinfo {author} {\bibfnamefont {G.~V.}\ \bibnamefont
  {Assche}}, \bibinfo {author} {\bibfnamefont {J.}~\bibnamefont {Wenger}},
  \bibinfo {author} {\bibfnamefont {R.}~\bibnamefont {Brouri}}, \bibinfo
  {author} {\bibfnamefont {N.~J.}\ \bibnamefont {Cerf}}, \ and\ \bibinfo
  {author} {\bibfnamefont {P.}~\bibnamefont {Grangier}},\ }\href@noop {}
  {\bibfield  {journal} {\bibinfo  {journal} {Nature}\ }\textbf {\bibinfo
  {volume} {421}},\ \bibinfo {pages} {238} (\bibinfo {year}
  {2003})}\BibitemShut {NoStop}%
\bibitem [{\citenamefont {Pirandola}\ \emph {et~al.}(2008)\citenamefont
  {Pirandola}, \citenamefont {Mancini}, \citenamefont {Lloyd},\ and\
  \citenamefont {Braunstein}}]{PML08}%
  \BibitemOpen
  \bibfield  {author} {\bibinfo {author} {\bibfnamefont {S.}~\bibnamefont
  {Pirandola}}, \bibinfo {author} {\bibfnamefont {S.}~\bibnamefont {Mancini}},
  \bibinfo {author} {\bibfnamefont {S.}~\bibnamefont {Lloyd}}, \ and\ \bibinfo
  {author} {\bibfnamefont {S.~L.}\ \bibnamefont {Braunstein}},\ }\href@noop {}
  {\bibfield  {journal} {\bibinfo  {journal} {Nat. Photon.}\ }\textbf {\bibinfo
  {volume} {4}},\ \bibinfo {pages} {726} (\bibinfo {year} {2008})}\BibitemShut
  {NoStop}%
\bibitem [{\citenamefont {Madsen}\ \emph {et~al.}(2012)\citenamefont {Madsen},
  \citenamefont {Usenko}, \citenamefont {Lassen}, \citenamefont {Filip},\ and\
  \citenamefont {Andersen}}]{MUL12}%
  \BibitemOpen
  \bibfield  {author} {\bibinfo {author} {\bibfnamefont {L.~S.}\ \bibnamefont
  {Madsen}}, \bibinfo {author} {\bibfnamefont {V.~C.}\ \bibnamefont {Usenko}},
  \bibinfo {author} {\bibfnamefont {M.}~\bibnamefont {Lassen}}, \bibinfo
  {author} {\bibfnamefont {R.}~\bibnamefont {Filip}}, \ and\ \bibinfo {author}
  {\bibfnamefont {U.~L.}\ \bibnamefont {Andersen}},\ }\href@noop {} {\bibfield
  {journal} {\bibinfo  {journal} {Nat. Commun.}\ }\textbf {\bibinfo {volume}
  {3}},\ \bibinfo {pages} {1083} (\bibinfo {year} {2012})}\BibitemShut
  {NoStop}%
\bibitem [{\citenamefont {Jouguet}\ \emph {et~al.}(2013)\citenamefont
  {Jouguet}, \citenamefont {Kunz-Jacques}, \citenamefont {Leverrier},
  \citenamefont {Grangier},\ and\ \citenamefont {Diamanti}}]{JJL13}%
  \BibitemOpen
  \bibfield  {author} {\bibinfo {author} {\bibfnamefont {P.}~\bibnamefont
  {Jouguet}}, \bibinfo {author} {\bibfnamefont {S.}~\bibnamefont
  {Kunz-Jacques}}, \bibinfo {author} {\bibfnamefont {A.}~\bibnamefont
  {Leverrier}}, \bibinfo {author} {\bibfnamefont {P.}~\bibnamefont {Grangier}},
  \ and\ \bibinfo {author} {\bibfnamefont {E.}~\bibnamefont {Diamanti}},\
  }\href@noop {} {\bibfield  {journal} {\bibinfo  {journal} {Nat. Photon.}\
  }\textbf {\bibinfo {volume} {7}},\ \bibinfo {pages} {378} (\bibinfo {year}
  {2013})}\BibitemShut {NoStop}%
\bibitem [{\citenamefont {Pirandola}\ \emph {et~al.}(2015)\citenamefont
  {Pirandola}, \citenamefont {Ottaviani}, \citenamefont {Spedalieri},
  \citenamefont {Weedbrook}, \citenamefont {Braunstein}, \citenamefont {Lloyd},
  \citenamefont {Gehring}, \citenamefont {Jacobsen},\ and\ \citenamefont
  {Andersen}}]{POS15}%
  \BibitemOpen
  \bibfield  {author} {\bibinfo {author} {\bibfnamefont {S.}~\bibnamefont
  {Pirandola}}, \bibinfo {author} {\bibfnamefont {C.}~\bibnamefont
  {Ottaviani}}, \bibinfo {author} {\bibfnamefont {G.}~\bibnamefont
  {Spedalieri}}, \bibinfo {author} {\bibfnamefont {C.}~\bibnamefont
  {Weedbrook}}, \bibinfo {author} {\bibfnamefont {S.~L.}\ \bibnamefont
  {Braunstein}}, \bibinfo {author} {\bibfnamefont {S.}~\bibnamefont {Lloyd}},
  \bibinfo {author} {\bibfnamefont {T.}~\bibnamefont {Gehring}}, \bibinfo
  {author} {\bibfnamefont {C.~S.}\ \bibnamefont {Jacobsen}}, \ and\ \bibinfo
  {author} {\bibfnamefont {U.~L.}\ \bibnamefont {Andersen}},\ }\href@noop {}
  {\bibfield  {journal} {\bibinfo  {journal} {Nat. Photon.}\ }\textbf {\bibinfo
  {volume} {9}},\ \bibinfo {pages} {397} (\bibinfo {year} {2015})}\BibitemShut
  {NoStop}%
\bibitem [{\citenamefont {Lo}\ \emph {et~al.}(2014)\citenamefont {Lo},
  \citenamefont {Curty},\ and\ \citenamefont {Tamaki}}]{LCT14}%
  \BibitemOpen
  \bibfield  {author} {\bibinfo {author} {\bibfnamefont {H.-K.}\ \bibnamefont
  {Lo}}, \bibinfo {author} {\bibfnamefont {M.}~\bibnamefont {Curty}}, \ and\
  \bibinfo {author} {\bibfnamefont {K.}~\bibnamefont {Tamaki}},\ }\href@noop {}
  {\bibfield  {journal} {\bibinfo  {journal} {Nat. Photon.}\ }\textbf {\bibinfo
  {volume} {8}},\ \bibinfo {pages} {595} (\bibinfo {year} {2014})}\BibitemShut
  {NoStop}%
\bibitem [{\citenamefont {Ma}\ \emph {et~al.}(2013)\citenamefont {Ma},
  \citenamefont {Sun}, \citenamefont {Jiang},\ and\ \citenamefont
  {Liang}}]{MSJ13}%
  \BibitemOpen
  \bibfield  {author} {\bibinfo {author} {\bibfnamefont {X.-C.}\ \bibnamefont
  {Ma}}, \bibinfo {author} {\bibfnamefont {S.-H.}\ \bibnamefont {Sun}},
  \bibinfo {author} {\bibfnamefont {M.-S.}\ \bibnamefont {Jiang}}, \ and\
  \bibinfo {author} {\bibfnamefont {L.-M.}\ \bibnamefont {Liang}},\ }\href@noop
  {} {\bibfield  {journal} {\bibinfo  {journal} {Phys. Rev. A}\ }\textbf
  {\bibinfo {volume} {88}},\ \bibinfo {pages} {022339} (\bibinfo {year}
  {2013})}\BibitemShut {NoStop}%
\bibitem [{\citenamefont {Qi}\ \emph {et~al.}(2015)\citenamefont {Qi},
  \citenamefont {Lougovski}, \citenamefont {Pooser}, \citenamefont {Grice},\
  and\ \citenamefont {Bobrek}}]{PhysRevX.5.041009}%
  \BibitemOpen
  \bibfield  {author} {\bibinfo {author} {\bibfnamefont {B.}~\bibnamefont
  {Qi}}, \bibinfo {author} {\bibfnamefont {P.}~\bibnamefont {Lougovski}},
  \bibinfo {author} {\bibfnamefont {R.}~\bibnamefont {Pooser}}, \bibinfo
  {author} {\bibfnamefont {W.}~\bibnamefont {Grice}}, \ and\ \bibinfo {author}
  {\bibfnamefont {M.}~\bibnamefont {Bobrek}},\ }\href {\doibase
  10.1103/PhysRevX.5.041009} {\bibfield  {journal} {\bibinfo  {journal} {Phys.
  Rev. X}\ }\textbf {\bibinfo {volume} {5}},\ \bibinfo {pages} {041009}
  (\bibinfo {year} {2015})}\BibitemShut {NoStop}%
\bibitem [{Note1()}]{Note1}%
  \BibitemOpen
  \bibinfo {note} {Even if the $-\theta $-phase shift is performed on the
  signal pulse instead of $\theta $-phase shift on the LO pulse, the output of
  the implementation is unchanged. This is so because the measurement commutes
  with the operator of the total photon number in the signal pulse and the LO
  pulse.}\BibitemShut {Stop}%
\bibitem [{\citenamefont {Horodecki}\ \emph {et~al.}(2009)\citenamefont
  {Horodecki}, \citenamefont {Horodecki}, \citenamefont {Horodecki},\ and\
  \citenamefont {Horodecki}}]{HHH09}%
  \BibitemOpen
  \bibfield  {author} {\bibinfo {author} {\bibfnamefont {R.}~\bibnamefont
  {Horodecki}}, \bibinfo {author} {\bibfnamefont {P.}~\bibnamefont
  {Horodecki}}, \bibinfo {author} {\bibfnamefont {M.}~\bibnamefont
  {Horodecki}}, \ and\ \bibinfo {author} {\bibfnamefont {K.}~\bibnamefont
  {Horodecki}},\ }\href@noop {} {\bibfield  {journal} {\bibinfo  {journal}
  {Rev. Mod. Phys.}\ }\textbf {\bibinfo {volume} {81}},\ \bibinfo {pages} {865}
  (\bibinfo {year} {2009})}\BibitemShut {NoStop}%
\bibitem [{\citenamefont {Simon}(2000)}]{S00}%
  \BibitemOpen
  \bibfield  {author} {\bibinfo {author} {\bibfnamefont {R.}~\bibnamefont
  {Simon}},\ }\href@noop {} {\bibfield  {journal} {\bibinfo  {journal} {Phys.
  Rev. Lett.}\ }\textbf {\bibinfo {volume} {84}},\ \bibinfo {pages} {2724}
  (\bibinfo {year} {2000})}\BibitemShut {NoStop}%
\bibitem [{\citenamefont {Duan}\ \emph {et~al.}(2000)\citenamefont {Duan},
  \citenamefont {Giedke}, \citenamefont {Cirac},\ and\ \citenamefont
  {Zoller}}]{DGC00}%
  \BibitemOpen
  \bibfield  {author} {\bibinfo {author} {\bibfnamefont {L.-M.}\ \bibnamefont
  {Duan}}, \bibinfo {author} {\bibfnamefont {G.}~\bibnamefont {Giedke}},
  \bibinfo {author} {\bibfnamefont {J.~I.}\ \bibnamefont {Cirac}}, \ and\
  \bibinfo {author} {\bibfnamefont {P.}~\bibnamefont {Zoller}},\ }\href@noop {}
  {\bibfield  {journal} {\bibinfo  {journal} {Phys. Rev. Lett.}\ }\textbf
  {\bibinfo {volume} {84}},\ \bibinfo {pages} {2722} (\bibinfo {year}
  {2000})}\BibitemShut {NoStop}%
\bibitem [{\citenamefont {Hyllus}\ and\ \citenamefont {Eisert}(2006)}]{HE06}%
  \BibitemOpen
  \bibfield  {author} {\bibinfo {author} {\bibfnamefont {P.}~\bibnamefont
  {Hyllus}}\ and\ \bibinfo {author} {\bibfnamefont {J.}~\bibnamefont
  {Eisert}},\ }\href@noop {} {\bibfield  {journal} {\bibinfo  {journal} {New J.
  Phys.}\ }\textbf {\bibinfo {volume} {8}},\ \bibinfo {pages} {51} (\bibinfo
  {year} {2006})}\BibitemShut {NoStop}%
\bibitem [{\citenamefont {Serafini}(2006)}]{S06}%
  \BibitemOpen
  \bibfield  {author} {\bibinfo {author} {\bibfnamefont {A.}~\bibnamefont
  {Serafini}},\ }\href@noop {} {\bibfield  {journal} {\bibinfo  {journal}
  {Phys. Rev. Lett.}\ }\textbf {\bibinfo {volume} {96}},\ \bibinfo {pages}
  {110402} (\bibinfo {year} {2006})}\BibitemShut {NoStop}%
\bibitem [{\citenamefont {Garc\'ia-Patr\'on}\ and\ \citenamefont
  {Cerf}(2006)}]{GC06}%
  \BibitemOpen
  \bibfield  {author} {\bibinfo {author} {\bibfnamefont {R.}~\bibnamefont
  {Garc\'ia-Patr\'on}}\ and\ \bibinfo {author} {\bibfnamefont {N.~J.}\
  \bibnamefont {Cerf}},\ }\href@noop {} {\bibfield  {journal} {\bibinfo
  {journal} {Phys. Rev. Lett.}\ }\textbf {\bibinfo {volume} {97}},\ \bibinfo
  {pages} {190503} (\bibinfo {year} {2006})}\BibitemShut {NoStop}%
\bibitem [{\citenamefont {Devetak}\ and\ \citenamefont
  {A.Winter}(2004)}]{DW04}%
  \BibitemOpen
  \bibfield  {author} {\bibinfo {author} {\bibfnamefont {I.}~\bibnamefont
  {Devetak}}\ and\ \bibinfo {author} {\bibnamefont {A.Winter}},\ }\href@noop {}
  {\bibfield  {journal} {\bibinfo  {journal} {Phys. Rev. Lett.}\ }\textbf
  {\bibinfo {volume} {93}},\ \bibinfo {pages} {080501} (\bibinfo {year}
  {2004})}\BibitemShut {NoStop}%
\bibitem [{\citenamefont {Yuen}\ and\ \citenamefont {Chan}(1983)}]{Yuen:83}%
  \BibitemOpen
  \bibfield  {author} {\bibinfo {author} {\bibfnamefont {H.~P.}\ \bibnamefont
  {Yuen}}\ and\ \bibinfo {author} {\bibfnamefont {V.~W.~S.}\ \bibnamefont
  {Chan}},\ }\href {\doibase 10.1364/OL.8.000177} {\bibfield  {journal}
  {\bibinfo  {journal} {Opt. Lett.}\ }\textbf {\bibinfo {volume} {8}},\
  \bibinfo {pages} {177} (\bibinfo {year} {1983})}\BibitemShut {NoStop}%
\bibitem [{\citenamefont {Schumaker}(1984)}]{Schumaker:84}%
  \BibitemOpen
  \bibfield  {author} {\bibinfo {author} {\bibfnamefont {B.~L.}\ \bibnamefont
  {Schumaker}},\ }\href {\doibase 10.1364/OL.9.000189} {\bibfield  {journal}
  {\bibinfo  {journal} {Opt. Lett.}\ }\textbf {\bibinfo {volume} {9}},\
  \bibinfo {pages} {189} (\bibinfo {year} {1984})}\BibitemShut {NoStop}%
\end{thebibliography}%

\end{document}